\documentclass[twocolumn]{aastex62}

\graphicspath{{./}{figures/}}

\submitjournal{ApJ}

\def\ra#1#2#3{#1$^{\rm h}$#2$^{\rm m}$#3$^{\rm s}$}
\def\dec#1#2#3{$#1^\circ#2'#3''$}

\shorttitle{Double-peaked Ca-rich gap transient iPTF\,16hgs}
\shortauthors{K. De et al.}

\begin{document}

\title{\MakeLowercase{i}PTF \MakeLowercase{16hgs}: A double-peaked C\MakeLowercase{a}-rich gap transient in a metal poor, star forming dwarf galaxy}

\correspondingauthor{Kishalay De}
\email{kde@astro.caltech.edu}

\author{Kishalay De}
\affil{Cahill Centre for Astrophysics, California Institute of Technology, 1200 East California Boulevard, Pasadena, CA 91125, USA.}

\author{Mansi M. Kasliwal}
\affil{Cahill Centre for Astrophysics, California Institute of Technology, 1200 East California Boulevard, Pasadena, CA 91125, USA.}

\author{Therese Cantwell}
\affil{Jodrell Bank Centre for Astrophysics, Alan Turing Building, School of Physics and Astronomy, The University of Manchester, Oxford Road, Manchester M139PL, UK.}

\author{Yi Cao}
\affil{Department of Astronomy, University of Washington, Box 351580, Seattle, WA 98195-1580, USA.}

\author{S. Bradley Cenko}
\affil{Astrophysics Science Division, NASA Goddard Space Flight Center, Mail Code 661, Greenbelt, MD 20771, USA.}
\affil{Joint Space-Science Institute, University of Maryland, College Park, MD 20742, USA.}

\author{Avishay Gal-Yam}
\affil{Department of Particle Physics and Astrophysics, Faculty of Physics, The Weizmann Institute of Science, Rehovot 76100, Israel.}

\author{Joel Johansson}
\affil{Department of Physics and Astronomy, Division of Astronomy and Space Physics, Uppsala University, Box 516, SE 751 20 Uppsala, Sweden.}

\author{Albert Kong}
\affil{Institute of Astronomy and Department of Physics, National Tsing Hua University, Hsinchu 30013, Taiwan.}

\author{Shrinivas R. Kulkarni}
\affil{Cahill Centre for Astrophysics, California Institute of Technology, 1200 East California Boulevard, Pasadena, CA 91125, USA.}

\author{Ragnhild Lunnan}
\affil{Oskar Klein Centre, Department of Astronomy, Stockholm University, 106 91 Stockholm, Sweden.}

\author{Frank Masci}
\affil{Infrared Processing and Analysis Center, California Institute of Technology, MS 100-22, Pasadena, CA 91125, USA.}

\author{Matt Matuszewski}
\affil{Cahill Centre for Astrophysics, California Institute of Technology, 1200 East California Boulevard, Pasadena, CA 91125, USA.}

\author{Kunal P. Mooley}
\affil{Cahill Centre for Astrophysics, California Institute of Technology, 1200 East California Boulevard, Pasadena, CA 91125, USA.}

\author{James D. Neill}
\affil{Cahill Centre for Astrophysics, California Institute of Technology, 1200 East California Boulevard, Pasadena, CA 91125, USA.}

\author{Peter E. Nugent}
\affil{Lawrence Berkeley National Laboratory, Berkeley, California 94720, USA.}
\affil{Department of Astronomy, University of California, Berkeley, CA, 94720-3411, USA.}

\author{Eran O. Ofek}
\affil{Department of Particle Physics and Astrophysics, Faculty of Physics, The Weizmann Institute of Science, Rehovot 76100, Israel.}

\author{Yvette Perrott}
\affil{Astrophysics Group, Cavendish Laboratory, 19 J. J. Thomson Avenue, Cambridge CB3 0HE, UK.}

\author{Umaa D. Rebbapragada}
\affil{Jet Propulsion Laboratory, California Institute of Technology, Pasadena, CA 91109, USA.}

\author{Adam Rubin}
\affil{Department of Particle Physics and Astrophysics, Faculty of Physics, The Weizmann Institute of Science, Rehovot 76100, Israel.}

\author{Donal O' Sullivan}
\affil{Cahill Centre for Astrophysics, California Institute of Technology, 1200 East California Boulevard, Pasadena, CA 91125, USA.}

\author{Ofer Yaron}
\affil{Department of Particle Physics and Astrophysics, Faculty of Physics, The Weizmann Institute of Science, Rehovot 76100, Israel.}

\begin{abstract}

Calcium rich gap transients represent an intriguing new class of faint and fast evolving supernovae that exhibit strong [Ca II] emission in their nebular phase spectra. In this paper, we present the discovery and follow-up observations of iPTF\,16hgs -- an intermediate luminosity and fast evolving transient that exhibited a double peaked light curve. Exhibiting a typical Type Ib spectrum in the photospheric phase and an early transition to a [Ca II] dominated nebular phase, we show that iPTF\,16hgs shows properties consistent with the class of Ca-rich gap transients, with two interesting exceptions. First, while the second peak of the light curve is similar to other Ca-rich gap transients (suggesting $M_{ej} \approx 0.4$ M$_{\odot}$ and peak luminosity $\approx 3 \times 10^{41}$ ergs s$^{-1}$), we show that the first blue and fast declining (over $\approx 2$ days) peak is unique to this source. Second, with Integral Field Unit observations of the host galaxy, we find that iPTF\,16hgs occurred in the outskirts (projected offset of $\approx 6$ kpc $\approx 1.9$ R$_{\textrm{eff}}$) of a low metallicity ($\approx 0.4$ Z$_{\odot}$), star forming, dwarf spiral galaxy. Using deep late-time VLA and uGMRT observations, we place stringent limits on the local environment of the source, ruling out a large parameter space of circumstellar densities and mass loss environments of the progenitor. If iPTF\,16hgs shares explosion physics with the class of Ca-rich gap transients, we suggest that the presence of the first peak can be explained by enhanced mixing of $0.01$ M$_\odot$ of $^{56}$Ni into the outer layers the ejecta, reminiscent of some models of He-shell detonations on WDs. On the other hand, if iPTF\,16hgs is physically unrelated to the class, the first peak is consistent with shock cooling emission (of an envelope with a mass of $\approx 0.08$ M$_{\odot}$ and radius of $\approx 13$ R$_{\odot}$) associated with a core-collapse explosion of a highly stripped massive star in a close binary system.

\end{abstract}


\section{Introduction}

The luminosity `gap' between novae and supernovae (SNe) has been populated with a variety of new classes of transients with the advent of wide field and high cadence transient surveys in recent years \citep{Kasliwal2012b}. In particular, Calcium-rich gap transients have emerged as an intriguing new class of faint explosions, proposed to be defined by their (1) intermediate luminosity (`gap' transients), (2) faster photometric evolution (rise and decline) than normal SNe, (3) photospheric velocities comparable to those of SNe, (4) rapid evolution to the nebular phase, and (5) a nebular spectrum dominated by calcium emission \citep{Kasliwal2012a}. Although not used as a defining characteristic of this class, these transients have been almost exclusively found at very large projected offsets from their host galaxies, potentially suggesting their association with old progenitor populations that have traveled far away from their host galaxies \citep{Lyman2014,Lunnan2017}.\\

Following the discovery of the prototype event of this class SN2005E (\citealt{Perets2010}), five additional events have been reported by the Palomar Transient Factory (PTF; \citealt{Law2009}): PTF 09dav (\citealt{Sullivan2011}; \citealt{Kasliwal2012a}), PTF 10iuv, PTF 11bij (\citealt{Kasliwal2012a}, PTF 11kmb and PTF 12bho (\citealt{Lunnan2017}); one event reported by PESSTO (SN 2012hn; \citealt{Valenti2014}) as well as one source found from archival observations (SN 2007ke; \citealt{Kasliwal2012a}). Additionally, \citealt{Milisavljevic2017} reported follow-up observations of iPTF 15eqv, a Type IIb SN discovered in a star forming galaxy, and classified it as `Calcium-rich' based on its nebular spectrum, although photometrically it is not a member of this class owing to its high peak luminosity and slow light curve evolution (with an unconstrained rise time to peak). \\

While the properties of all confirmed Ca-rich gap transients are consistent with the aforementioned criteria, there remain several differences in other observed properties of these sources. For example, the photospheric spectra of these sources show significant diversity, although all but one of these events show He-rich spectra akin to Type Ib SNe at peak light \citep{Filippenko1997,Gal-Yam2017}. Such diversity is potentially indicative of heterogeneity in the progenitor channels, as previously suggested in several studies \citep{Sell2015,Lunnan2017}. Nevertheless, \citealt{Lunnan2017} show that the light curves of these sources form a fairly uniform class, showing very similar rise and decay characteristics. However, the small number of known events leave considerable uncertainty on the spread of the intrinsic properties of this unique class of events.\\

The progenitor systems of Ca-rich gap transients remain largely uncertain, although it has been generally suggested that they are associated with very old progenitor systems. The majority of evidence arises from their remote locations, as well as associated stringent limits on the presence of stellar associations at the locations of these transients \citep{Lyman2014,Lyman2016b,Lunnan2017}. Additionally, the old stellar populations of their host galaxies as well as the lack of nearby star formation has been used to argue against massive star progenitor channels \citep{Kasliwal2012a,Lunnan2017}. Thus, the several proposed explosion channels for these events arise from old binary progenitor systems, such as tidal detonations of low mass He WDs by neutron stars or black holes \citep{Metzger2012,Sell2015,Macleod2016,Margalit2016}, and He shell detonations on the surface of C/O white dwarfs (WDs) (also known as .Ia detonations; \citealt{Bildsten2007,Shen2010,Perets2010,Waldman2011,Dessart2015a}), possibly induced by hardening of a WD-WD binary due to gravitational interactions with a central super-massive black hole in the host galaxy \citep{Foley2015}. Although their old environments argue against scenarios involving the collapse of a massive star, those found in star forming environments could be associated with highly stripped massive star progenitors that collapse to produce fast transients with ejecta masses of $\approx 0.2 - 0.4 $ M$_{\odot}$ \citep{Tauris2015,Moriya2017}. \\

In this paper, we present the discovery and follow-up observations of a unique Ca-rich gap transient, iPTF\,16hgs, which exhibited a double peaked light curve. The paper is organized as follows. We describe the discovery and follow-up observations of the transient in Section \ref{sec:observations}. We analyze the photometric and spectroscopic properties of the source in the context of Ca-rich gap transients in Section \ref{sec:analysis}. We model the unique light curve of the transient in Section \ref{sec:doubleModel}. Section \ref{sec:radioAn} presents an analysis of the radio observations of the source in the context of models of a SN shock interacting with a CSM as well as a tidal detonation event. We discuss the host galaxy and local explosion environment of the transient in Section \ref{sec:hostEnv}. We end with a broader discussion of the properties of Ca-rich gap transients in the context of this discovery in Section \ref{sec:discussion}. Calculations in this paper assume a $\Lambda$CDM cosmology with $H_0$ = 70 km s$^{-1}$ Mpc$^{-1}$, $\Omega_M$=0.27 and $\Omega_{\Lambda}$=0.73 \citep{Komatsu2011}.

\section{Observations}
\label{sec:observations}
\subsection{Discovery and classification}
iPTF 16hgs (= SN 2016hgs) was discovered by the intermediate Palomar Transient Factory (iPTF; \citealt{Law2009,Rau2009,Cao2016,Masci2017}) and was detected first in $r$ band photometry taken with the CFH12K 96-Megapixel camera \citep{Rahmer2008, Law2010} mounted on the 48 inch Samuel Oschin Telescope at Palomar Observatory (P48), on 2016 October 20.32 (MJD 57681.32) \footnote{UTC times are used throughout this paper} at J2000 coordinates $\alpha =$ \ra{00}{50}{51.39}. $\delta =$ +\dec{27}{22}{48.0}. The source was discovered at an apparent magnitude of $r \approx 18.9$ mag, while it was not detected on 2016 October 8.01 (MJD 57669.01; 12.31 days before discovery) up to a limiting magnitude of $r \geq$ 20.8. \\

\begin{figure}
\includegraphics[width=\columnwidth]{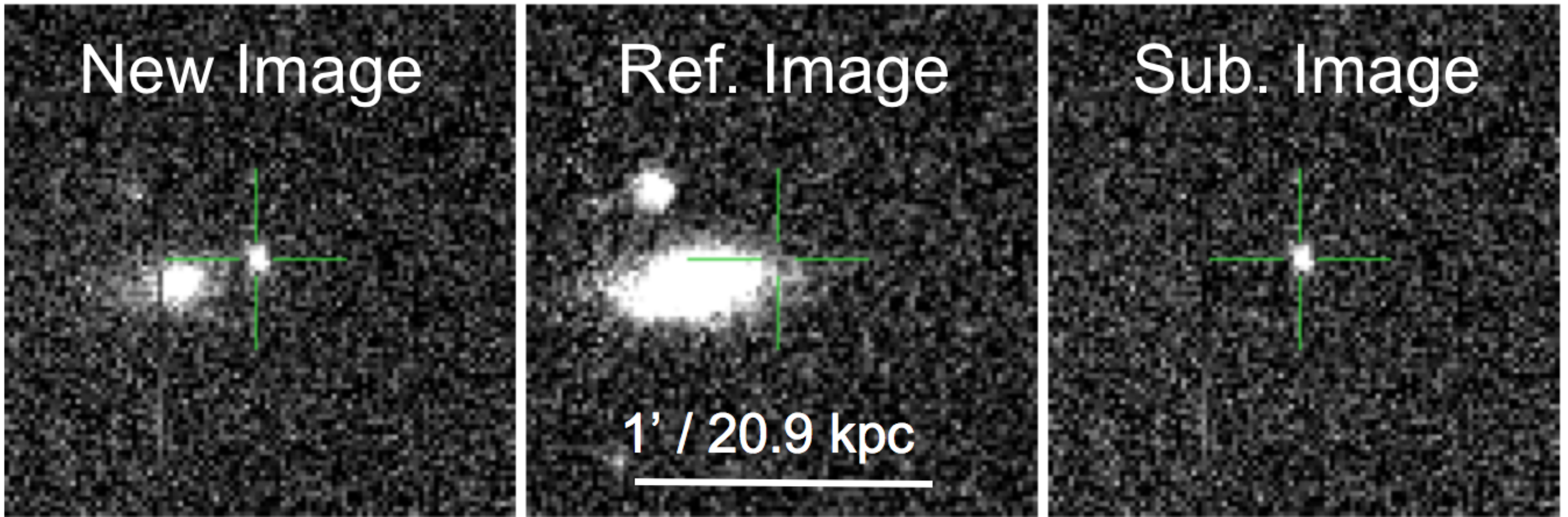}
\caption{Discovery field of iPTF\,16hgs. Shown here are the discovery (left), reference (middle) and subtracted transient image (right) of iPTF\,16hgs from the Palomar 48-inch telescope in $r$ band. The location of the transient is highlighted by the green cross.\\}
\label{fig:16hgs_discovery}
\end{figure}

The transient was found in the outskirts of a nearly edge-on spiral host galaxy with a photo-z of 0.017, and at a projected offset of $\approx 17\arcsec$ ($\approx 1.9$ R$_{eff}$) from the nucleus (Figure \ref{fig:16hgs_discovery}). We obtained a spectrum of the source on 2016 October 22 with the Discovery Channel Telescope (DCT), to find a Type Ib like SN spectrum, similar to the pre-peak photospheric spectrum of the Ca-rich transient PTF 10iuv for the assumed photo-z of the host galaxy. A subsequent spectrum of the apparent host galaxy confirmed a redshift of z = 0.017, corresponding to a luminosity distance of D$_L$ = 73.8 Mpc.  Subsequent follow-up photometry and spectroscopy of the transient revealed that the source exhibited a faint peak absolute magnitude ($M_r \approx -15.5$) and early transition (at $\approx +30$ days) to a nebular phase dominated by [Ca II] emission, thus leading to its classification as a Ca-rich gap transient.\\

\subsection{Photometry}

We obtained $r$ and $g$ band photometry of iPTF 16hgs with the P48 CFH12K camera, along with $gri$ band photometry with the automated 60-inch telescope at Palomar (P60; \citealt{Cenko2006}). PTF image reduction is presented in \citealt{Laher2014} and its photometric calibration and magnitude system is described in \citealt{Ofek2012}. P48 images were reduced with the Palomar Transient Factory Image Differencing and Extraction (PTFIDE) pipeline \citep{Masci2017}, which performs host subtracted point spread function (PSF) photometry, while the P60 images were reduced using the pipeline described in \citealt{Fremling2016}. We correct all our photometry for galactic extinction for $E(B-V) = 0.056$ from the maps of \citealt{Schlafly2011}. We do not correct for any additional host extinction since we do not detect any Na D absorption at the host redshift in our spectra (Section \ref{sec:spectroscopy}).\\

\begin{figure}
\centering
\includegraphics[width=\columnwidth]{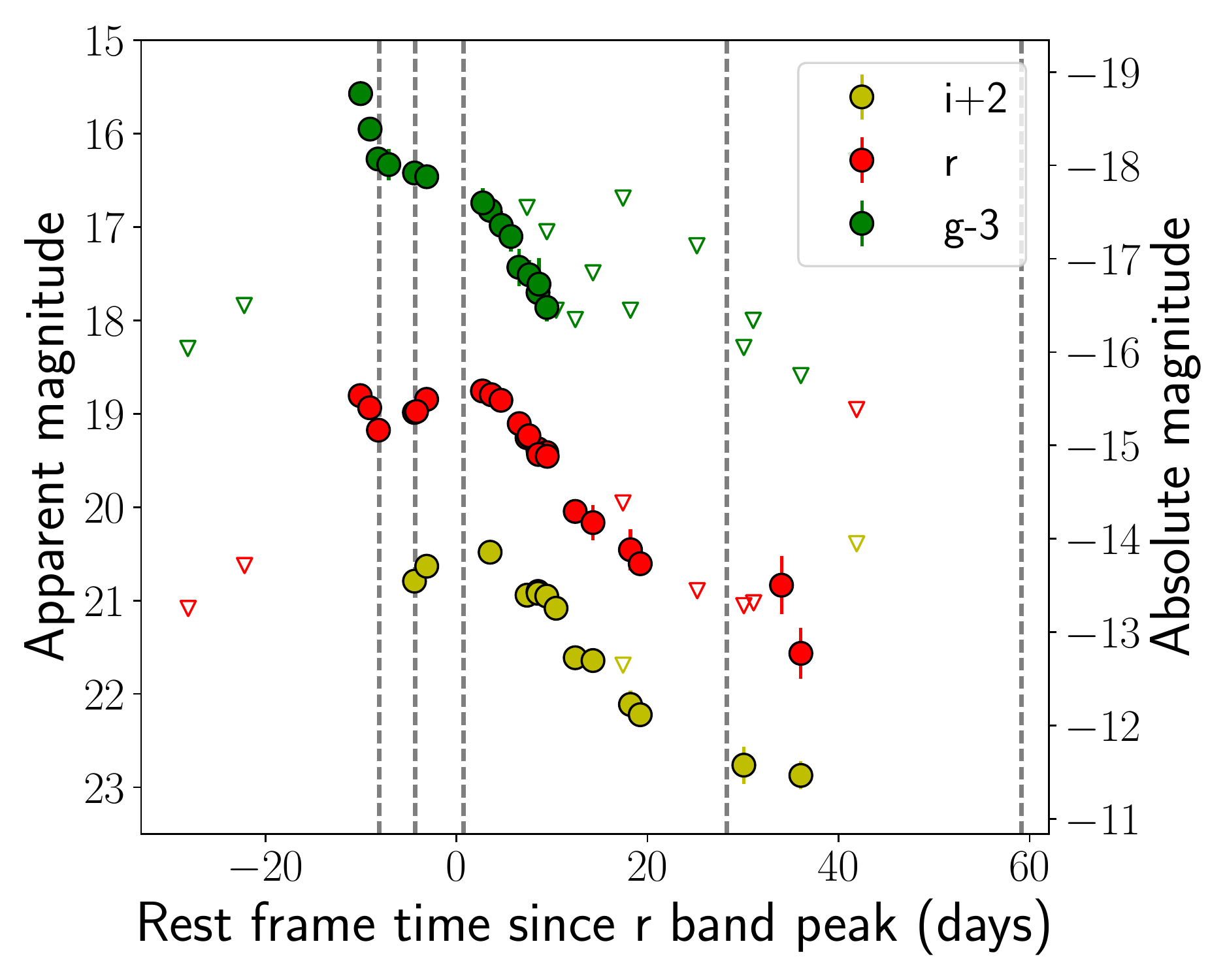}
\caption{Multicolor light curves of iPTF 16hgs. The filled circles denote magnitudes where the source was detected while the open triangles indicate epochs where only upper limits where obtained. The dashed lines denote epochs of spectroscopy.\\}
\label{fig:16hgs_photometry}
\end{figure}

We show the multi-color light curves for iPTF 16hgs in Figure \ref{fig:16hgs_photometry}, while the data are presented in Table \ref{tab:lc}. As shown, the source exhibited a double peaked light curve in all photometric bands where we had early time coverage. For all subsequent discussions, we refer to the second peak of the light curve as the main peak as well as specify all observation phases with respect to this peak.

\subsection{Spectroscopy}
\label{sec:spectroscopy}

\begin{figure}
\centering
\includegraphics[width=\columnwidth]{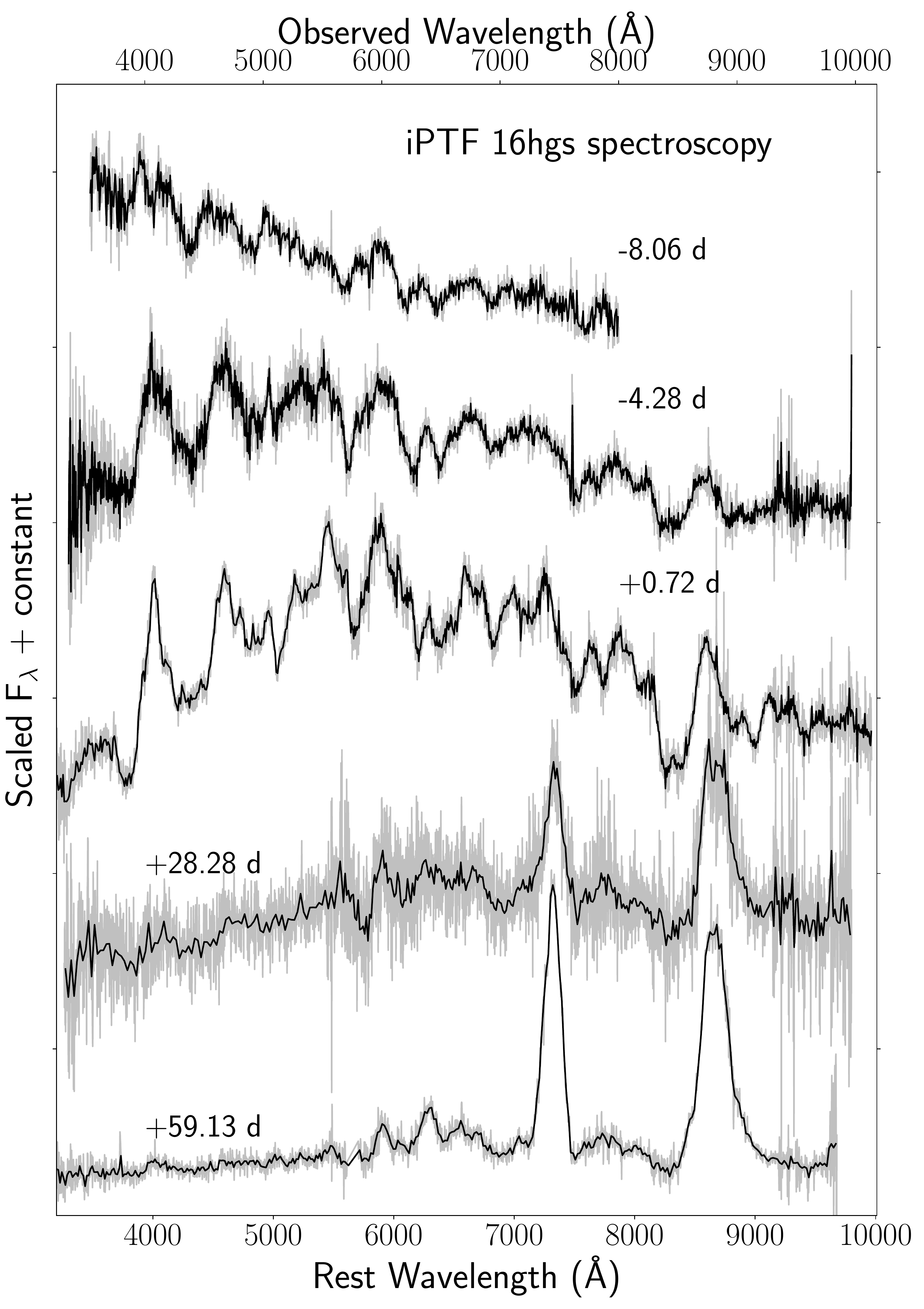}
\caption{Spectroscopic sequence for iPTF 16hgs. The black lines indicate binned spectra while the gray lines show the unbinned spectra. The phase of the light curve at the time of the spectrum (with respect to $r$ band peak) is indicated along side each spectrum.}
\label{fig:16hgs_spectra}
\end{figure}

\begin{figure}
\centering
\includegraphics[width=\columnwidth]{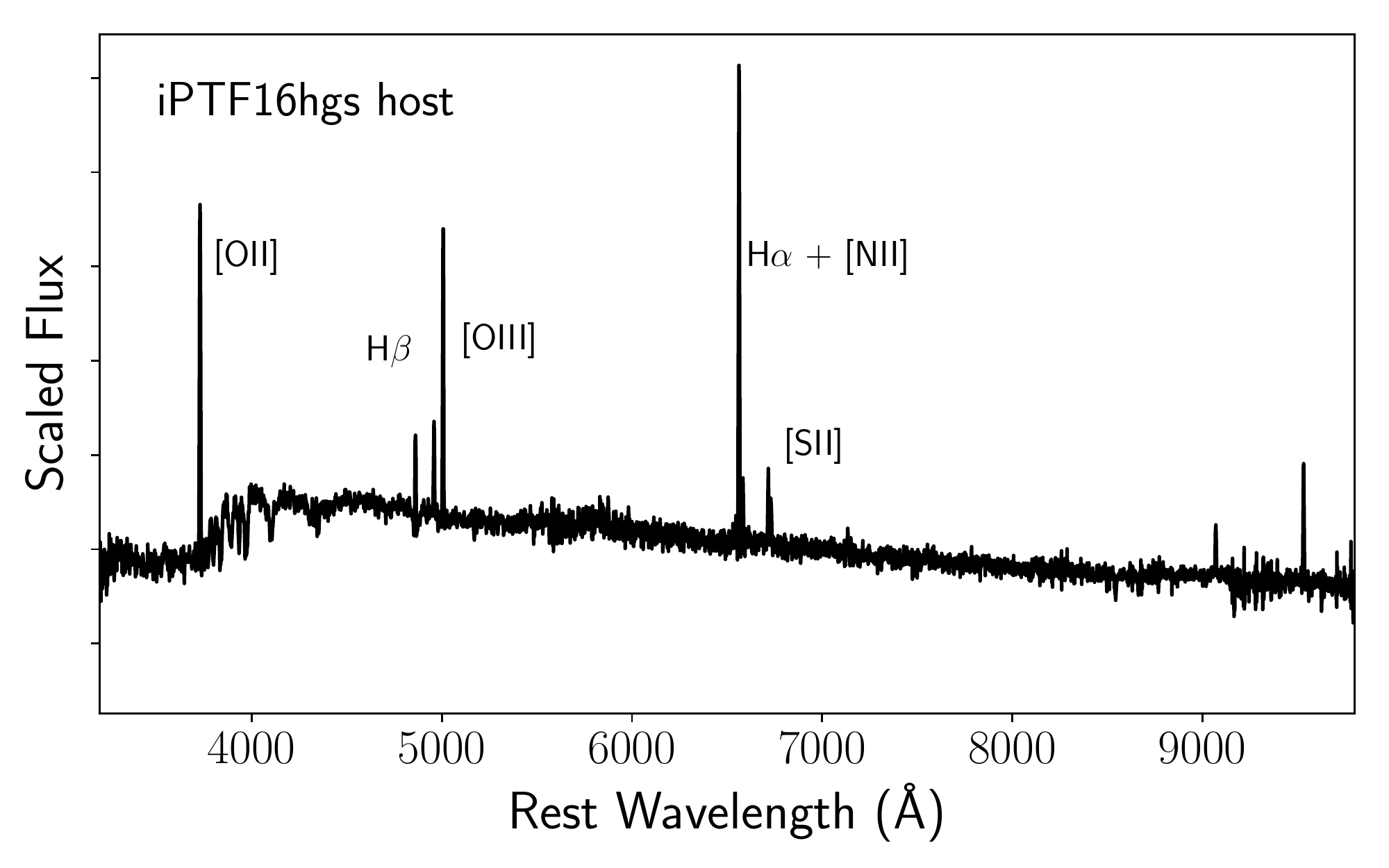}
\caption{Spectrum of the core of the host galaxy of iPTF 16hgs. Prominent emission lines are marked.}
\label{fig:16hgs_hostSpectrum}
\end{figure}

We obtained spectroscopic follow-up of the transient starting from $\approx -8$ d to $\approx$ +59 d after $r$ band peak using the DeVeny spectrograph on the Discovery Channel Telescope \citep{Bida2014}, the Double Beam Spectrograph (DBSP) on the 200-inch Hale telescope \citep{Oke1982} and the Low Resolution Imaging Spectrograph (LRIS) on the Keck-I telescope \citep{Oke1995}. Our last spectrum of the source obtained at +59 days after $r$ band peak was in the form of a spectroscopic mask observation aimed to characterize the host environment of the transient. We present our sequence of spectra in Figure \ref{fig:16hgs_spectra} (the spectroscopy epochs are indicated as dashed lines in Figure \ref{fig:16hgs_photometry}), while the spectroscopic observations are summarized in Table \ref{tab:spectra}. We discuss the spectroscopic evolution of the source in Section \ref{sec:specProp}. We also obtained a spectrum of the nucleus of the host galaxy of iPTF 16hgs with LRIS, which was found to exhibit prominent emission lines of H$\alpha$, H$\beta$, [NII], [SII], [OII] and [OIII], as shown in Figure \ref{fig:16hgs_hostSpectrum}.\\

All spectra and photometry will be made available by the WISeREP repository \citep{Yaron2012} \footnote{\url{https://wiserep.weizmann.ac.il}}. 

\subsection{Swift Observations}
We obtained X-ray follow-up of the transient with the \emph{Swift} X-ray telescope (XRT; \citealt{Burrows2005}). The source was observed on 2017 February 17 (MJD 57801; Phase $\sim$ +107 days) for a total exposure time of 5 ks, and the data was processed with the {\tt HEAsoft} package\footnote{\url{http://heasarc.nasa.gov/lheasoft/}}. No source was detected at the location of the transient to a 3$\sigma$ upper limit of $2.2 \times \, 10^{-3}$ counts s$^{-1}$, corresponding to an unabsorbed 0.3 - 10 keV flux upper limit of $7.5 \times \, 10^{-14}$ ergs cm$^{-2}$ s$^{-1}$ for a photon index of $\Gamma = 2$\footnote{The WebPIMMS intreface at \url{https://heasarc.gsfc.nasa.gov/cgi-bin/Tools/w3pimms/w3pimms.pl} was used for this calculation}. This constrains the unabsorbed X-ray luminosity from the source to $< 4.9 \times 10^{40}$ ergs s$^{-1}$. 

The Swift Ultraviolet / Optical telescope (UVOT; \citealt{Roming2005}) also simultaneously observed the field in the UVW2 band. No source was detected at the transient location up to a $5 \sigma$ limiting AB magnitude of 23.05.

\subsection{Host observations}

\subsubsection{Host environment spectroscopy}
\label{sec:specMask}
Our last spectrum of the source (where the transient was detected) was obtained as a part of a spectroscopic mask observation with LRIS, where we placed additional slits on a number of extended sources classified as galaxies in Sloan Digital Sky Survey (SDSS) catalog. The aim of the mask observation was to measure redshifts of galaxies near the transient region, in order to ascertain if the host galaxy was a part of a galaxy group or cluster (as is typically found for Ca-rich gap transient host galaxies; \citealt{Lunnan2017}). These measurements were then combined with previously measured redshifts of galaxies in NED within a projected offset of 1 Mpc from the apparent host galaxy. \\

We selected a total of 37 sources classified as galaxies in SDSS within $\approx 6$\arcmin\, of the transient location, with the source selection prioritized by the projected offset from the location of the transient. The data were reduced with standard routines in IRAF. Details of the spectroscopic mask observation are given in Table \ref{tab:spectra}, while the measured redshifts of the galaxies are given in Table \ref{tab:16hgs_hostSpec}. Figure \ref{fig:16hgs_environment} shows the positions of the galaxies whose redshifts could be measured from the spectra. As shown, only one of all the sources placed in the spectroscopic mask was found to be at the same redshift as that of the apparent host galaxy. However, we caution that this galaxy was coincident with the diffuse outskirts of the brighter apparent host galaxy, and hence this source may be an H-II region in the outskirts of the host galaxy instead.\\

The faintest source placed in the spectroscopic mask had a $r$-band magnitude of 25.4, while the faintest source for which a redshift could be identified had a $r$-band magnitude of 24.1. The faintest source measured was also the nearest in terms of  projected offset from the transient location ($\approx 4$\arcsec), and coincident with the position of the unidentified radio source mentioned in Section \ref{sec:radioObs}. Based on sources classified as galaxies in SDSS, we estimate that our redshift identification procedure was complete up to an apparent magnitude of $r \approx 22.08$ within a projected offset of 50 kpc ($\approx 143$\arcsec) of the transient, corresponding to an absolute magnitude of $M_r \approx -12.3$ at the redshift of the transient.

\subsubsection{Late-time imaging and spectroscopy}
We undertook deep imaging of the transient region in $g$ and $r$ bands with Keck LRIS on 2017 September 13 (MJD 58009.4) for a total exposure time of 1500s and 600s respectively. The data was reduced with standard procedures in \texttt{lpipe}\footnote{http://www.astro.caltech.edu/~dperley/programs/lpipe.html}. No source was detected at the transient location up to a 3$\sigma$ limiting magnitude of 27.0 mag and 25.5 in the $g$ and $R$ bands respectively. At the distance of the transient, these limits correspond to a extinction corrected absolute magnitude limit of $M_g = -7.6$ and $M_R = -9.0$.\\

We also obtained one late-time spectrum of the transient region with LRIS on 2017 September 13 for a total exposure time of 3600s. We did not detect any continuum or broad nebular emission features at the transient location, although narrow galaxy emission features from the host galaxy are clearly detected. These emission features were also detected on top of the SN continuum in deep spectroscopy taken $\approx 60$ days after peak light when the transient had faded significantly.

\subsubsection{Host IFU observations}

iPTF\,16hgs is unique in that it has the smallest host offset (both in terms of absolute distance and host normalized offset) of all known Ca-rich gap transients (Section \ref{sec:hostEnv}). Hence, iPTF\,16hgs provides a unique opportunity to study the local ISM environments of a Ca-rich gap transient in detail. We thus observed the host region of iPTF 16hgs with the Palomar Cosmic Web Imager (PCWI) on 2017 October 18 to measure spatially resolved metallicity, star formation rate and ISM electron density for the host galaxy. \\

The PCWI is an integral field spectrograph mounted on the Cassegrain focus of the 200-inch Hale telescope at Palomar observatory \citep{Matuszewski2010}. The instrument has a field of view of 40" x 60 " divided across 24 slices with dimensions of 40" x 2.5" each. The spectrograph uses an image slicer and volume phase holographic gratings. For our observations, we selected the red, R $\sim$ 5000 grating and red filter to achieve an instantaneous bandwidth of $\approx 550$ \AA. A complete description of the instrument, observing approach and data analysis methodology can be found in \citealt{Martin2014}.\\

The PCWI field of view matches the projected dimensions of the apparent host galaxy on the sky ($\approx$ 1\arcmin\, along the major axis), and hence this observation was carried out to characterize spatially resolved properties of the environment of the transient, as well as the overall host galaxy (Section \ref{sec:hostEnv}). The instrument was used with its red grating and filter, and configured to a central wavelength of 6630 \AA, covering the wavelength range from approximately 6400 \AA\, to 6900 \AA\,. This specific wavelength range was chosen to include a number of redshifted emission lines from the host galaxy, including H$\alpha$, [N II] $\lambda$6584, and [S II] $\lambda\lambda$6716, 6731, that can be used as tracers of star formation, metallicity and the local electron density of the medium.\\

We obtained a total of 12 dithered exposures of the host galaxy (centered around its nucleus), each with an exposure time of 1200 s. In order to obtain similar spatial sampling in two directions, half of these exposures were obtained with the IFU slices oriented in North-South direction while the other half had slices oriented in the East-West direction. The host galaxy exposures were interleaved with exposures of a nearby empty sky region to subtract out emission features from the sky. We also obtained calibration images including arc lamp spectra, dome flats and a standard star spectrum (GD 248). The two dimensional spectra were sliced, rectified, spatially aligned and wavelength calibrated using the calibration images to produce data cubes for each sky exposure, sampled at (RA, Dec., $\lambda$) intervals of (2.6\arcsec, 0.6\arcsec, 0.22 \AA).\\
 
The sky background cubes were then subtracted from the source cubes to remove the sky emission lines, followed by flux calibration using the standard star GD 248. The flux calibrated (and dithered) spectral cubes for each source exposure were then combined spatially to produce a final spectral cube covering a sky area of $\approx 70$\arcsec $\times$ 70\arcsec, and with a spatial sampling of 0.6 \arcsec\, along both axes. The spatial resolution is thus completely seeing limited ($\approx 1.2$ \arcsec on the night of the observation), and corresponds to a projected resolution of $\approx 0.4$ kpc at the redshift of the host galaxy.

\subsection{Radio Observations}
\label{sec:radioObs}

Owing to its apparent close location to its host galaxy and potential proximity to relatively dense ISM, we initiated deep radio follow-up of the transient to constrain the presence of a radio counterpart, as potentially expected in some proposed models for Ca-rich gap transients (e.g. tidal detonations of white dwarfs that produce a relativistic jet, or a core-collapse explosion of a massive star).

\subsubsection{AMI observations}
We observed iPTF 16hgs with the Arcminute MicroKelvin Imager Large Array (AMI-LA; \citealt{zwart2008, Hickish2018}) radio telescope.
The data have 4096 frequency channels across a 5 GHz bandwidth between 13--18 GHz. The observations were made on 12th March (MJD 57824.65) for a duration of three hours, with the phase calibrator J0057+3021 observed every 10 minutes for a duration of two minutes. The AMI-LA data were binned to 8$\times$0.625 GHz channels and processed (RFI excision and calibration) with a fully-automated pipeline, AMI-REDUCE \citep[e.g.][]{Davies2009}. Daily measurements of 3C48 and 3C286 were used for the absolute flux calibration, which is good to about 10\%. The calibrated and RFI-flagged data were then imported into CASA and imaged with the task {\it clean} to produce 512$\times$512 pix$^2$ (4\arcsec pix$^{-1}$).
We do not detect iPTF 16hgs in the resulting image, and although the RFI was substantial, we can place a stringent upper limit to the flux density at 15.5 GHz of 210 $\mu$Jy (3$\sigma$). This constrains the 15 GHz radio luminosity of the source to $\lesssim 1.4 \times 10^{27}$ ergs s$^{-1}$ Hz$^{-1}$.

\subsubsection{JVLA observations}
We obtained radio observations of the transient with the Very Large Array (VLA) under the Director's Discretionary Time (DDT) program (17A-427; PI: De). The VLA observed the source on 2017 June 24 (in C configuration; $\sim$ 250 days after $r$ band peak) at X band (centered at 10 GHz) for a total on source time of $\approx$ 1.8 hours, with the Wideband Interferometric Digital Architecture (WIDAR) correlator configured in continuum mode with 4 GHz bandwidth. The data were flagged and calibrated with the VLA calibration pipeline, while deconvolution and imaging was performed with standard routines in CASA. 3C48 was used as the flux and bandpass calibrator while the source J0042+2320 was used as the phase calibrator. The final processed image has a noise RMS of $\approx 2.6 \, \mu$Jy/beam. \\

Although we find a faint radio source very close to the transient location, its position is offset by 5" from the source and coincident with a background galaxy in the late-time LRIS image, and hence not associated with the transient. We also obtained a spectrum of the background galaxy in our spectroscopic mask observation (Section \ref{sec:specMask}), and found it to be consistent with an Active Galactic Nucleus (AGN) at a redshift of 0.362, and clearly unrelated to the transient. No other source is detected at the transient location up to a 3$\sigma$ limiting flux density of $\approx 7.8 \, \mu$Jy (Figure \ref{fig:16hgs_radioImage}). At the redshift of the host galaxy, the flux upper limit constrains the 10 GHz radio luminosity to $\lesssim 5.1 \times 10^{25}$ ergs s$^{-1}$ Hz$^{-1}$.

\subsubsection{uGMRT Observations}
We obtained radio follow-up of the source with the upgraded Giant Metrewave Radio Telescope (uGMRT) under the DDT program (DDTB272; PI: De). The source was observed on 2017 July 19 ($\sim$ 275 days after $r$ band peak) at L band (centered at 1.2 GHz) for a total on source time of $\approx$ 4.5 hours. We used the GMRT Wideband Backend (GWB; \citealt{Reddy2017}) configured in the continuum interferometric mode with 400 MHz bandwidth. 3C48 was used as the flux density and bandpass calibrator, while the source J0029+349 was used as the phase calibrator. The data were analyzed using the Astronomical Image Processing System (AIPS). After flagging the original data set for non-working antennas and Radio Frequency Interference (RFI), data from a single frequency channel of the flux and phase calibrators were used to construct time-based amplitude and phase calibrations, while bandpass calibration was done with the flux calibrator.  \\

\begin{figure*}[!ht]
\centering
\includegraphics[width=0.49\textwidth]{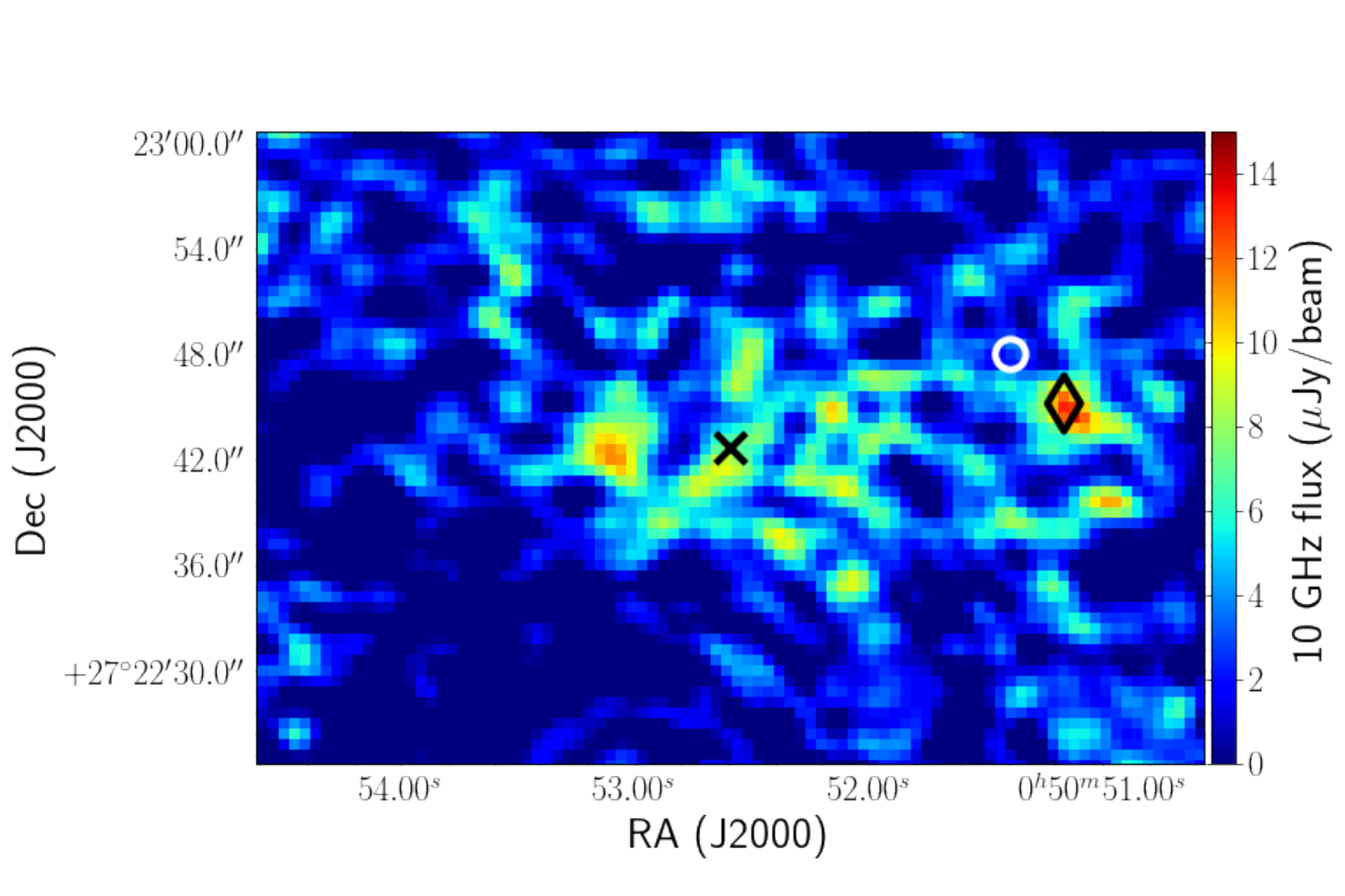}
\includegraphics[width=0.49\textwidth]{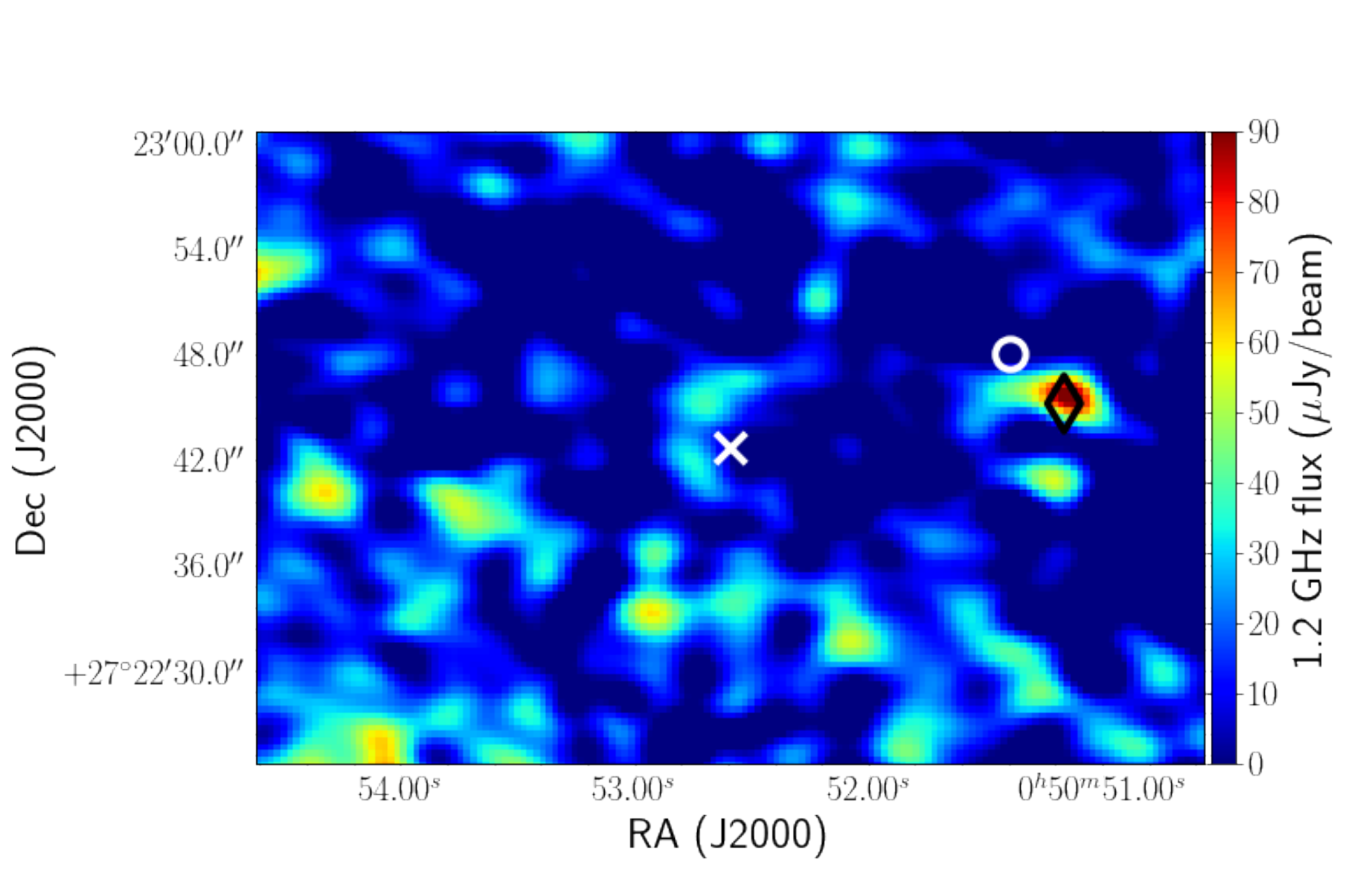}
\caption{Radio images of the host region of iPTF 16hgs from observations with the VLA (10 GHz; Left) and the uGMRT (1.2 GHz; Right). The nucleus of the host is indicated by the cross, while the location of the transient is indicated by the white circle. While a radio source is detected very close to the transient in both these observations (indicated by the black diamond), its position is consistent with a background AGN identified in the spectroscopic mask observation, and hence unrelated to the transient. The Image RMS is $\approx 2.6 \mu$Jy for the VLA X band image, and $\approx 17 \mu$Jy for the GMRT L band image.}
\label{fig:16hgs_radioImage}
\end{figure*}

No source is detected at the location of the optical transient (Figure \ref{fig:16hgs_radioImage}) up to a 3$\sigma$ limiting flux of $\approx 50 \mu$Jy, corresponding to a 1.2 GHz radio luminosity of $3.3 \times 10^{26}$ ergs s$^{-1}$ Hz$^{-1}$. We note that the faint radio source detected at a 5" offset in the VLA image is also well detected in the GMRT image at L band, coincident with a background galaxy in the late-time LRIS image.\\

\section{Analysis}
\label{sec:analysis}
In this section, we analyze the photometric and spectroscopic properties of iPTF\,16hgs and show that it is a member of the class of Ca-rich gap transients as defined by \citealt{Kasliwal2012a}.

\subsection{Light curve properties}
\label{sec:lightCurve}
We first analyze the properties of the light curve of iPTF 16hgs, which we show to be unique in comparison to other known Ca-rich gap transients. The multi-color light curves of iPTF 16hgs are shown in Figure \ref{fig:16hgs_photometry}. The light curve of iPTF\,16hgs shows clear evidence for two distinct components -- an early declining phase (which was caught at discovery), followed by re-brightening to a second (main) peak which was followed up to late times. The early declining phase was detected in both $r$ and $g$ bands, and was characterized by significantly bluer colors than the rest of the light curve, as evident from the rapid early decline in $g$ band. \\

\begin{figure*}[!ht]
\centering
\includegraphics[width=0.9\columnwidth]{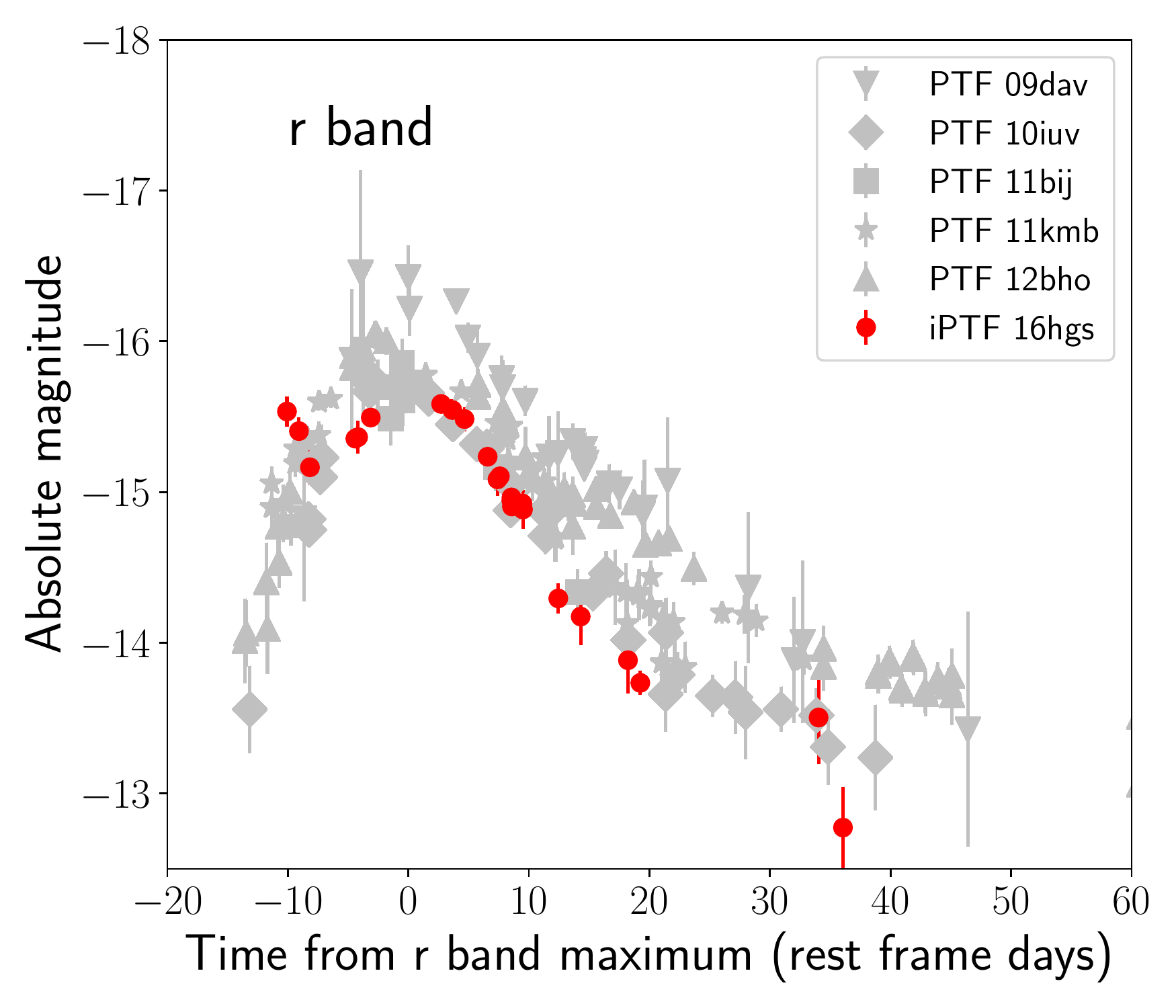}
\includegraphics[width=0.9\columnwidth]{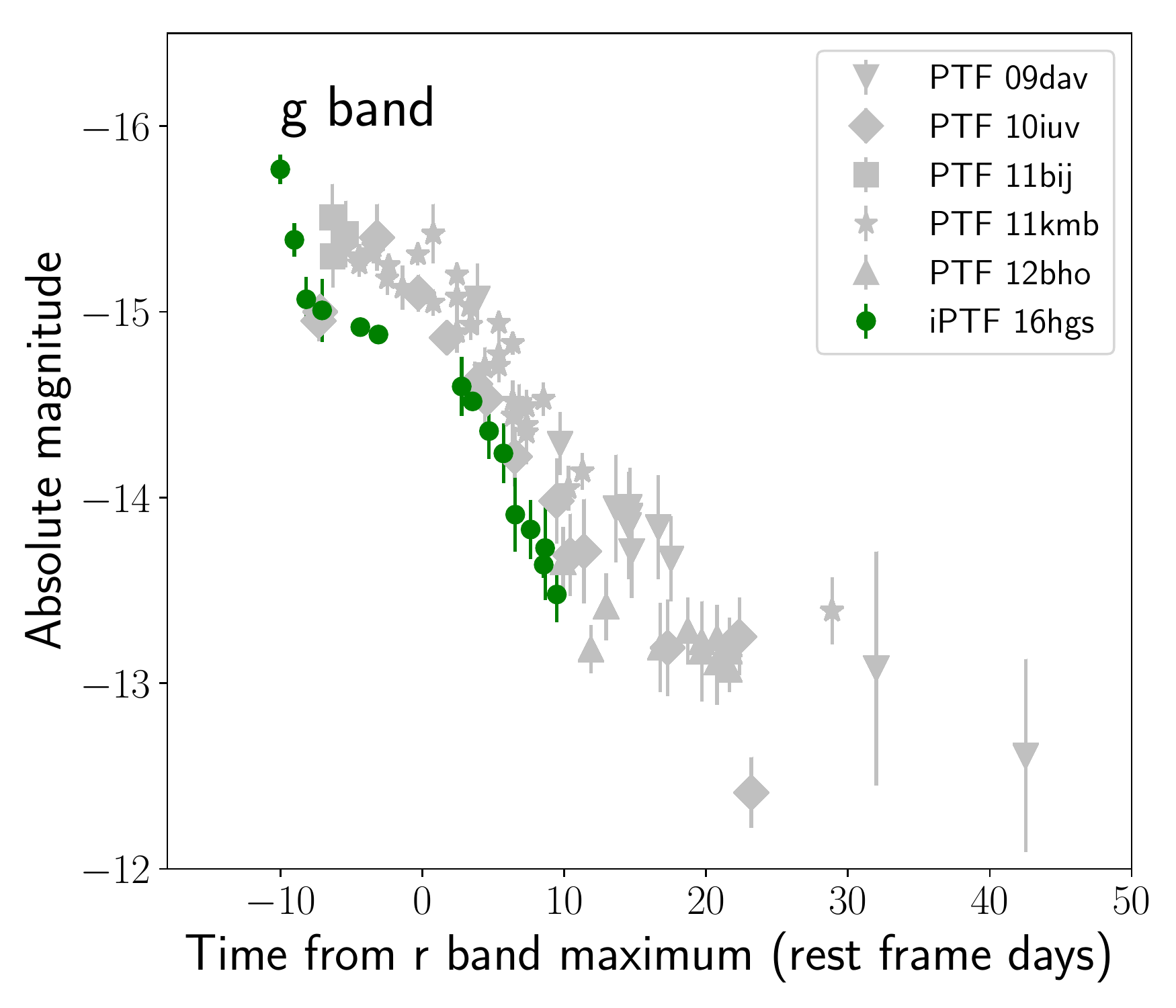}\\
\includegraphics[width=0.9\columnwidth]{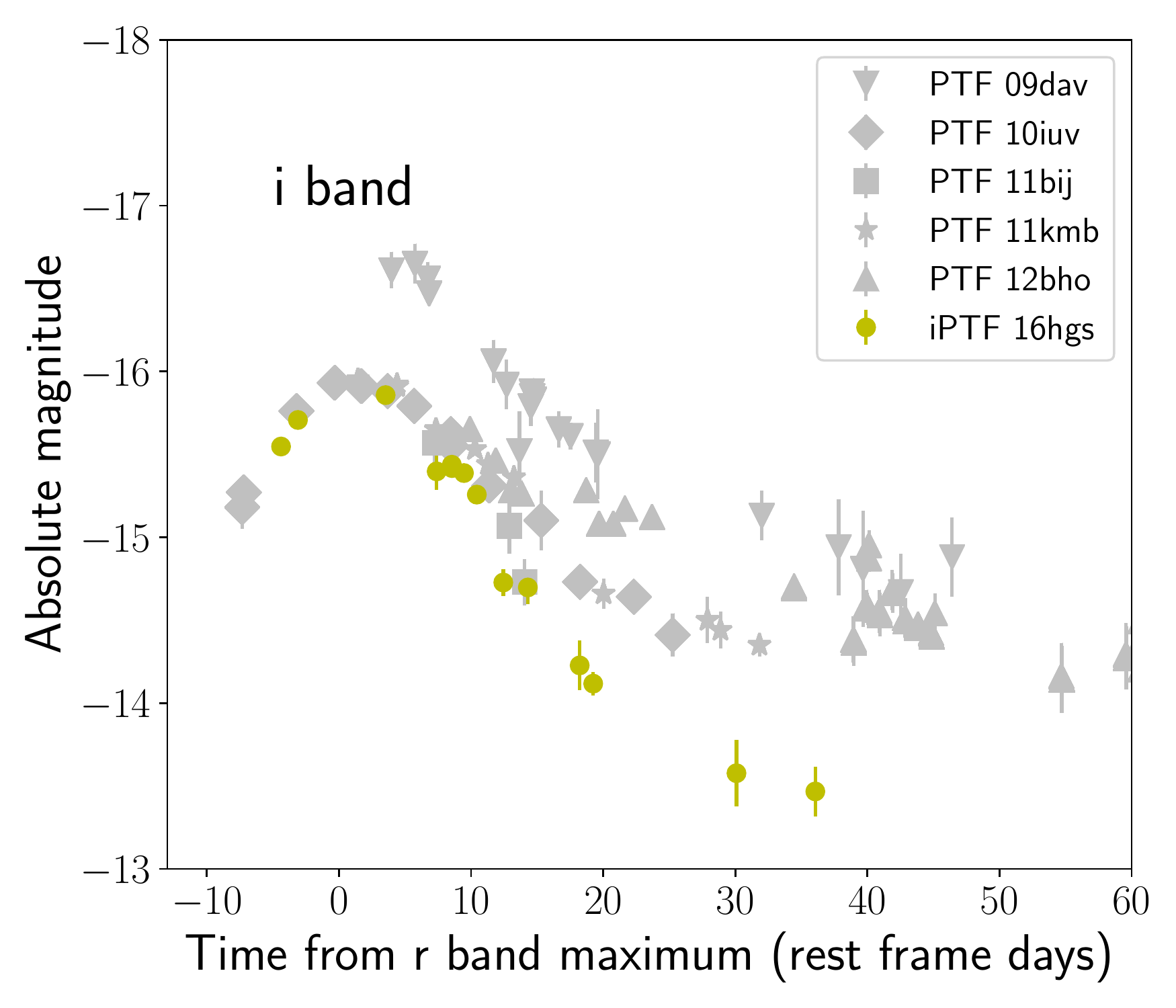}
\includegraphics[width=0.9\columnwidth]{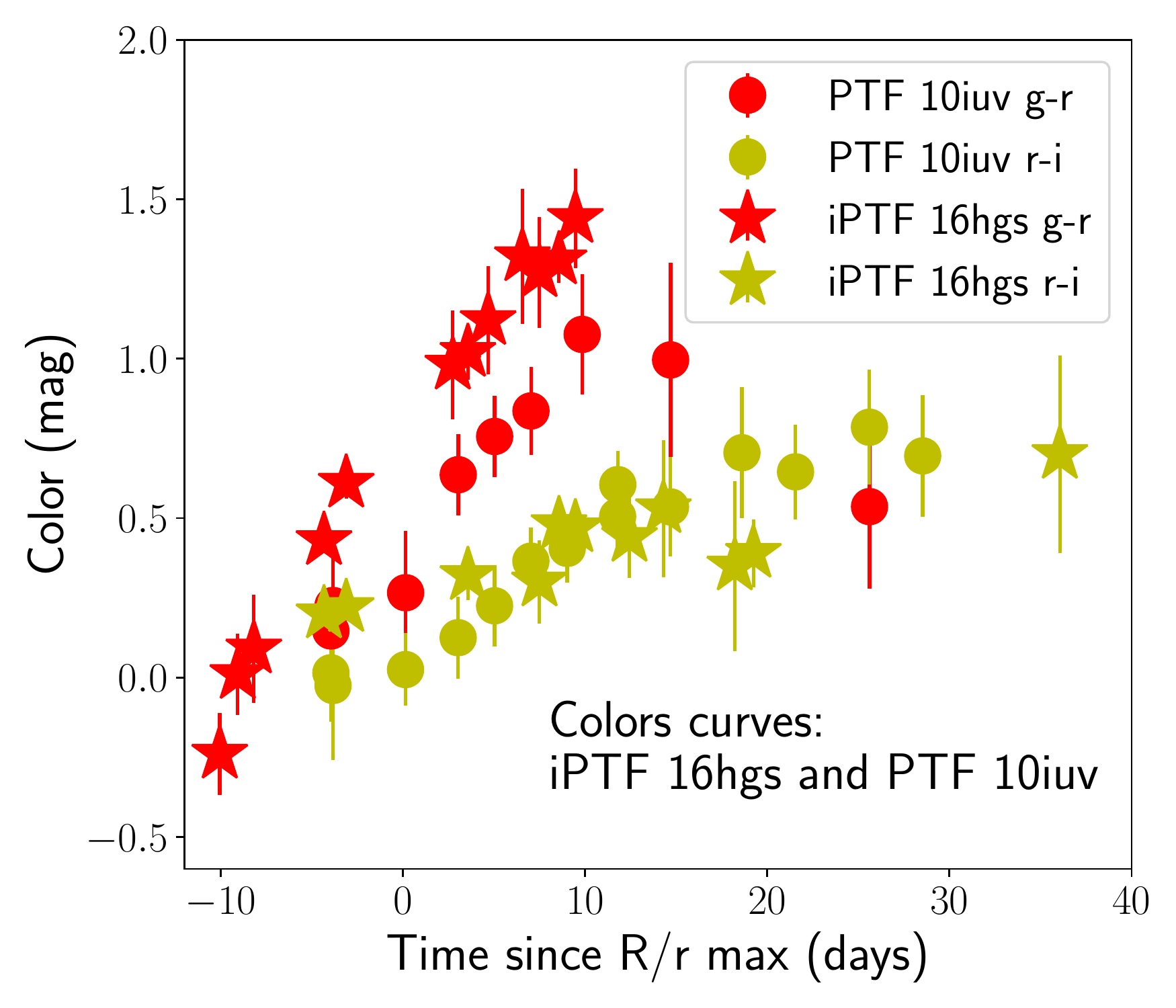}\\
\caption{Comparison of the multi-color light curves (in $r$, $g$ and $i$ bands) of iPTF 16hgs to that of other Ca-rich gap transients discovered by PTF -- PTF 09dav, PTF 10iuv, PTF 11bij, PTF 11kmb and PTF 12bho. The lower right panel shows a comparison of the $g-r$ and $r-i$ color evolution of iPTF\,16hgs compared to that of PTF 10iuv.}
\label{fig:16hgs_compareCaLC}
\end{figure*}

Figure \ref{fig:16hgs_compareCaLC} compares the multi-color light curves of iPTF\,16hgs to that of other known Ca-rich gap transients from PTF \citep{Kasliwal2012a, Lunnan2017}. As shown, the second light curve peak of iPTF 16hgs is very well matched to that of the light curves of the other Ca-rich transients, although the rapid decline from the first peak distinguishes it from this entire sample. Since some of the previously known transients have very well sampled early light curves (particularly in $r$ band for PTF\,10iuv and PTF\,12bho), we can rule out the possibility that a first peak at similar timescales was missed in the case of the previous transients. Figure \ref{fig:16hgs_compareCaLC} also shows the $g-r$ and $r-i$ color evolution of iPTF\,16hgs compared to the Ca-rich transient PTF\,10iuv (which had good multi-color photometric coverage). As shown, iPTF\,16hgs exhibited very blue colors at early times after discovery (with $g-r \approx -0.2$), while it subsequently exhibited rapid reddening over the next $\approx 20$ days, evolving to $g-r \approx 1.5$ at $\approx$ 10 days after the $r$ band peak. The $r-i$ color also exhibits reddening with time, although the evolution is much slower. For comparison, the color curves of PTF\,10iuv also exhibited similar but less rapid reddening with time, although iPTF\,16hgs remained redder at similar light curve phases.\\

We fit a third order polynomial to the main peak of the $r$ band light curve (which is best sampled) to find a peak magnitude of $M_r = -15.65$ and a peak time of MJD 57691.59. All phases mentioned in this paper are with respect to this epoch. We note that the peak absolute magnitude is similar to that of several previously confirmed Ca-rich gap transients. Determining the explosion time for this transient is non-trivial due to the presence of the early declining emission, thus precluding a conventional $t^2$ fit to the early light curve. Additionally, the first rise is not sampled due to a gap in coverage of $\approx$ 12 days between the last non-detection and the first detection. Nonetheless, we try to estimate the rise time by fitting a parabolic function to the main peak of the light curve in flux space, and find a best fit rest frame rise time of $\approx$ 9.9 days. Based on the last non-detection, we can put an upper limit of $\approx$ 12.3 days on the rise time of the first peak.\\

\subsection{Spectroscopic properties}
\label{sec:specProp}

\begin{figure*}[!ht]
\centering
\includegraphics[width=\columnwidth]{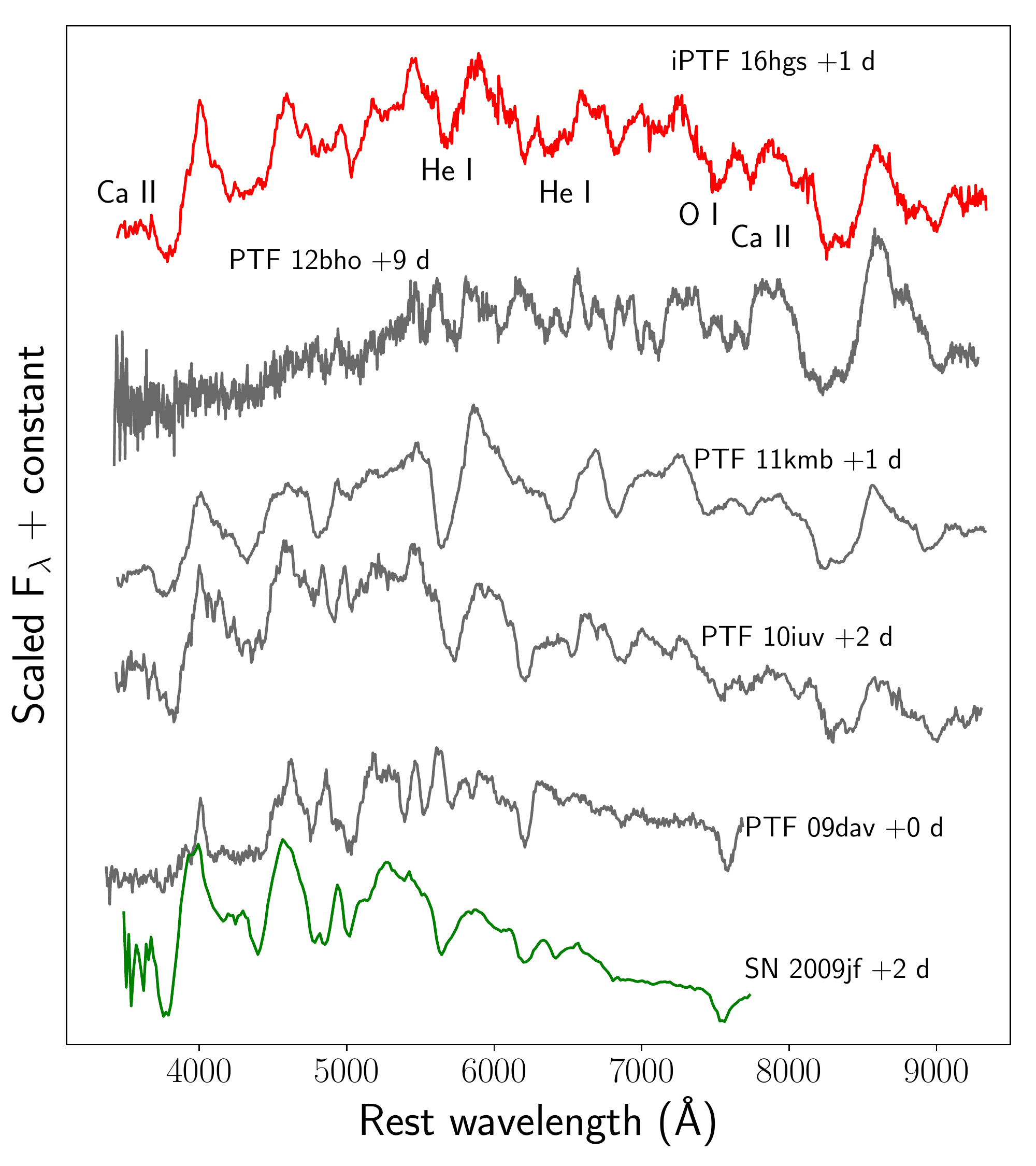}
\includegraphics[width=\columnwidth]{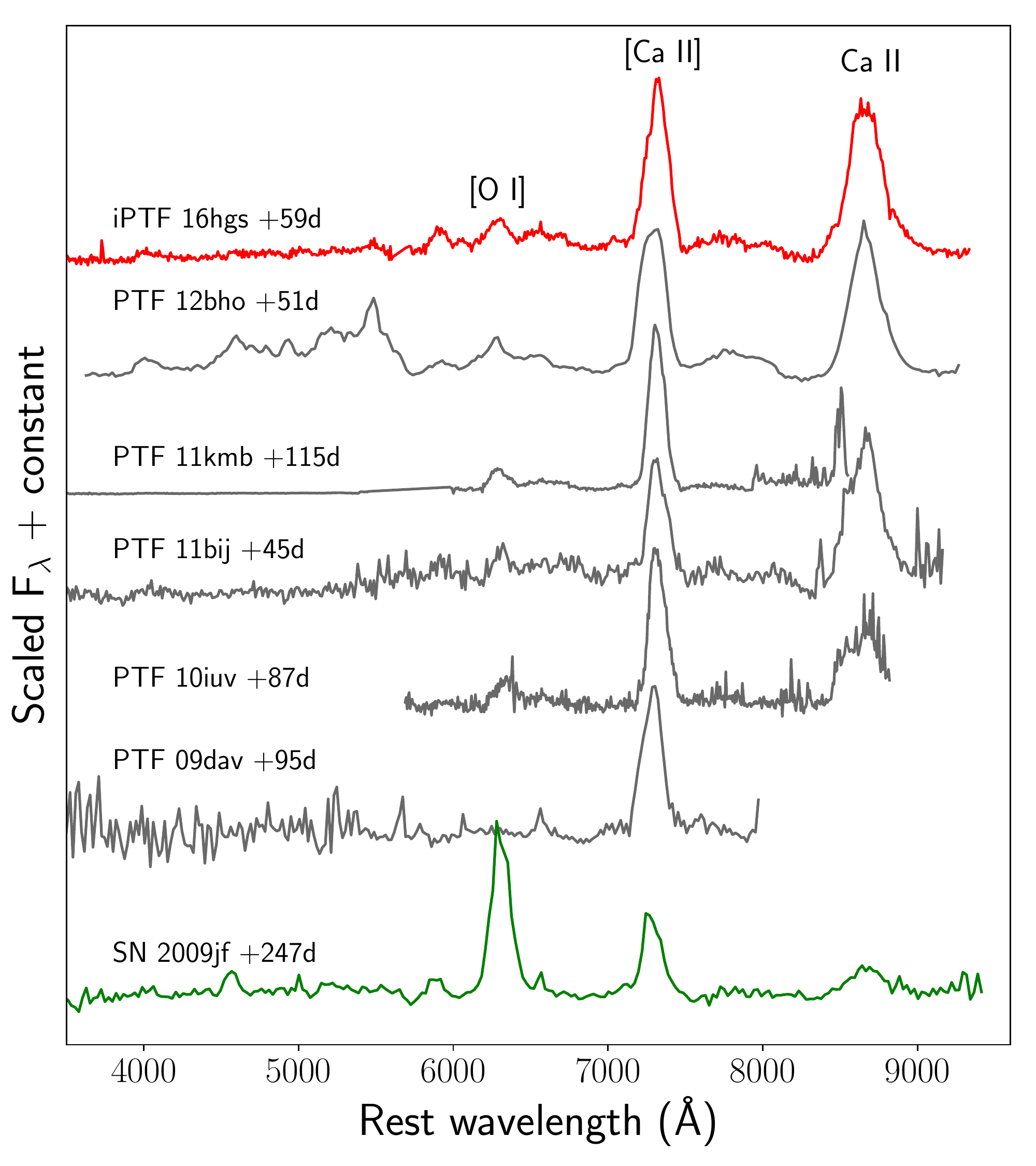}\\
\caption{Comparison of the spectra of iPTF 16hgs to that of other Ca-rich gap transients PTF\,09dav, PTF\,10iuv, PTF\,11bij \citep{Kasliwal2012a}, and PTF\,11kmb and PTF\,12bho \citep{Lunnan2017}. We also show photospheric and nebular phase spectra of the normal Type Ib SN 2009jf (in green; from \citealt{Valenti2011}) for comparison.}
\label{fig:16hgs_compareCaSpectra}
\end{figure*}

Photospheric phase spectra taken near the second peak show typical features of Type Ib SNe, most notably prominent lines of He I, O I, Mg II and Ca II. We compare the photospheric phase spectra of iPTF 16hgs to example spectra of other Ca-rich transients in Figure \ref{fig:16hgs_compareCaSpectra}. The photospheric spectrum of iPTF 16hgs exhibits a number of good similarities to that of PTF 11kmb and PTF 10iuv, most notably in the strong He features. He features in photospheric spectra are indeed common in many Ca-rich transients (PTF 11kmb, PTF 10iuv, SN 2005E and SN 2007ke), although it is not used to exclusively define this class. Indeed, as noted by \citealt{Lunnan2017}, the photospheric phase diversity may point to different progenitor channels of this observationally defined class. From the peak photospheric spectrum (taken at 0.72 days after $r$ band peak in the source rest frame), we measure a photospheric velocity of $\approx 10,000$ km s$^{-1}$ for this source.\\

\begin{figure}
\includegraphics[width=\columnwidth]{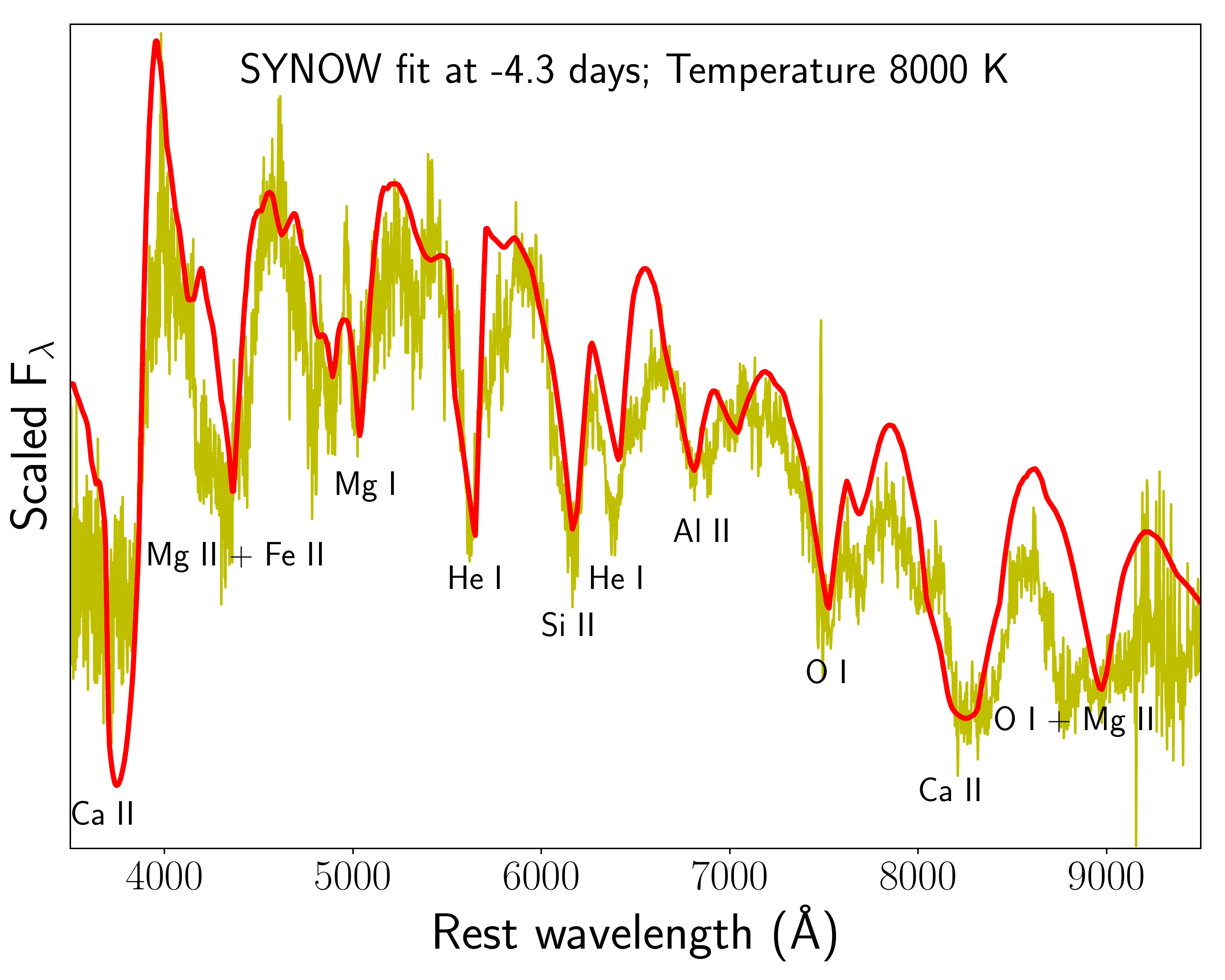}
\caption{Identification of spectral lines in the $-4.28$ days spectrum of iPTF 16hgs using SYNOW. The yellow line indicates the observed spectrum while the red line indicates the SYNOW fit. The elements corresponding to the absorption features are marked.}
\label{fig:16hgs_synow}
\end{figure}

We identify prominent features in the $-4.28$ d spectrum of iPTF 16hgs with SYNOW \citep{Fisher2000} in Figure \ref{fig:16hgs_synow} (note that SYNOW may not be self-consistent and is only suggestive for line identification). We constrain the fit by matching the visible features along with the relative strengths of the features in order to constrain the composition of the ejecta as well as the photospheric velocities of the lines . The most prominent features include He I, Mg II, Si II, O I and Ca II, along with weaker features of Fe II and Al II. The SYNOW model uses a continuum temperature of 8000 K and velocities in the range of 8,000 - 12,000 km s$^{-1}$, with He I found to be at the highest velocity of $\approx 12,000$ km s$^{-1}$. The velocities of other prominent ions include O I at 10000 km s$^{-1}$, Mg II at 8000 km s$^{-1}$, Si II at 9000 km s$^{-1}$, Ca II at 8000 km s$^{-1}$, Al II at 10000 km s$^{-1}$ and Fe II at 8000 km s$^{-1}$. Overall, the SYNOW fit fairly reproduces all the absorption features in the spectrum. \\

Figure \ref{fig:16hgs_compareCaSpectra} also shows a comparison of a nebular spectrum of iPTF\,16hgs to that of other Ca-rich gap transients. These spectra exhibit weak continua superimposed with strong forbidden and permitted lines of Ca and O. In particular, iPTF\,16hgs exhibits the characteristic nebular features of this class, i.e., strong [Ca II] $\lambda\lambda$ 7291, 7324 emission combined with relatively weak [O I] $\lambda\lambda$ 6300, 6363 emission (with a [Ca II]/[O I] ratio of $\approx 7$). For comparison, we also show the photospheric and nebular phase spectra of SN 2009jf \citep{Valenti2011} which exhibits significantly stronger [O I] emission than [Ca II] in the nebular phase. The [Ca II]/[O I] ratio has been used as a defining feature of the class of `Ca-rich transients', e.g. \citealt{Milisavljevic2017} suggest that a [Ca II]/[O I] ratio of $>$ 2 separates the class of Ca-rich transients from other Type Ib/c SNe. We analyze this issue further in Section \ref{sec:discussion}.\\

\begin{figure}
\centering
\includegraphics[width=0.9\columnwidth]{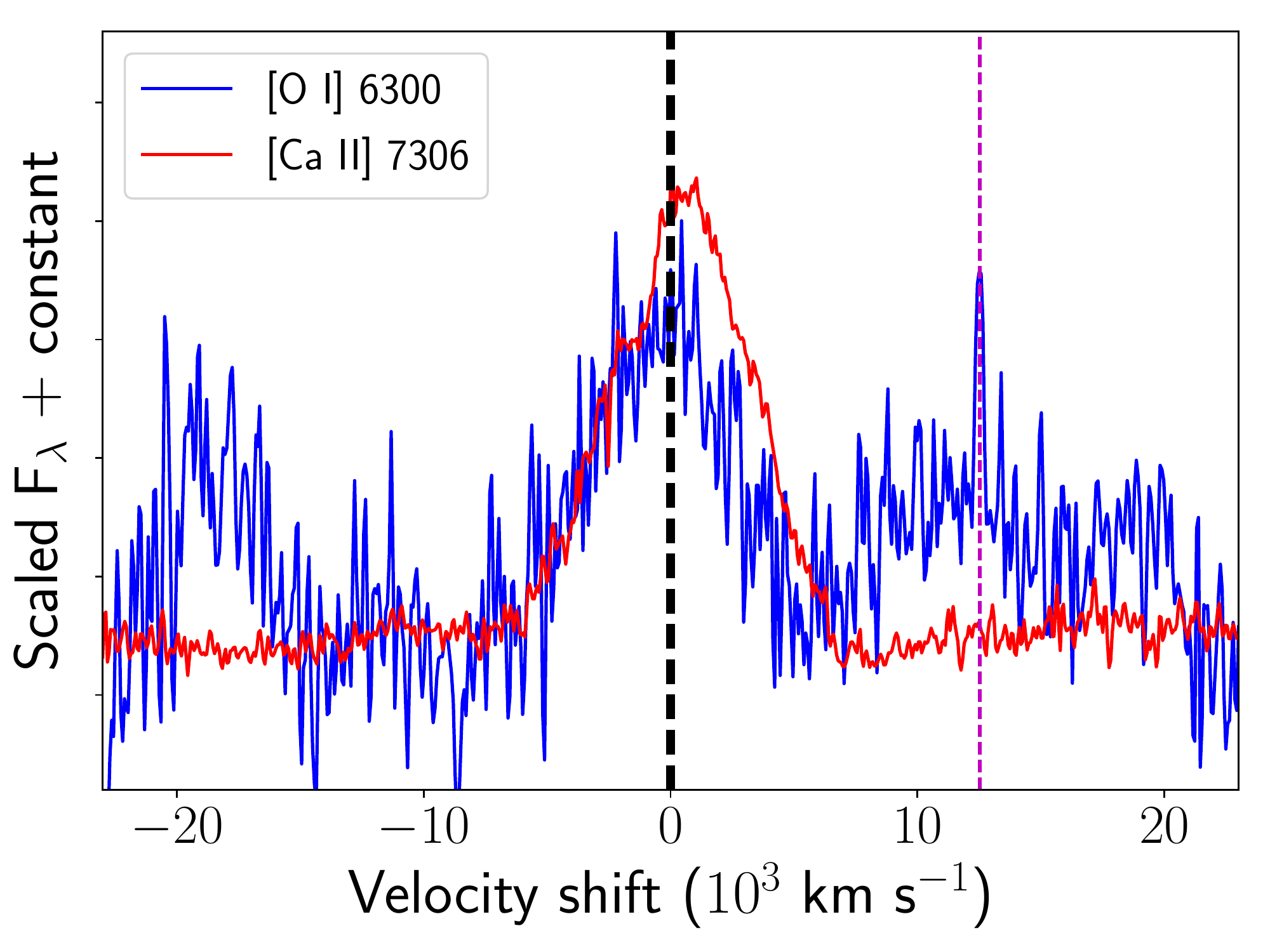}
\caption{Velocity profiles of the late-time nebular [Ca II] $\lambda\lambda$ 7291, 7324 (red) and [O I] $\lambda\lambda$ 6300, 6364 (blue) lines in iPTF\,16hgs. There is clear evidence of a narrow galaxy H$\alpha$ feature red-wards of the [O I] feature (marked with the magenta dashed line), along with an underlying broader feature. The broad nebular feature blue-wards of the [O I] line is likely from He.}
\label{fig:16hgs_caOProfiles}
\end{figure}

In Figure \ref{fig:16hgs_caOProfiles}, we show a comparison of the velocity profiles of the nebular [Ca II] and [O I] emission lines. Interestingly, this last nebular spectrum taken at $\approx +60$ days shows clear evidence of host galaxy emission features  of H$\alpha$ and [O II] $\lambda\lambda$3727, 3729 (on the blue side). These were also detected in a late-time Keck LRIS spectrum taken after the transient faded away. Hence, as opposed to other Ca-rich gap transients that are found in the far outskirts of their host galaxies \citep{Kasliwal2012a, Lunnan2017}, iPTF\,16hgs shows evidence of being located inside its host galaxy. However, we cannot rule out a scenario where the transient was located significantly behind the host galaxy, so that the galaxy emission features arise from the foreground. We note that the narrow H$\alpha$ feature is found superimposed on a broader underlying component (Figure \ref{fig:16hgs_caOProfiles}), that could be potentially associated with hydrogen in the ejecta. A similar H feature was also observed in the nebular spectrum of PTF\,09dav \citep{Kasliwal2012a}, and potentially in PTF 11kmb as well \citep{Milisavljevic2017}. However, \citealt{Milisavljevic2017} suggest that the feature could also be associated with Ca I] $\lambda$6572. \\

Taken together, we have thus far demonstrated that iPTF\,16hgs exhibited (1) a low peak luminosity of $M_r \approx -15.6$, (2) rapid photometric evolution similar to other known Ca-rich gap transients, (3) normal photospheric velocities ($\sim 10,000$ km s$^{-1}$), (4) early transition to a nebular phase (at $\approx 30$ days after $r$ band peak) and (5) a nebular phase spectrum dominated by [Ca II] emission. Thus, despite its unique light curve, iPTF\,16hgs falls squarely in the class of Ca-rich gap transients.\\
 
\section{Modeling the double-peaked light curve}

\label{sec:doubleModel}

\subsection{Bolometric light curve}
We begin our modeling by constructing a bolometric light curve for the transient. For epochs where we have contemporaneous photometry in $gri$ bands, we compute a pseudo-bolometric luminosity by performing a trapezoidal integration of the multi-color fluxes till the edges of the $g$ and $i$ band. In order to account for additional flux below 4000 \AA\, and above 8000 \AA\, we use the peak photospheric spectrum (at $+0.72$ days) to determine the fraction of flux missed between 3000 \AA\, and 10000 \AA\,. We find that the $gri$ trapezoidal integration misses 23 \% of the total flux, and hence we scale all the photospheric phase fluxes the peak by a factor of 1.3. We also add a 5\% uncertainty to the computed luminosities to account for potential uncertainties in the fraction of flux missed. \\

Although the bolometric luminosity for the first peak is important to understand its origin, we do not have photometry in $i$ band for the early peak. Hence, we use the only spectrum taken within the early decline (at $-8.06$ days) to find the fraction of flux missed between 3500 \AA\, and 8000 \AA\ in a $gr$ trapezoidal flux estimate ($\approx 33$\%), and scale all the $gr$ trapezoidal fluxes within the first peak. Lastly, we have one epoch of contemporaneous $ri$ detection at $\approx +35$ days. For this epoch, we use the $+28$ d spectrum to similarly estimate the fraction of flux missed between 3000 \AA\, and 10000 \AA\, ($\approx 50$\%), and scale the $ri$ trapezoidal luminosity to account for it. Note that the pseudo-bolometric luminosity estimates above are strict lower limits on the total bolometric luminosity from the source.\\

\begin{figure}
\includegraphics[width=\columnwidth]{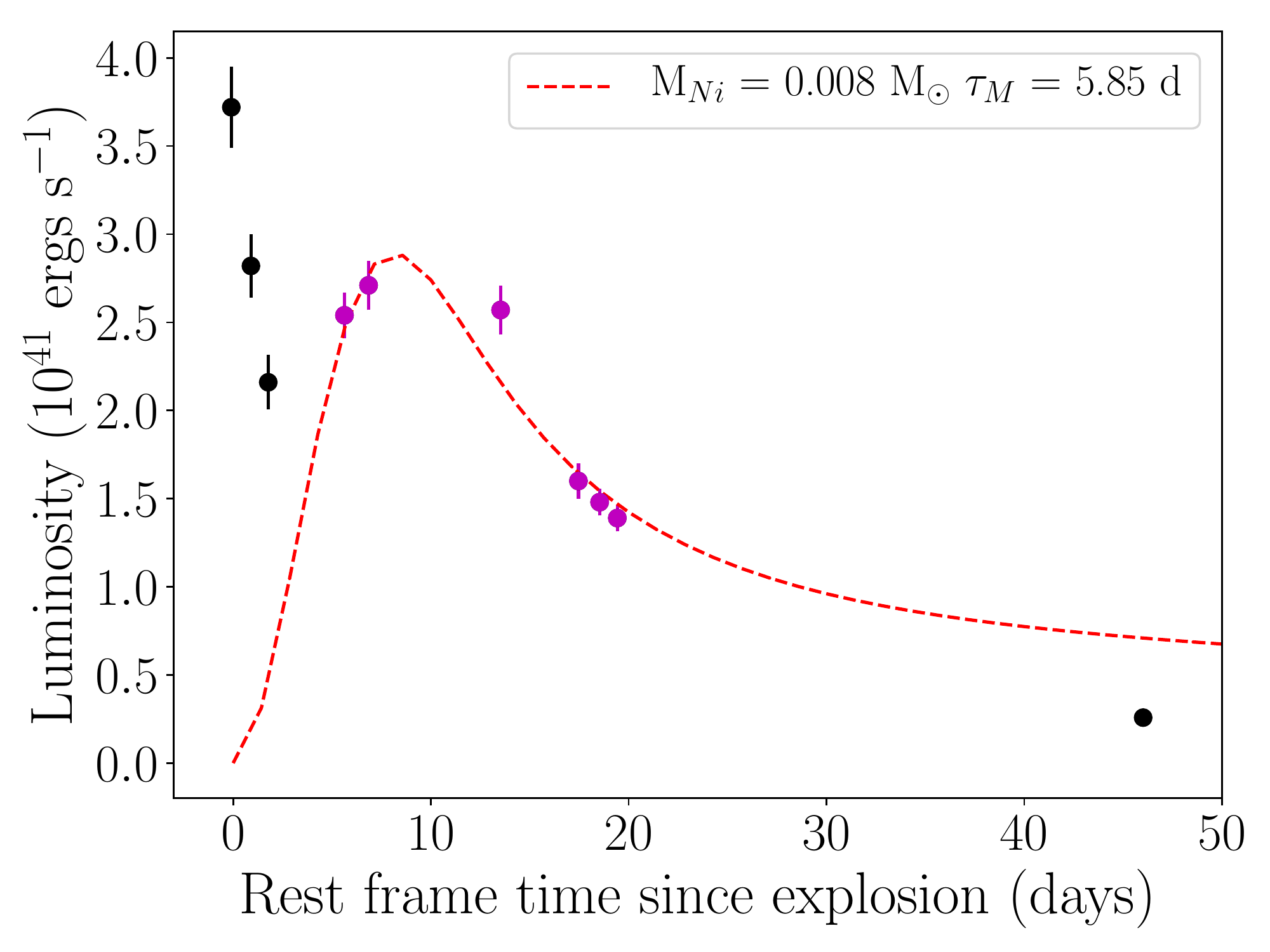}
\caption{Pseudo-bolometric light curve of iPTF\,16hgs (as described in the text) along with a Arnett model fit to the main peak (red dashed line). Only the magenta colored points were used in the Arnett model fitting. The best-fit $^{56}$Ni mass and diffusion time $\tau_M$ are shown.}
\label{fig:16hgs_boloArnett}
\end{figure}

The bolometric light curve thus obtained is shown in Figure \ref{fig:16hgs_boloArnett}, clearly exhibiting a rapidly declining and relatively luminous early peak, followed by a rise to a second (main) peak and subsequent decline. The main peak reaches a peak bolometric luminosity of $\approx 3 \times 10^{41}$ ergs s$^{-1}$, while the peak luminosity for the first component was at least $\approx 4 \times 10^{41}$ ergs s$^{-1}$. The last photometric data point around 30 days after $r$-band peak indicates that the source faded to $\lesssim 4 \times 10^{40}$ ergs s$^{-1}$ by this epoch.\\

\subsection{Radioactively powered main peak}
We first try to understand the power source for the main peak of the light curve. Since the main peak is very similar to that observed in other Ca-rich gap transients, we consider a radioactively powered light curve for the main peak. We fit a simple Arnett model to the bolometric light curve assuming that the decay of $^{56}$Ni powers the main peak of the light curve. Note that the Arnett model has several simplifications, including that of homologous expansion, spherical symmetry, constant opacity and centrally located $^{56}$Ni in the ejecta. We do not include data points within the first peak (at $< -5$ days from $r$-band maximum), as well as the last data point since the Arnett model is only valid in the optically thick photospheric phase. \\

We use the analytic relations presented in \citealt{Lyman2016a} and \citealt{Valenti2014} for this fitting, for which the only free parameters are the diffusion time through the ejecta $\tau_M$ and the Ni mass $M_{Ni}$. Keeping the explosion time as an additional free parameter, we get the best fit Arnett model as shown by the red dashed line in Figure \ref{fig:16hgs_boloArnett}. As shown, the model does a reasonably reproduces the bolometric evolution of the main peak, although there are clear discrepancies near peak light. This is not unexpected, as the Arnett model is very simplified, for instance, it ignores potential effects of $^{56}$Ni mixing in the outer layers of the ejecta, that can significantly affect the rise of the light curve.  The best-fit model indicates a $^{56}$Ni Mass of $8 \times 10^{-3}$ M$_{\odot}$ and diffusion time of $\approx 5.9$ days. The best-fit explosion time is 9.96 days before $r$-band peak, which is very close to our initial estimate based on the $t^2$ law fitting to the $r$ band light curve. \\
 
We note that the last luminosity estimate from $\approx 45$ days after the explosion is clearly much fainter than the predicted Arnett model luminosity. This is expected, as $\gamma$-ray trapping is likely to be inefficient at these late phases given the low ejecta mass, and hence the Arnett model is not applicable at late times \citep{Valenti2014}. Using an optical opacity of $\kappa = 0.07$ cm$^{2}$ g$^{-1}$ \citep{Cano2013,Taddia2017} and ejecta velocity of 10,000 km s$^{-1}$ for the second peak, we derive an ejecta mass of 0.38 M$_{\odot}$ and explosion kinetic energy of $\approx 2.3 \times 10^{50}$ ergs. Note that these estimates do not include the (yet) unknown power source of the first peak, which is clearly not consistent with the Arnett model presented here. Given that such an early emission component has never been observed in Ca-rich gap transients, we discuss several potential power sources for the first peak in the following sections.\\

\subsection{Radioactively powered first peak}
Since the second peak of the light curve peak of iPTF\,16hgs can be well understood by radioactive decay, we first consider a radioactively powered scenario for the first peak. In this case, the shape of the early emission with respect to the main peak puts strong constraints on the radial distribution of the radioactive material. \citealt{Dessart2012} and \citealt{Piro2016} show that a radial monotonically decreasing distribution of $^{56}$Ni in the ejecta lead to smoothly rising light curves for radioactively powered Type Ib/c SNe. As the early decline in iPTF\,16hgs is distinctly separated from the main peak, it is likely that the relevant radioactive isotope was strongly mixed into the surface of the progenitor, and separated from the radioactive material powering the second peak.\\

We can obtain approximate estimates for the amount of radioactive material, and the ejecta mass above it by analyzing the early bolometric light curve. Taking the peak luminosity of the first peak to be $> 4 \times 10^{41}$ ergs s$^{-1}$, we estimate that $\gtrsim 7 \times 10^{-3}$ M$_{\odot}$ of $^{56}$Ni in the outer layers would be required if the early component was powered by $^{56}$Ni decay. However, we do not have strong constraints on the rise time of the first peak since the transient was discovered on the early declining phase. Given that the Arnett model fitting of the main peak suggests that the explosion occured at $\approx 10 $ days before the $r$ band peak (which is almost at the epoch when the transient was first discovered), we consider the case where the observed width of the first component of $\approx 2$ days corresponds to the diffusion time through the ejecta above this radioactive layer. In such a case, the mass above this layer would be $\gtrsim 0.05$ M$_{\odot}$ taking an opacity of $\kappa = 0.07$ cm$^2$ g$^{-1}$ and ejecta velocity of $12,000$ km s$^{-1}$ (as measured from the first spectrum of the source). We note that the derived estimates are similar to that of some other Type Ib/c SNe with early excess blue emission \citep{Drout2016,Bersten2013}, and where a radioactivity powered first peak was also suggested. \\

\subsection{Interaction with a companion}

A possible explanation for the early peak could be due to interaction of the SN ejecta with a non-degenerate companion \citep{Kasen2010}. Such interaction signatures depend sensitively on the binary separation of the companion as well as the viewing angle of the observer, with the most prominent signatures arising when the source is viewed along the direction of the companion. These signatures have been previously suggested in Type Ia SNe (e.g. \citealt{Cao2015,Hosseinzadeh2017} ), allowing one to estimate the separation of the companion and its radius (assuming Roche lobe overflow). For viewing angles oriented close to the direction of the companion, the analytic equations in \citealt{Kasen2010} yield good estimates of the expected emission.\\

We attempted to fit the early $g$ and $r$ band light curves with the \citealt{Kasen2010} models but could not get a reasonable match to the data. This is particularly because the colors of iPTF\,16hgs on the declining phase are markedly different than that predicted in the companion interaction models. For instance, we observe colors of $g - r \approx 0$ about one day after discovery when the bolometric luminosity is $\gtrsim 2.5 \times 10^{41}$ ergs. Taking equation (22) in \citealt{Kasen2010} for $t_{day} = 1$, $v_9 = 1.2$ and $M_c = 0.3$, the bolometric luminosity requires the separation $a$ to be $\gtrsim 2.5 \times 10^{11}$ cm. At the same time, the $g - r$ color suggests that the color temperature of the emission is $\sim 5500$ K assuming a blackbody spectrum, which gives $a \sim 2.4 \times 10^{10}$ cm using equation Kasen's equation (25). Note that this result is insensitive to the exact epoch of this observation since the model luminosity and temperature scale similarly ($\propto t^{-1/2}$) with time after explosion.\\ 

Hence, we find that a companion interaction scenario is unable to account for the early declining emission within the framework of the analytic equations presented in \citealt{Kasen2010}. However, we note that viewing angle dependencies may affect our conclusions. For example, Figure 2 in \citealt{Kasen2010} shows that both the peak luminosity and morphology of the companion interaction light curve may be significantly affected along lines of sight away from the companion. In fact, the early drop of a factor of $\approx 1.7$ in luminosity over the first 2 days in iPTF\,16hgs is indeed reminiscent of the bolometric evolution predicted along directions $\sim 90^{\circ}$ to the companion, as shown in \citealt{Kasen2010}. As the simulations with varying viewing angles (in \citealt{Kasen2010}) were specifically for Type Ia SN ejecta, future modeling will be required to understand if these signatures would be similar in ejecta with different compositions as in iPTF\,16hgs.\\

\subsection{Interaction with circumstellar material}

We now consider if interaction with dense external circumstellar material (CSM) can explain the early blue emission in iPTF\,16hgs. Given the projected location of iPTF\,16hgs inside its host galaxy (and hence its likely proximity to dense CSM), such a scenario can potentially explain the uniqueness of the light curve of iPTF\,16hgs with respect to the other members of this class. We can obtain rough estimates of the characteristics of this CSM by using the early bolometric light curve and the methods presented in \citealt{Smith2016}. The interaction luminosity can be estimated using,
\begin{equation}
L = \frac{1}{2} 4 \pi R^2 \rho V^3
\end{equation}
which depends on the velocity of the ejecta material $V$ and the density of the medium $\rho$.\\

Assuming a constant density CSM, and using $L \gtrsim 3 \times 10^{41}$ ergs s$^{-1}$, $V \sim 12000$ km s$^{-1}$ at $t \sim 1$ day after the explosion, we find $\rho \sim 3 \times 10^{-15}$ g cm$^{-3}$. This corresponds to a particle density of $5 \times 10^8$ cm$^{-3}$ if the CSM was dominated by He at a distance of $\sim 10^{14}$ cm from the progenitor. Instead, if we assume a constant mass loss wind like CSM density profile, using the same values as above, we estimate a $\dot{M} \sim 7 \times 10^{-5} \frac{v_{CSM}}{100\,\textrm{km/s}}$ M$_{\odot}$ yr$^{-1}$. However, we note that there is no evidence for spectral signatures of circumstellar interaction (as seen in Type IIn and Type Ibn SNe) in iPTF\,16hgs, which would argue against a CSM interaction scenario. Nevertheless, such signatures may be hidden in the case of asymmetric CSM configurations (e.g. in the form of a disk) where the interaction region is hidden by the expanding ejecta \citep{Smith2016}.\\

\subsection{Shock cooling of an extended progenitor}

\begin{figure}
\includegraphics[width=\columnwidth]{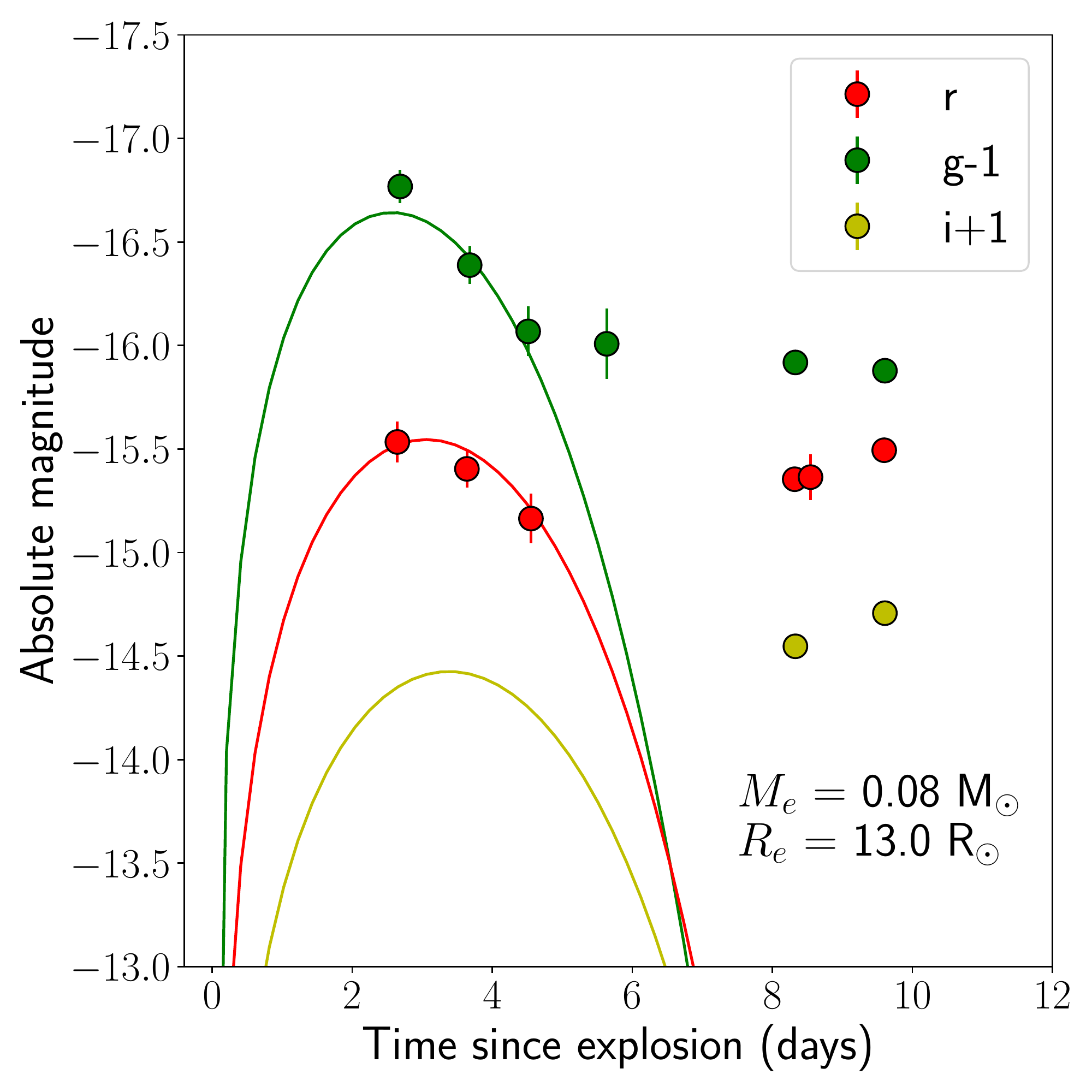}
\caption{The early light curve of iPTF\,16hgs fit with a shock cooling model of an extended envelope, as presented in \citealt{Piro2015}. The best-fit model indicates an extended mass of $M_e = 0.08$ M$_{\odot}$, $R_e = 13.0$ R$_{\odot}$ and explosion time of 12.7 days before the main peak of the $r$ band light curve.}
\label{fig:16hgs_shockCool}
\end{figure}

The double peaked light curves of some stripped envelope core-collapse SNe have been explained using shock cooling emission of an extended envelope around the progenitor. Such extended envelopes have been shown to be particularly relevant for sources that exhibit an early peak in the redder $r$ and $i$ bands, since `normal' progenitors (i.e. progenitors without an extended envelope) cannot reproduce such light curves \citep{Nakar2014,Piro2015,Sapir2017}. Since iPTF\,16hgs was found in a star forming host galaxy (indicating a core-collapse supernova origin is a possibility given the presence of a young stellar population), we examine if the early declining emission (which was detected in $g$ and $r$ bands) can be explained by shock cooling emission of an extended progenitor. \\

We use the extended envelope models of \citealt{Piro2015} to fit the first peak of iPTF\,16hgs. Adopting the ejecta mass and explosion energy as derived in the Arnett model and an optical opacity of $\kappa = 0.2$ cm$^2$ g$^{-1}$, the only other free parameters in the model is the mass in the extended envelope $M_e$ and the radius of the envelope $R_e$. Keeping the explosion time $t_0$ as an additional free parameter, we obtain the best-fit model as shown in Figure \ref{fig:16hgs_shockCool}. As shown, an extended envelope model with $M_e = 0.08$ M$_{\odot}$ and $R_e = 13.0$ R$_{\odot}$ is able to well reproduce the early peak for an explosion time $t_0 = -12.7$ days before the main peak of the $r$ band light curve. Although the model used here is very simplified in that it ignores the crucial density structure of the envelope, the numbers derived are expected to be correct to an order of magnitude (as shown in more realistic simulations including density profiles; \citealt{Piro2017}).\\

\section{Constraints on radio emission}
\label{sec:radioAn}

Although the progenitors for Ca-rich gap transients remain elusive, we expect a number of proposed progenitor channels to be associated with potentially bright radio counterparts if the explosion took place in a dense CSM. For example, radio emission from core-collapse and thermonuclear SNe can arise from synchrotron radiation produced by electrons accelerated in the forward shock of the SN explosion \citep{Chevalier1998}. On the other hand, if Ca-rich gap transients are associated with tidal disruptions of WDs, one would expect bright radio emission arising from the interaction of a relativistic collimated jet or a fast wind outflow with the surrounding CSM \citep{Macleod2016,Metzger2012,Margalit2016}. \\

However, their preference for remote locations far away from their host galaxies suggest that they also explode in likely low CSM density environments where such radio emission would be easily suppressed. Given its location close to its host galaxy (and thus potentially in a dense ISM environment), iPTF\,16hgs is thus useful to constrain models that predict significant radio emission from these transients. We thus use our radio limits to constrain models of radio emission associated with both spherical SN shocks as well as afterglows expected with collimated jet-like outflows in tidal disruption events.\\

\subsection{Radio emission from a spherical shock}

\begin{figure}[!ht]
\includegraphics[width=\columnwidth]{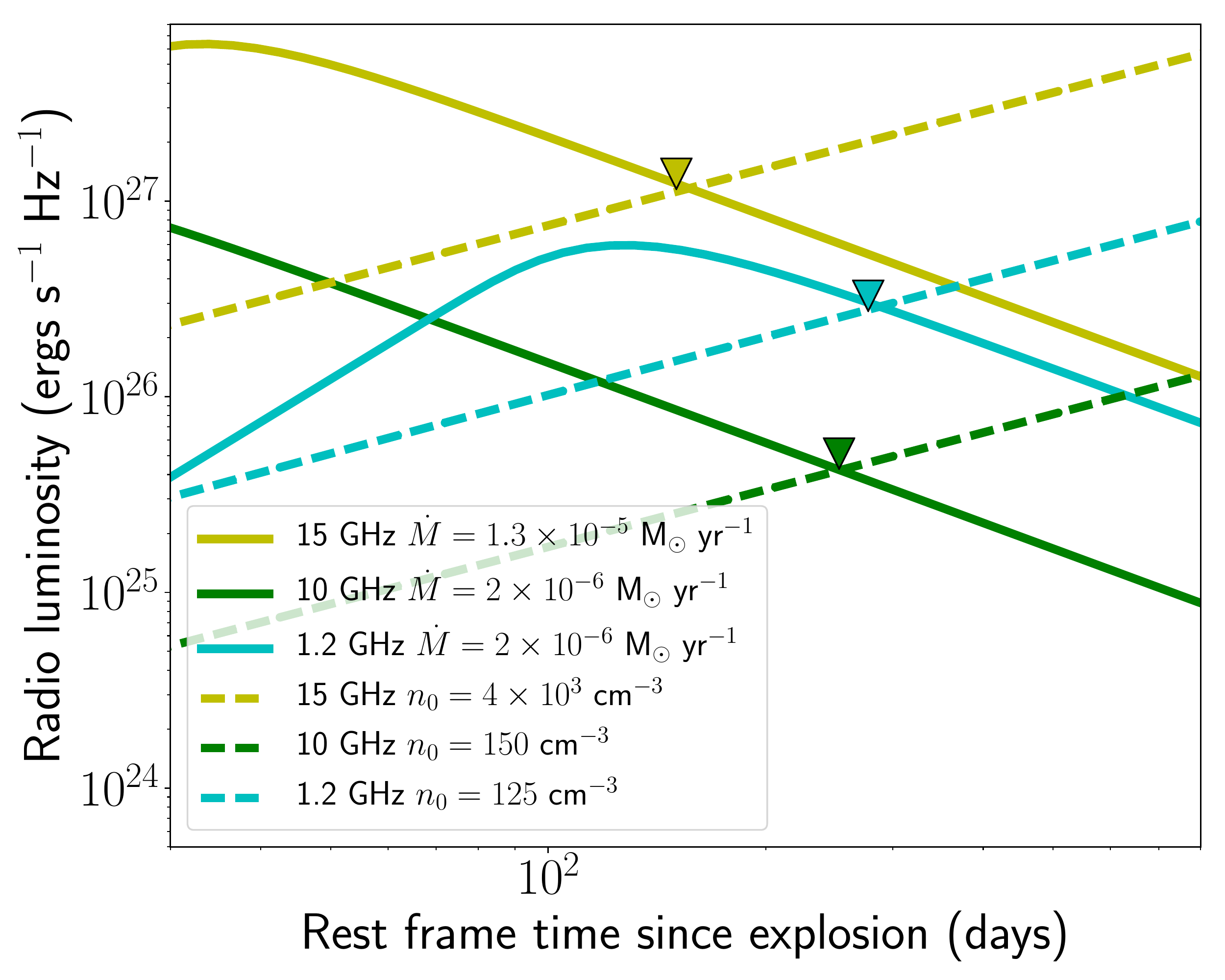}
\caption{Model radio light curves compared with radio limits on iPTF\,16hgs. The solid lines correspond to the limiting models consistent with the radio non-detections, for a wind-like CSM density profile ($\rho \propto r^{-2}$) at 15 GHz (yellow), 10 GHz (green) and 1.2 GHz (cyan) respectively. The corresponding limiting mass loss rates for $\epsilon_B = 0.1$ and $v_w = 100$ km s$^{-1}$ are indicates in the legend. If we adopt $\epsilon_B = 0.01$ instead, the limiting mass loss rates are $5.5 \times 10^{-5} \frac{v_w}{100 \textrm{km/s}}$ M$_{\odot}$ yr$^{-1}$, $8 \times 10^{-6} \frac{v_w}{100 \textrm{km/s}}$ M$_{\odot}$ yr$^{-1}$ and $8 \times 10^{-6} \frac{v_w}{100 \textrm{km/s}}$ M$_{\odot}$ yr$^{-1}$ for the AMI, VLA and uGMRT observations respectively. The dashed lines (with the same color coding) indicate the light curve models for a constant density environment, with $\epsilon_B = 0.1$. The corresponding limiting densities for $\epsilon_B = 0.01$ are $n_0 \lesssim 2.5 \times 10^4$ cm$^{-3}$, $n_0 \lesssim 10^3$ cm$^{-3}$ and $n_0 \lesssim 8 \times 10^2$ cm$^{-3}$ for the AMI, VLA and uGMRT observations respectively. }
\label{fig:16hgs_radioLC}
\end{figure}

We use the synchrotron self-absorption model of \citealt{Chevalier1998} to generate analytic radio light curves for a range of circumstellar environments. We follow the prescription given in \citealt{Chomiuk2016}, who present analytic equations for the expected radio light curves of Type Ia SNe based on \citealt{Chevalier1998} (but are also applicable to other hydrogen-poor SNe). We generate these light curves for both a constant wind mass loss environment (where $\rho = K r^{-2}$, with $K = \frac{\dot{M}}{4 \pi v_w}$) and a constant density environment ($\rho =$ constant), using an outer ejecta density profile of $\rho \propto r^{-10}$, as appropriate for compact progenitor stars \citep{Matzner1999}. By comparing the predicted radio light curves to those of our upper limits, we constrain the wind mass loss parameter $K$ and the external circumstellar density $n_0$ by obtaining the limiting cases for a 3$\sigma$ detection, as shown in Figure \ref{fig:16hgs_radioLC}. \\

In the case of a wind mass loss environment, the strongest constraints arise from the VLA (10 GHz) and uGMRT (1.2 GHz) observations, which we use to constrain the mass loss rate to $\dot{M} \lesssim 2 \times 10^{-6} \frac{v_w}{100 \textrm{km/s}}$ M$_{\odot}$ yr$^{-1}$ for $\epsilon_B = 0.1$. Adopting $\epsilon_B = 0.01$, the VLA and uGMRT observations constrain the mass loss rate to $\dot{M} \lesssim 8 \times 10^{-6} \frac{v_w}{100 \textrm{km/s}}$ M$_{\odot}$ yr$^{-1}$. For the case of a constant density environment, the strongest constraints arise from the uGMRT observations, which limit the circumstellar density to $n_0 \lesssim 125$ cm$^{-3}$ for $\epsilon_B = 0.1$, and to $n_0 \lesssim 800$ cm$^{-3}$ for $\epsilon_B = 0.01$.  \\

\subsection{Radio emission from a relativistic jet}
\label{sec:radioJet}

\citealt{Sell2015} proposed that Ca-rich gap transients could arise from tidal disruptions of low mass He WDs by intermediate mass black holes, based on the work of \citealt{Rosswog2009} (see also \citealt{Rosswog2008}; \citealt{Macleod2014}). In this scenario, when a low mass WD comes within the tidal radius of a massive compact object (with a mass of $\lesssim 10^{5}$ M$_{\odot}$), the WD is tidally crushed leading to a runaway thermonuclear detonation powering an optical transient. The accretion of the WD on to the compact object would then also power a super-Eddington X-ray flare, potentially leading to the launch of a relativistic jet \citep{Sell2015, Macleod2016}. One direct prediction of such a model is that these transients should then also be associated with prominent X-ray and radio emission for suitably oriented observing angles. \\

\citealt{Macleod2016} presented simulations of disruptions of 0.6 M$_{\odot}$ WDs by an intermediate mass black hole, including predictions for expected light curves and spectra of the thermonuclear transient. They show that the disruption of the WD together with the explosive detonation leads to less than half of the WD mass being accreted on to the BH. They also predict several characteristics of the radio emission that would be expected if these events produced relativistic jets that eventually interact with the surrounding interstellar medium (ISM). We thus use our deep radio limits on iPTF\,16hgs to constrain the phase space of jet energy and ISM density for different viewing angles of the observer. \\

The super-Eddington flare of accretion in such a disruption event can lead to the launching of a relativistic jet that carries away some fracion of the rest mass energy of the accreted WD \citep{Macleod2016}. We thus consider a range of jet energies from $10^{47}$ ergs to $2 \times 10^{51}$ ergs. If the disruption event in iPTF\,16hgs involved a 0.4 M$_{\odot}$ He WD (as expected from the He-rich spectra), the range of jet energies corresponds to a fraction of $\sim 10^{-6} - 10^{-2}$ of the rest mass energy of the accreted half of the WD. We then use the \texttt{BOXFIT} code \citep{vanEerten2010} to generate simulated multi-frequency radio light curves for different surrounding ISM densities (in the range between $10^{-6} - 1$ cm$^{-3}$). As suggested in \citealt{Macleod2016}, we also assume that the jet has an initial starting Lorentz factor of $\Gamma \sim 10$ and an opening angle of 0.2 rad (i.e. a jet beaming factor of 0.02). \\

\begin{figure*}
\centering
\includegraphics[width=\columnwidth]{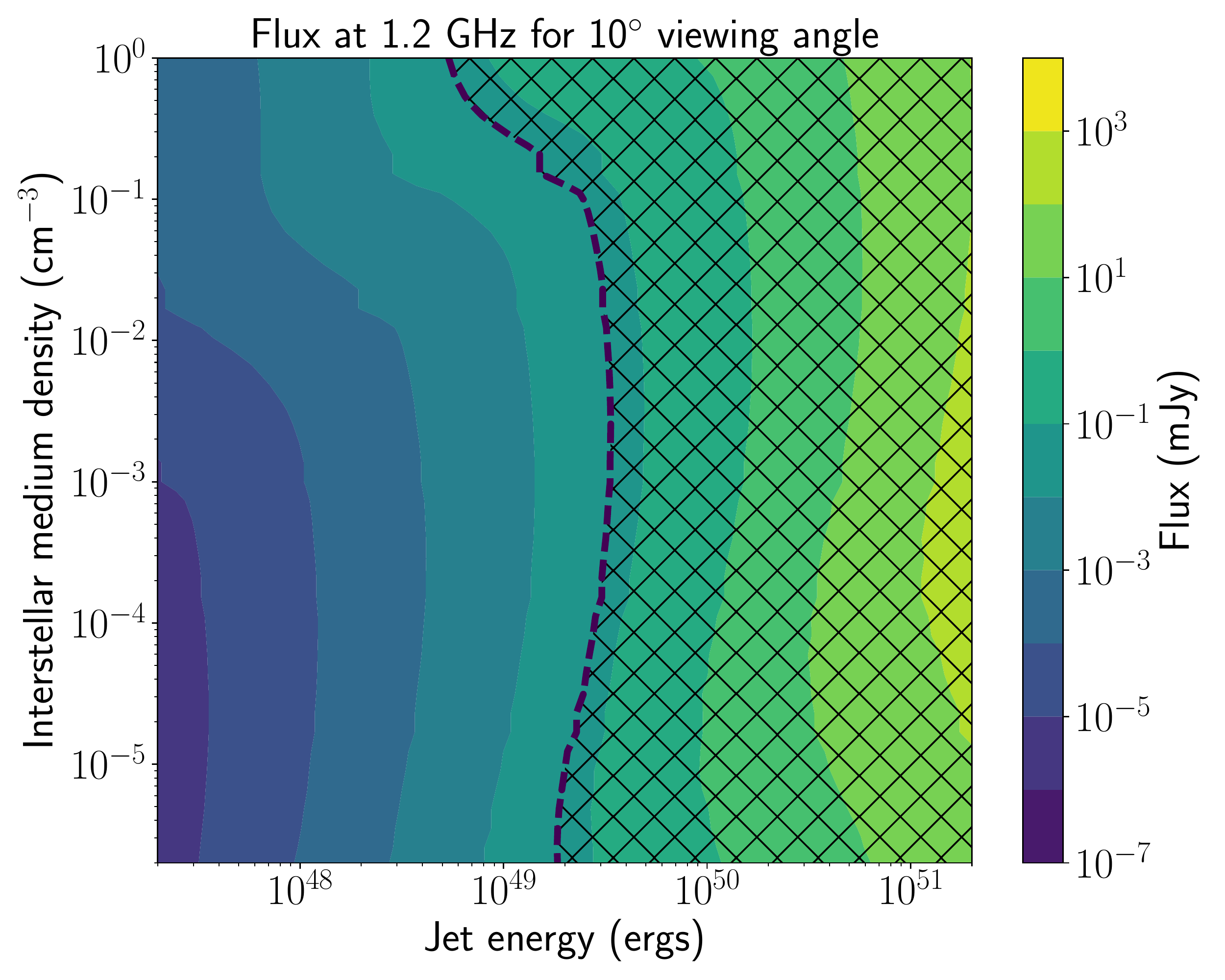}
\includegraphics[width=\columnwidth]{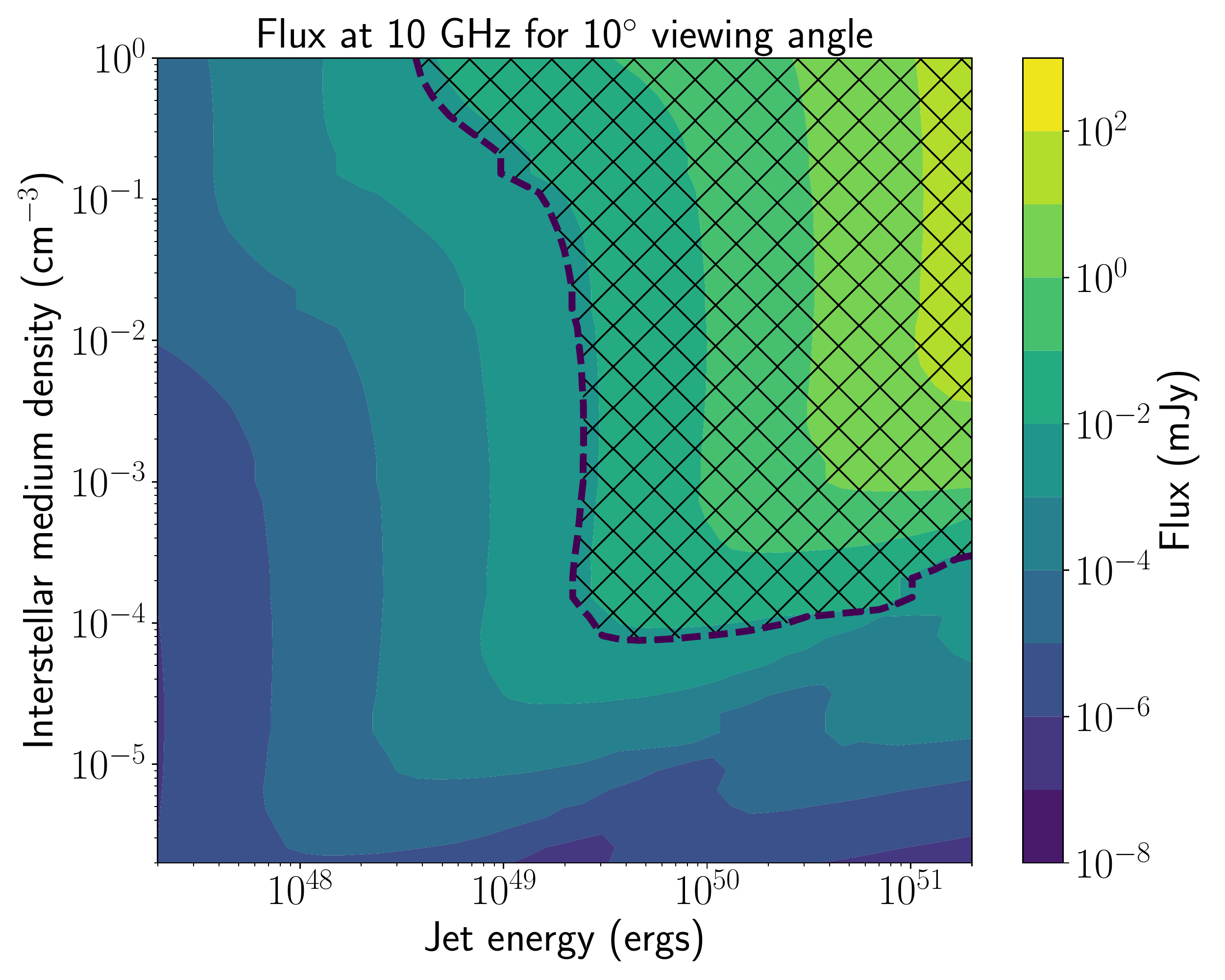}\\
\includegraphics[width=\columnwidth]{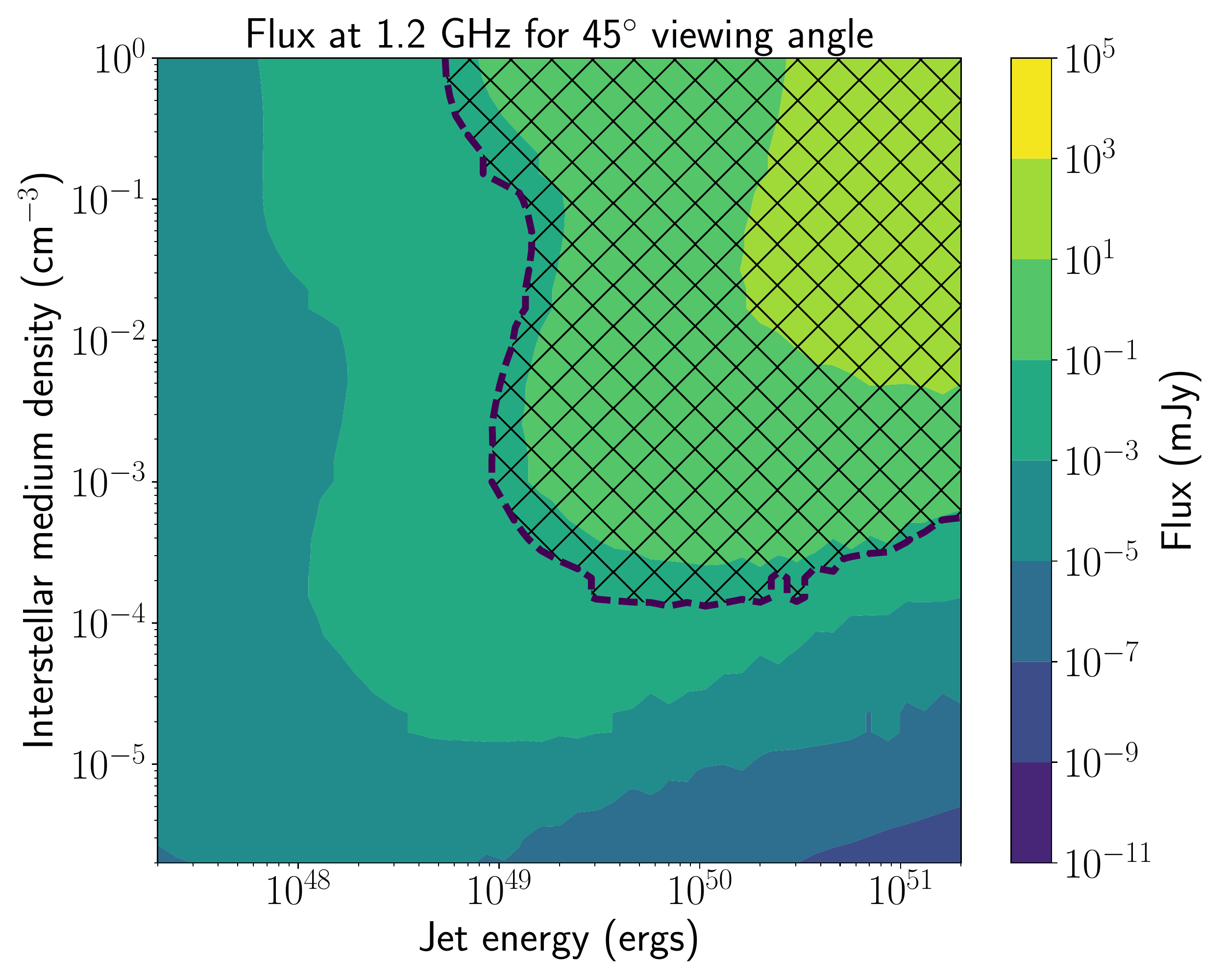}
\includegraphics[width=\columnwidth]{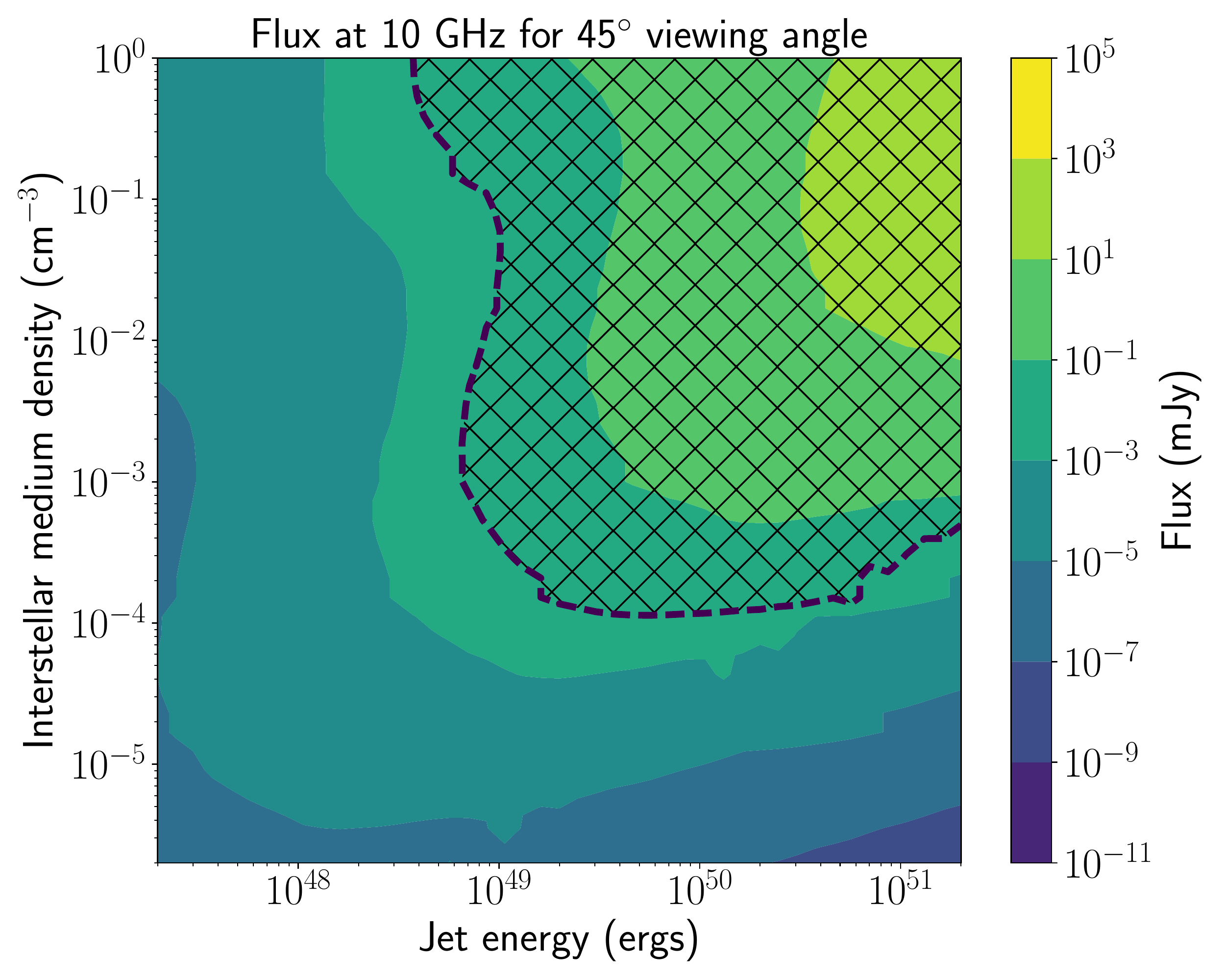}\\
\includegraphics[width=\columnwidth]{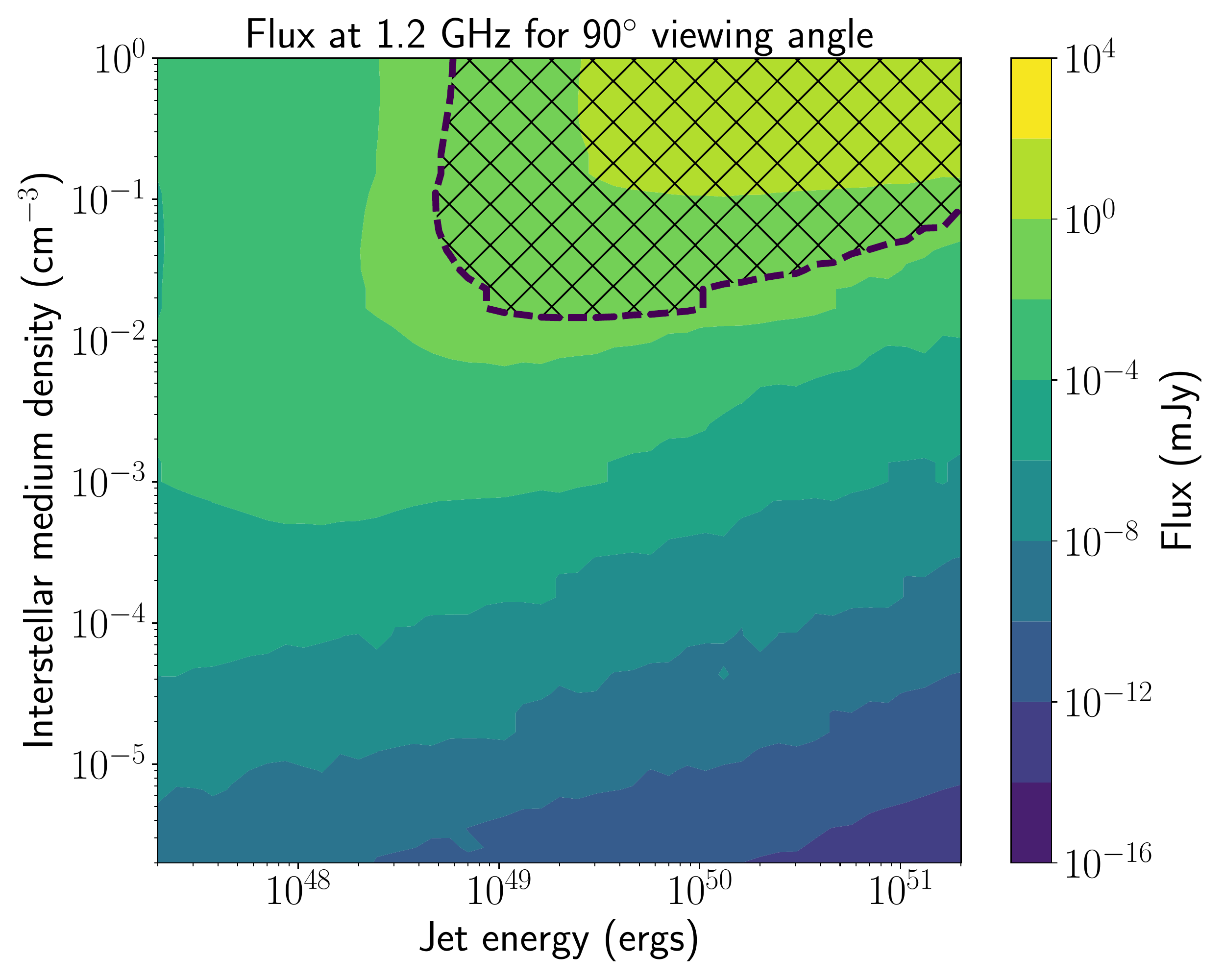}
\includegraphics[width=\columnwidth]{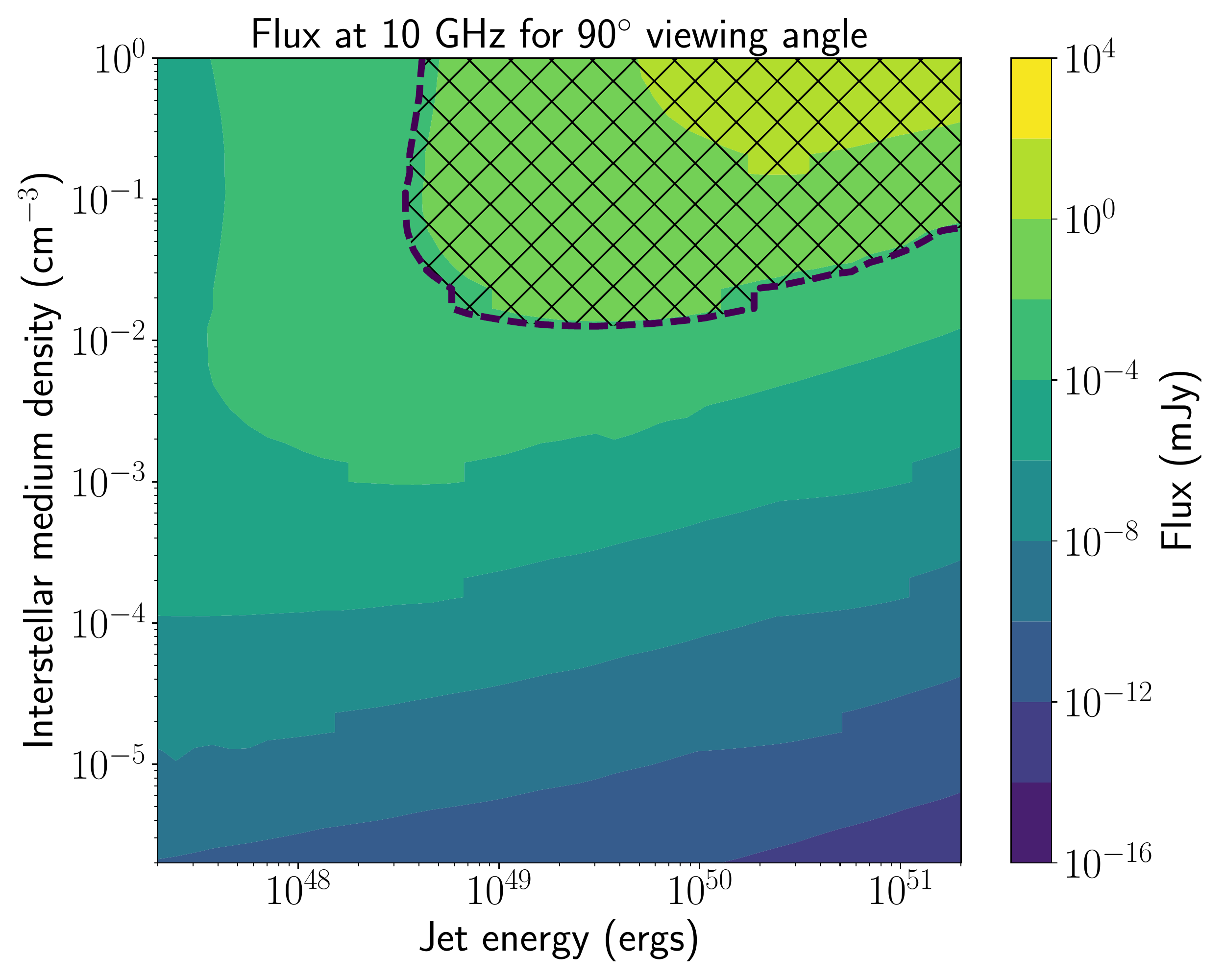}\\
\caption{Constrained parameter space for a collimated jet outflow in iPTF\,16hgs (for different viewing angles), as suggested by the radio non-detection with the uGMRT and VLA at late times. In each of the panels, we show a contour plot of the expected fluxes in the GMRT (1.2 GHz) and VLA (10 GHz) bands for a phase space of jet energy and ISM electron densities at a fixed observing angle (see text). The black dotted line in each panel shows the location of the 3$\sigma$ radio upper limit on iPTF\,16hgs. \textit{Only the phase space with a lower flux than the dotted line in each panel is allowed by the observations.} The hatched region is ruled out by our observations under the assumptions of the jetted outflow model.}
\label{fig:16hgs_radioTidal}
\end{figure*}

We show the contour plots (in the phase space of jet energy and circumstellar density) of the expected radio fluxes at the epochs of the VLA and GMRT observations for different viewing angles, together with our limits on the radio emission of this source in Figure \ref{fig:16hgs_radioTidal}. In each of the panels, the phase space ruled out by our observations are indicated by the hatched region. There are several interesting factors to note from the allowed phase space. First, for a nearly along the line of sight jet ($\approx 10^{\circ}$), the GMRT upper limits completely rule out jet energies higher than about $10^{49}$ ergs for ISM densities as low as $10^{-6}$ cm$^{-3}$. Note that the critical density of the universe is $\sim 10^{-6}$ cm$^{-3}$ while the electron density inside the host galaxy should be at least an order of magnitude larger, ruling out the on-axis case completely. For a viewing angle of 45$^{\circ}$, the GMRT and VLA upper limits together rule out ISM densities $\sim 10^{-4}$ cm$^{-3}$ if the jet energy is larger than $10^{49}$ ergs. For lower jet energies, our radio limits do not constrain the ISM environment since the limits lie above the optically thick locus of the light curves. Lastly, for a 90$^{\circ}$ observing angle, the radio limits are least constraining but nevertheless rule out ISM densities $\sim 10^{-2}$ cm$^{-3}$ if the jet energy was larger than about $5 \times 10^{48}$ ergs, but are not constraining if the jet energy was lower.\\ 

\section{The host galaxy of iPTF 16hgs}
\label{sec:hostEnv}

The host environment and location of iPTF\,16hgs is interesting in the context of Ca-rich gap transients for a number of reasons. First, it is only the second Ca-rich gap transient after PTF\,09dav to be found in a star forming spiral galaxy, as indicated by our spectrum of the host galaxy (Section \ref{sec:spectroscopy}). Apart from their preference for old environments, Ca-rich gap transients have also been noted for their large offsets from their host galaxies \citep{Kasliwal2012a, Lunnan2017}. In fact, their preference for remote locations appears to be intrinsic even after accounting for potential PTF survey biases against finding faint transients on bright galaxy backgrounds \citep{Frohmaier2017,Frohmaier2018}. Thus, it is interesting to note that when compared to the offset distribution of Ca-rich gap transients from their host galaxies, iPTF\,16hgs exhibits the smallest projected host offset of $\approx 5.9$ kpc ($\approx 1.9$ R$_{eff}$) of all known Ca-rich gap transients, both in terms of physical and host-normalized offset.\\

\begin{figure*}[!ht]
\centering
\includegraphics[width=0.85\textwidth]{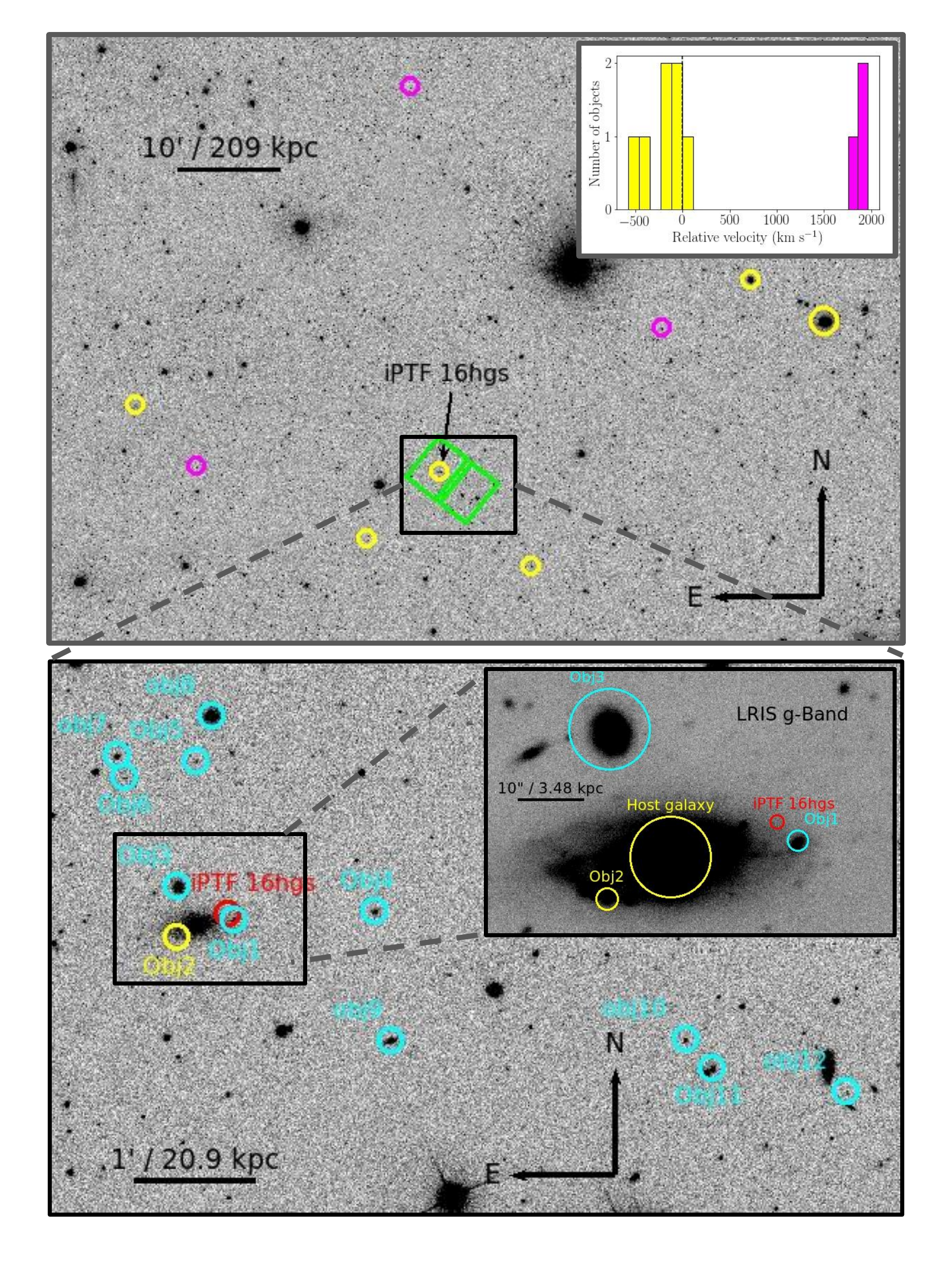}
\caption{Host environment of iPTF 16hgs. (Top) SDSS $r$-band image of the host region, with yellow circles indicating galaxies confirmed to be at a redshift consistent with that of the transient, and within 1 Mpc from its host galaxy. The magenta circles show galaxies with marginally different redshifts but within $\sim 1800$ km s$^{-1}$ of the host galaxy. The green boxes indicate the field of the view of Keck LRIS during the spectroscopic mask observation. The inset shows the velocity histogram (with respect to the host galaxy) of the galaxies found at a similar redshift as that of the apparent host galaxy. (Bottom) Galaxies with redshifts identified in the spectroscopic mask observation (within the green boxes in the top panel). Only the host galaxy $Obj2$ was found to be at the same redshift as that of the transient in this region (marked in yellow), while all other galaxies were found to be background sources (marked in cyan). The red circle marks the location of the transient. The inset shows the Keck LRIS $g$ band image of the host galaxy zoomed into the environment of the transient.}
\label{fig:16hgs_environment}
\end{figure*}

\citealt{Lunnan2017} show that Ca-rich gap transients also show a preference for group and cluster environments, as 7 out the 8 transients reported thus far were found in galaxy clusters or groups. In order to test such a scenario for iPTF\,16hgs, we searched NED for all galaxies within a projected offset of 1 Mpc from the host galaxy, and within a velocity of 3000 km s$^{-1}$, and found 9 such galaxies. 6 of these 9 galaxies had redshift within 500 km s$^{-1}$ of the host galaxy of iPTF\,16hgs, while an additional 3 galaxies were found to be clustered around a velocity offset of 2000 km s$^{-1}$ from the transient host galaxy. The locations of these galaxies are shown in Figure \ref{fig:16hgs_environment}, with yellow circles indicating the galaxies within 500 km s$^{-1}$ of the transient host galaxy, while magenta circles indicate the group offset by 2000 km s$^{-1}$.\\

Since none of these galaxies were at close projected offsets ($<$ 10\arcmin) from the host galaxy of iPTF\,16hgs, we undertook a spectroscopic mask observation of the region around the host galaxy with Keck LRIS to determine redshifts of nearby objects classified as galaxies in SDSS. The locations of the objects placed in the spectroscopic mask are shown in the lower panel of Figure \ref{fig:16hgs_environment}, while the redshifts are reported in Table \ref{tab:16hgs_hostSpec}. As shown, only one of the objects ($Obj2$) selected was found to be at the same redshift as the host galaxy. Although $Obj2$ is classified as a galaxy in SDSS, it lies right on top of the disk of the host galaxy, and is more likely to be a star forming region in the host galaxy itself.  We show the relative velocity distribution of the galaxy velocities of all objects found to be at the same redshift as iPTF\,16hgs in the histogram in Figure \ref{fig:16hgs_environment}. With a total of 6 - 10 objects (depending on whether the cluster at 2000 km s$^{-1}$ is associated to the host galaxy group) at the same redshift, we conclude that iPTF\,16hgs was located in a sparse galaxy group, consistent with other members of the class of Ca-rich gap transients.\\

\subsection{Global properties of the host galaxy}

We first estimate the gas phase metallicity of the host galaxy using the emission lines fluxes in the spectrum of its nucleus and the \texttt{pyMCZ} code \citep{Bianco2016}. The measured emission line fluxes are presented in Table \ref{tab:16hgs_hostFlux}. The code calculates the host oxygen metallicity, and is based on the original code of \citealt{Kewley2002} with updates from \citealt{Kewley2008}. Typical metallicity estimates derived using this method indicate 12 + $\log$(O/H) metallicity of 8.26$^{+0.03}_{-0.03}$ on the O3N2 scale of \citealt{Pettini2004} and $8.21^{+0.02}_{-0.02}$ on the O3N2 scale of \citealt{Marino2013}. In general, we note that all the derived oxygen metallicity indicators suggest a significantly sub-solar metallicity (where 12 + $\log$(O/H)$_{\odot} \approx 8.69$; \citealt{Asplund2009}) of $\approx 0.4$ Z$_{\odot}$ ($Z \approx 0.008$) for the spectrum of the nucleus. The low metallicity estimate places the host galaxy in the lowest 10\% of the distribution of host galaxy metallicities of Type Ib/c SNe, while it is on the lowest 30\% of the range of the host galaxies of Type Ic-BL SNe \citep{Sanders2012}.\\

Next, we use the integrated fluxes of the host galaxy to estimate the global properties of the stellar population in the host galaxy of iPTF\,16hgs. The photometric fluxes in the SDSS $ugriz$, 2MASS $JHK$ and GALEX FUV/NUV bands were fit using the \texttt{FAST} code \citep{Kriek2009}. The fitting was performed assuming a \citealt{Maraston2005} stellar population, an exponentially declining star formation history, a Salpeter IMF and a Milky Way like extinction law. We also constrain the models to be at a sub-solar metallicity (as indicated by the spectra of the host galaxy) of $Z = 0.01$ (0.5 Z$_{\odot}$), which is the model grid closest to the inferred metallicity. \\

Using this model, we obtain a best-fit stellar mass of $6.45^{+0.31}_{-0.29} \times 10^8$ M$_{\odot}$, mean stellar population age of $1.99^{+0.05}_{-0.17} \times 10^8$ years. The integrated star formation rate is poorly constrained from the photometry only, and hence we estimate it from the H$\alpha$ maps in our IFU observations (see Section \ref{sec:ifuProp}). Integrating the H$\alpha$ flux over the entire map where H$\alpha$ emission is detected, we find a total H$\alpha$ flux of $\approx 10^{-13}$ ergs cm$^{-2}$ s$^{-1}$. Converting this to an equivalent star formation rate using the redshift of the host galaxy (total H$\alpha$ luminosity of $\approx 6.5 \times 10^{40}$ ergs s$^{-1}$)  and the relations in \citealt{Kennicutt1998}, we get an integrated star formation rate of 0.5 M$_{\odot}$ yr$^{-1}$, placing this galaxy in the lower half of the distribution of star formation rates found in the hosts of stripped envelope SNe \citep{Galbany2014}. \\

\begin{figure*}
\includegraphics[width = \textwidth]{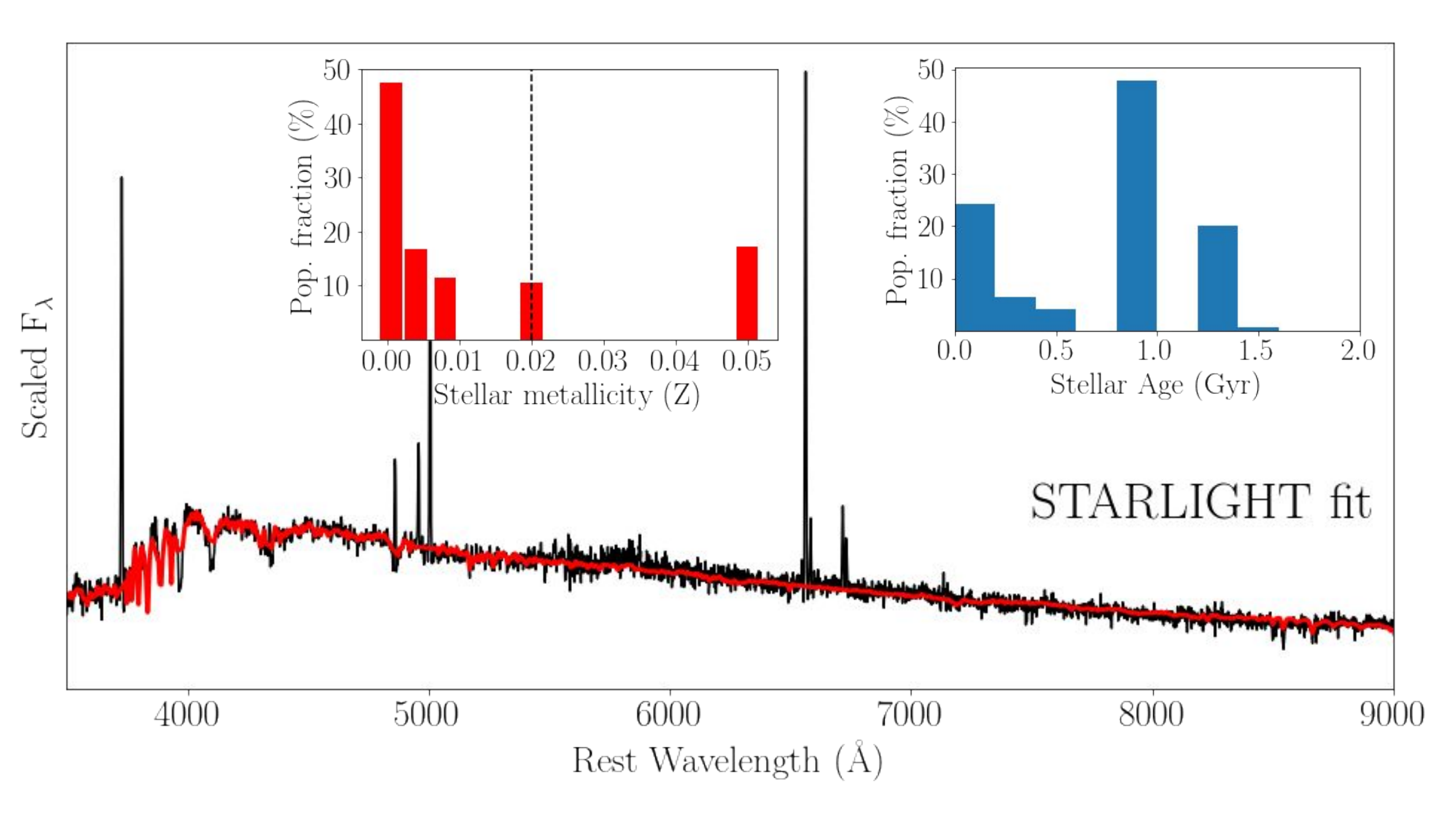}
\caption{The stellar continuum and absorption line spectrum of the nucleus of the host galaxy of iPTF\,16hgs fit with the STARLIGHT code \citep{CidFernandes2005}. The black line is the raw spectrum while the red line is the best-fit star light model. The histograms in the inset panels show the stellar population mixtures (in terms of age and metallicity) found in the best-fit model. As shown, the stellar continuum suggests a mixture of both old (age $\gtrsim 1$ Gyr) and young (age $\lesssim 0.5$ Gyr) stellar populations in the host galaxy. The majority of the stellar population is found to be at a sub-solar metallicity (solar metallicity is indicated by the dashed line in the metallicity panel), consistent with the low metallicity inferred from the emission line spectrum.\\}
\label{fig:16hgs_starlight}
\end{figure*}

We also use our Keck-LRIS spectrum of the nucleus of the host galaxy to estimate the stellar age and stellar phase metallicity of the older stellar populations in the galaxy. Similar to the analysis presented in \citealt{Galbany2016}, we fit the stellar continuum and absorption features in the host nucleus spectrum using the STARLIGHT code \citep{CidFernandes2005}. Using a \citealt{Cardelli1989} dust extinction law and \citealt{Bruzual2003} stellar population models at a range of metallicities (from Z = 0.001 to Z = 0.05), we find the best-fit spectrum as shown in Figure \ref{fig:16hgs_starlight}. The insets in Figure \ref{fig:16hgs_starlight} also show the best-fit stellar population mixture computed by STARLIGHT. As shown, the stellar continuum is well fit by a mixture of both old (age $\gtrsim$ 1 Gyr) and young (age $\lesssim 0.5$ Gyr) stellar populations, where $>$ 65 \% of the stellar population in the STARLIGHT fit is in the former category.\\

In terms of metallicity, more than 70\% of the stellar population in the best STARLIGHT fit is at a sub-solar metallicity ($Z < 0.02$), with $\approx 50$\,\% of the population at $Z < 0.001$. Note that the stellar metallicity reflects the galaxy metallicity when the stars were formed (which is likely to be lower than the current metallicity), while the emission line based estimates reflect the current gas-phase metallicity in the galaxy. Weighting over the entire population mixture produced by the model, we find a mean stellar age of $7.2 \times 10^8$ yrs and a mean stellar metallicity of $Z = 0.012$. Thus, the mean stellar age estimated from the nuclear spectrum is older than that inferred for the whole galaxy from the broadband photometry, while the mean stellar metallicity is similar to the gas phase metallicity estimated from the emission line fluxes in the nuclear spectrum. Taken together, we conclude that the host galaxy of iPTF\,16hgs is a metal-poor, star forming dwarf galaxy with a mixture of both young and old stellar populations.

\subsection{Spatially resolved properties of the host galaxy}

\label{sec:ifuProp}

\begin{figure*}[!ht]
\centering
\includegraphics[width=0.45\textwidth]{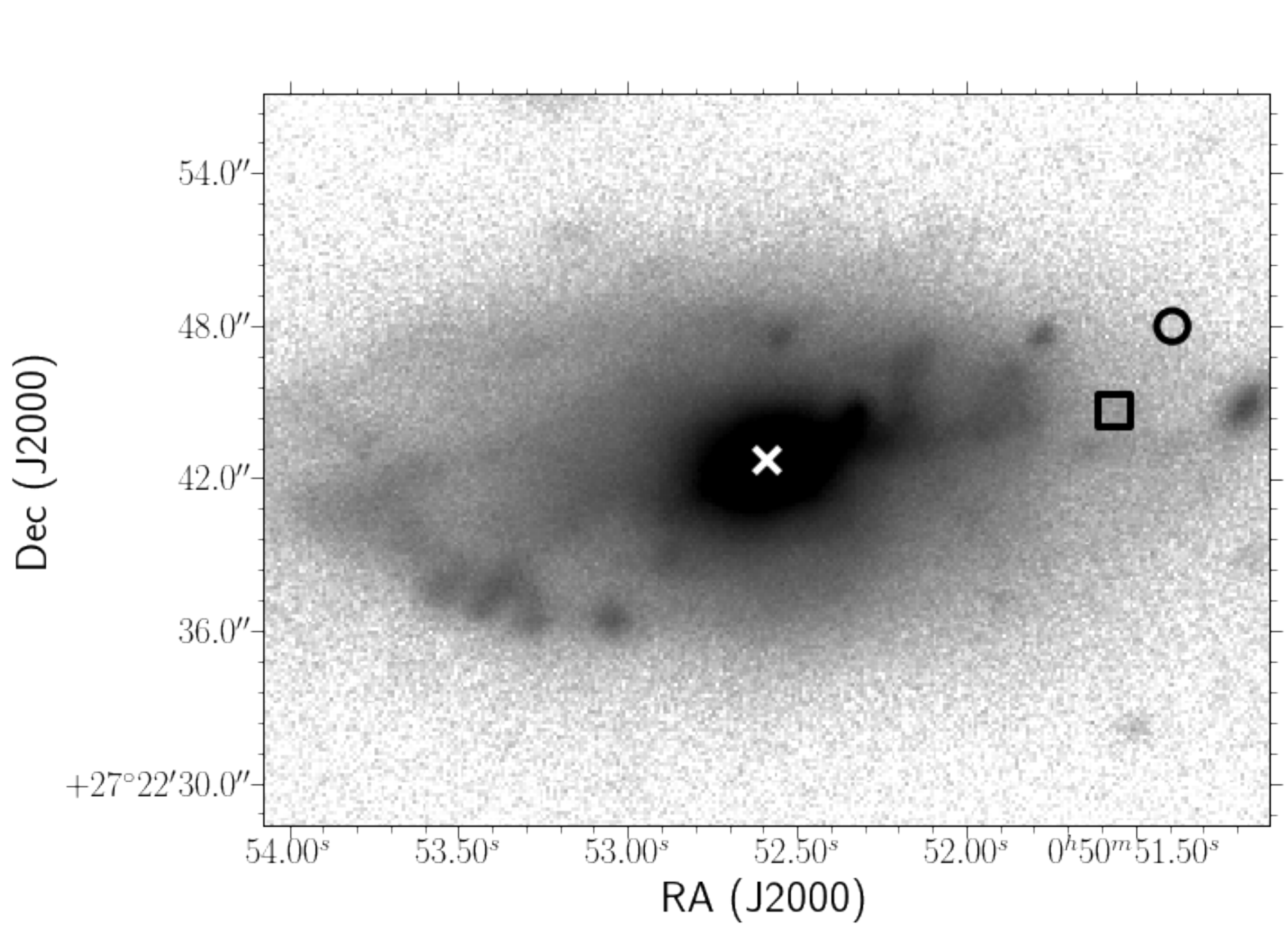}
\includegraphics[width=0.49\textwidth]{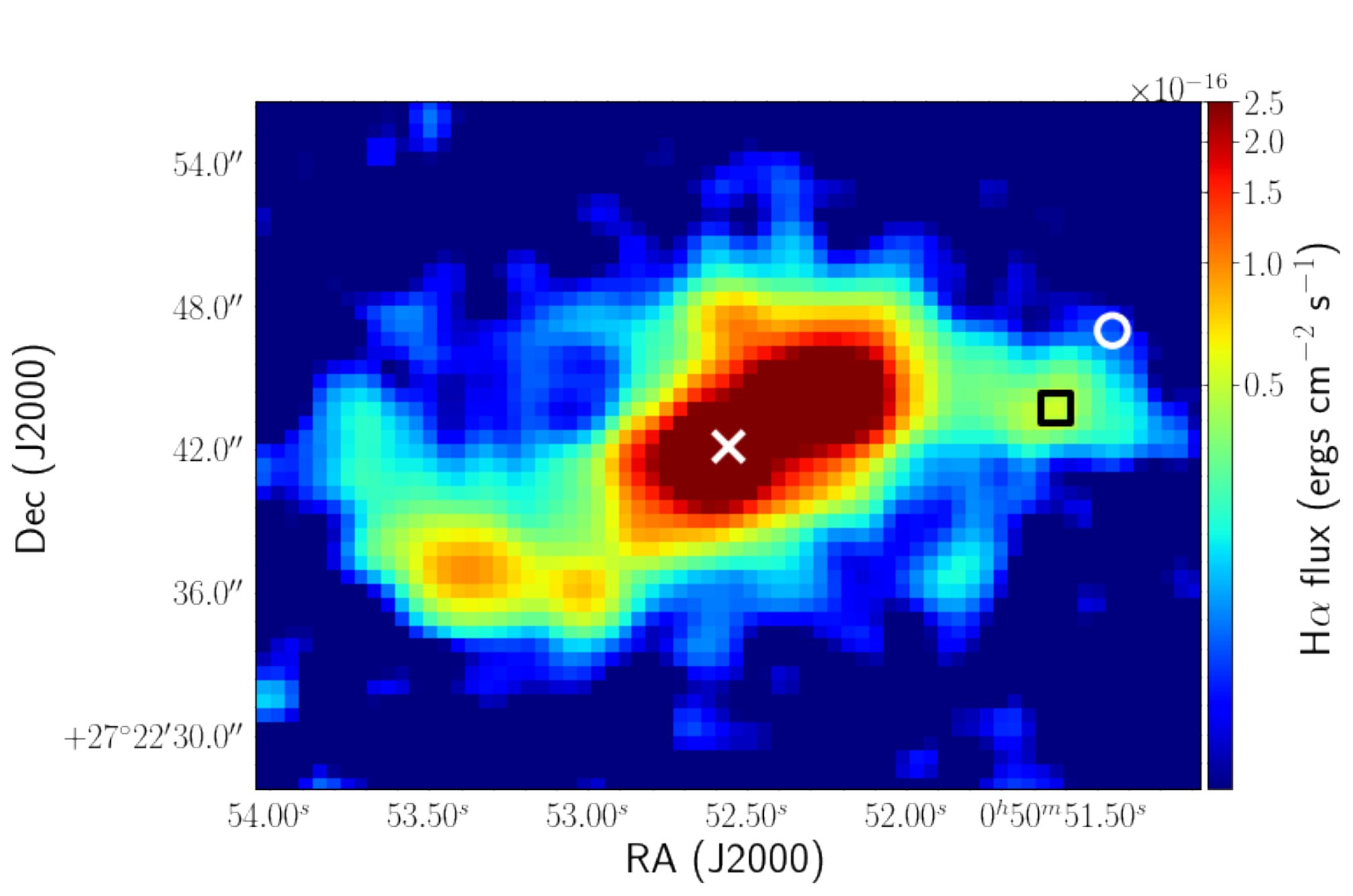}\\
\includegraphics[width=0.49\textwidth]{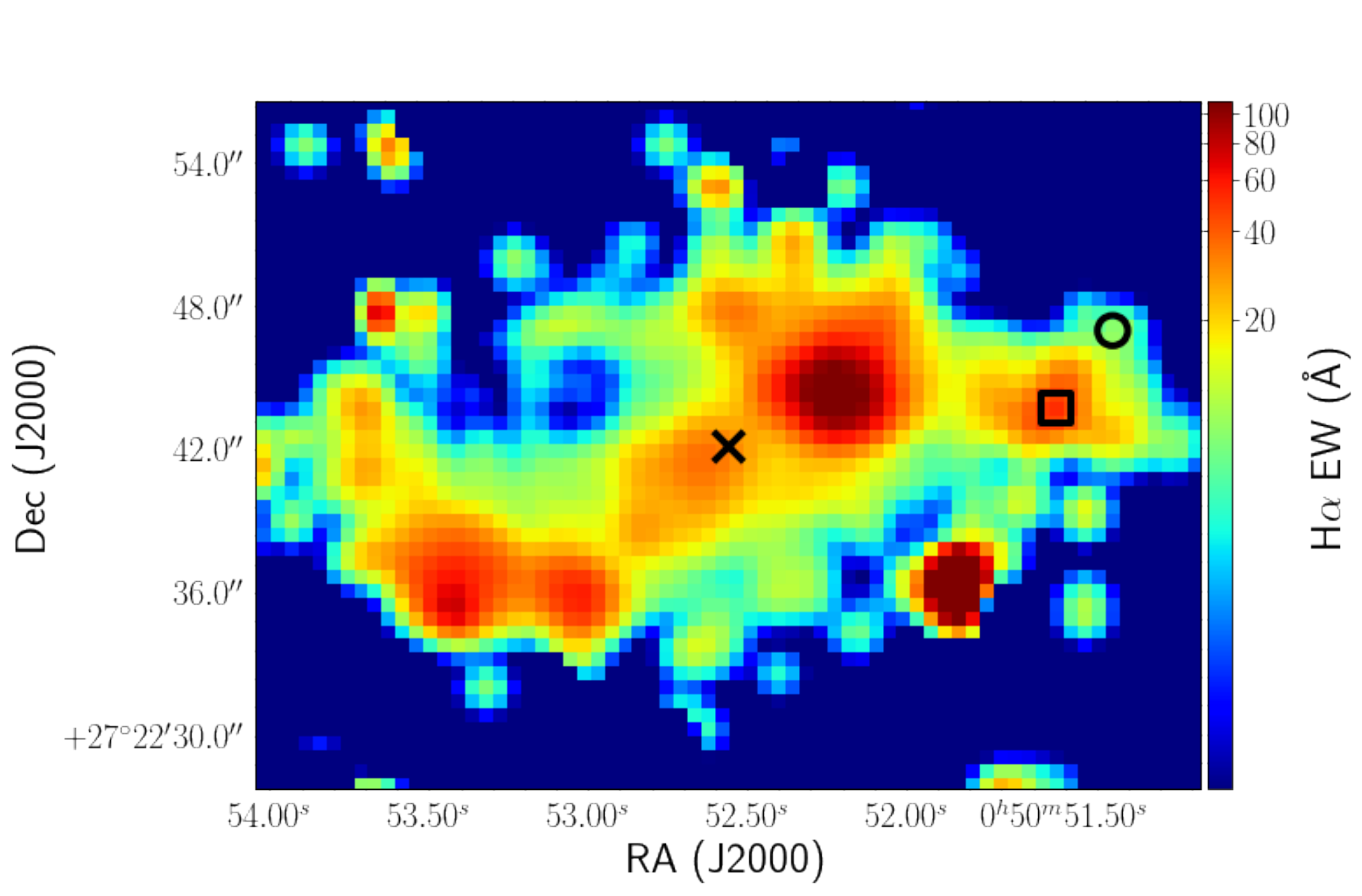}
\includegraphics[width=0.49\textwidth]{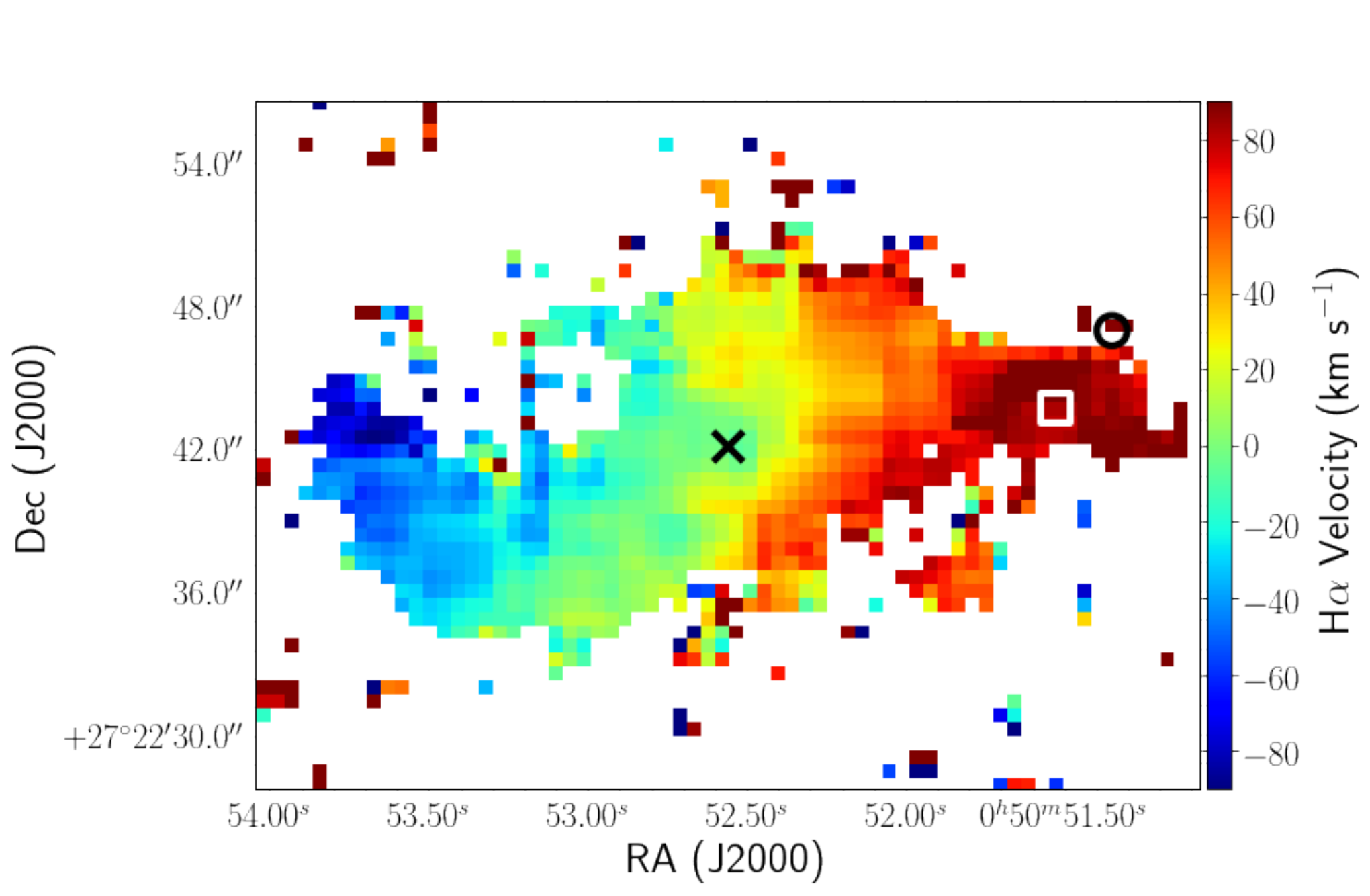}\\
\includegraphics[width=0.49\textwidth]{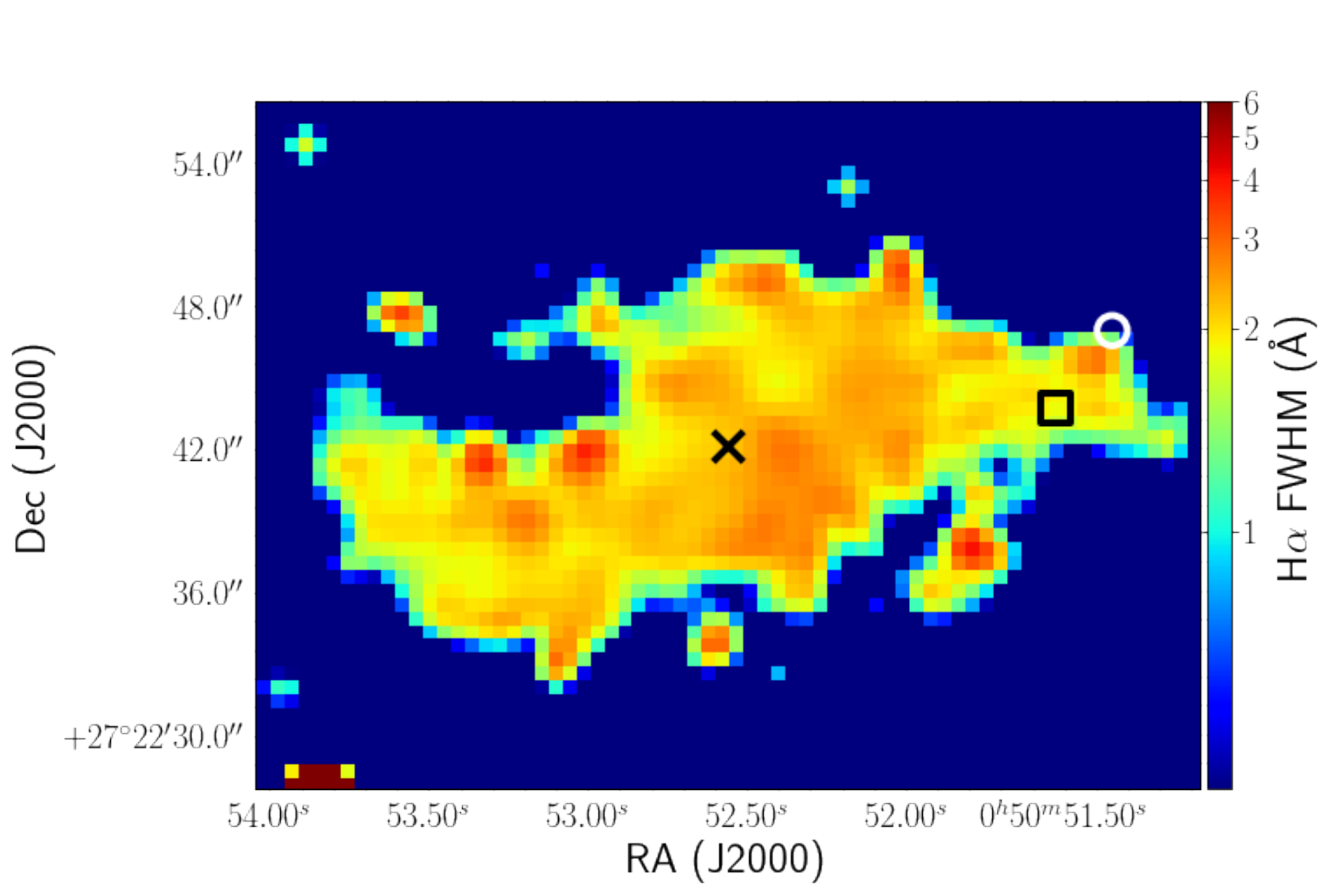}
\includegraphics[width=0.49\textwidth]{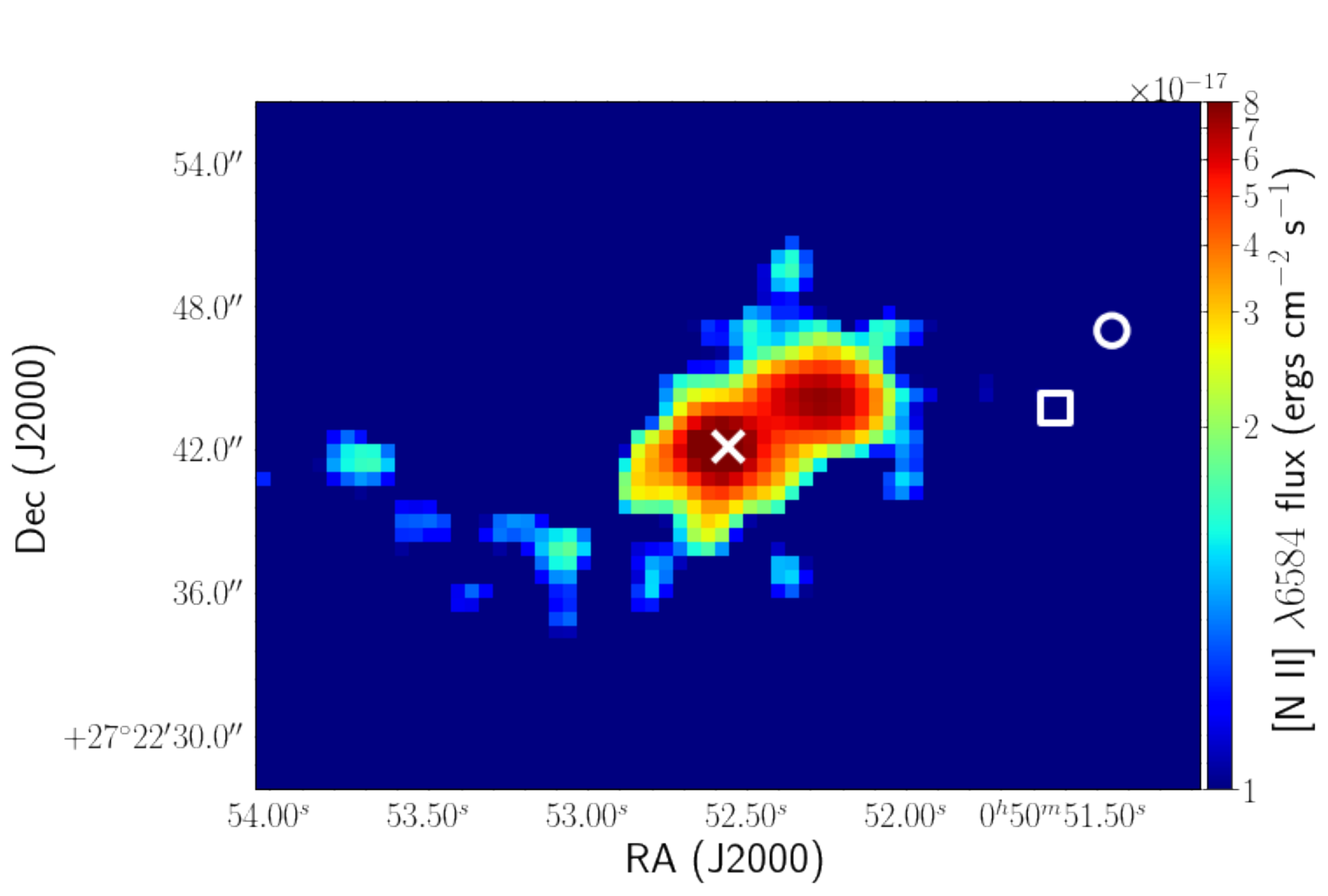}\\
\caption{Maps obtained from the late-time observations of the host galaxy of iPTF 16hgs. (Top: left) Late time LRIS $g$-band image of the host galaxy of iPTF 16hgs. (Top: right) H$\alpha$ flux map of the host galaxy. (Middle: left) H$\alpha$ equivalent width (EW) map of the host galaxy. (Middle: right) H$\alpha$ velocity map (with respect to the systemic velocity of the host galaxy) The rotation of the spiral galaxy is clearly visible. (Botttom: left) Map of the Full Width at Half Maximum (FWHM) of the H$\alpha$ line over the galaxy. (Bottom: right) [N II] $\lambda$6584 flux map of the host galaxy. (see text). In each of the panels, the cross denotes the cataloged position of the center of the host galaxy, the circle denotes the location of the transient and the square denotes the location of the nearest H-II region to the location of the transient.}
\label{fig:16hgs_IFUmaps1}
\end{figure*}

Due to its small offset from its host galaxy, we obtained IFU observations of the host galaxy with the PCWI to study the spatially resolved ISM of the host galaxy, and in particular, the local ISM environment of iPTF\,16hgs. For each pixel in the reduced and stacked spectral cube from the PCWI, we modeled the continuum emission using a low degree polynomial, and subtracted it out to measure the emission line fluxes of the most prominent lines from the host galaxy. The lines that were within the spectral cube include H$\alpha$, [N II] $\lambda$6584, [S II] $\lambda$6716 and $\lambda$6731. For the strongest H$\alpha$ emission line, we also measure the velocity by fitting a Gaussian profile to the continuum subtracted H$\alpha$ feature, its equivalent width and full width at half maximum (FWHM). The resulting maps are shown in Figure \ref{fig:16hgs_IFUmaps1}. For comparison, we also show a continuum image of the host galaxy, as obtained from the late-time LRIS observation in Figure \ref{fig:16hgs_IFUmaps1}.\\

A few features are readily apparent from the maps in Figure \ref{fig:16hgs_IFUmaps1}. iPTF\,16hgs occurred in a star forming spiral galaxy, as indicated by the morphology of the galaxy in the continuum image and the prominent H$\alpha$ emission extending through out the image. The multiple blobs of H$\alpha$ emission are likely to be individual H-II regions in the disk of the galaxy. The H$\alpha$ velocity map clearly shows evidence of ordered rotation of the spiral arms of the galaxy (reaching a velocity of $\approx 100$ km s$^{-1}$ near the edges) such that the location of iPTF\,16hgs is on the receding arm of the galaxy that is viewed close to the plane of the disk. In particular, we note that the object $Obj2$ in the spectroscopic mask (that was classified as a separate galaxy in SDSS) is consistent with being an individual H-II region in the host galaxy given that its velocity lies exactly on the rotation curve of the host.  \\

The H$\alpha$ equivalent width (EW) map also shows regions of very high EW ($\gtrsim$ 100 \AA), suggestive of very young stellar populations in the host galaxy.  The metal emission lines of S and N are significantly weaker than the bright H$\alpha$ emission (as would be expected from a metal-poor galaxy), and are thus detected only near the nucleus, and along the plane of the galaxy. In particular, we note that the metal emission line maps show evidence of bright emission regions co-located with the bright blobs of H$\alpha$ emission near the nucleus of the host galaxy.\\

\subsubsection{Star formation density}

We use the H$\alpha$ flux map from our observations to measure the equivalent star formation surface density using the relations in \citealt{Kennicutt1998}. Since the galaxy is viewed nearly edge on, we caution that the star formation density will be subject to a projection effect. Nevertheless, we show the spatially resolved star formation density in Figure \ref{fig:16hgs_IFUPhysMaps}. As shown, the host galaxy of iPTF\,16hgs exhibits multiple prominent blobs of star formation, reaching surface densities $\gtrsim 0.1$ M$_{\odot}$ yr$^{-1}$ kpc$^{-2}$. Interestingly, there is evidence for outlying H-II regions in the host galaxy, as evidenced by the bright H$\alpha$ emission blob located to the south-west of the nucleus, which is very faint in the continuum image (and hence has a very high EW). Given that this blob lies on the rotation curve of the galaxy, it likely represents a very young star forming H-II region as opposed to a companion dwarf galaxy. Additionally, there is clear evidence of a large star forming region close to the location of iPTF\,16hgs, at a projected offset of $\approx$ 3\arcsec ($\approx 1$ kpc at the redshift of the galaxy; denoted by the black square) from the location of the transient. \\

\subsubsection{Metallicity map}

Next, we examine the spatially resolved gas phase metallicity of the host galaxy of iPTF\,16hgs. The wavelength range of our observations include a number of important metallicity diagnostics, i. e., the [N II] $\lambda$6584 and the [S II] $\lambda\lambda$6716, 6731 lines. We thus compute the oxygen metallicity 12 + $\log$(O/H) using two different estimators. We use the N2 index calibration as presented in \citealt{Pettini2004}, which uses the ratio of the [N II] $\lambda$6584 and the H$\alpha$ line to estimate the metallicity. Additionally, we also use the calibration presented in the \citealt{Dopita2016} (hereafter D16), which is based on photo-ionization models and is robust to changes in ionization parameter. The D16 index is based on the [N II] $\lambda$6584 and H$\alpha$ lines, in addition to the [S II] $\lambda\lambda$6716, 6731 lines.\\

\begin{figure*}
\centering
\includegraphics[width=0.49\textwidth]{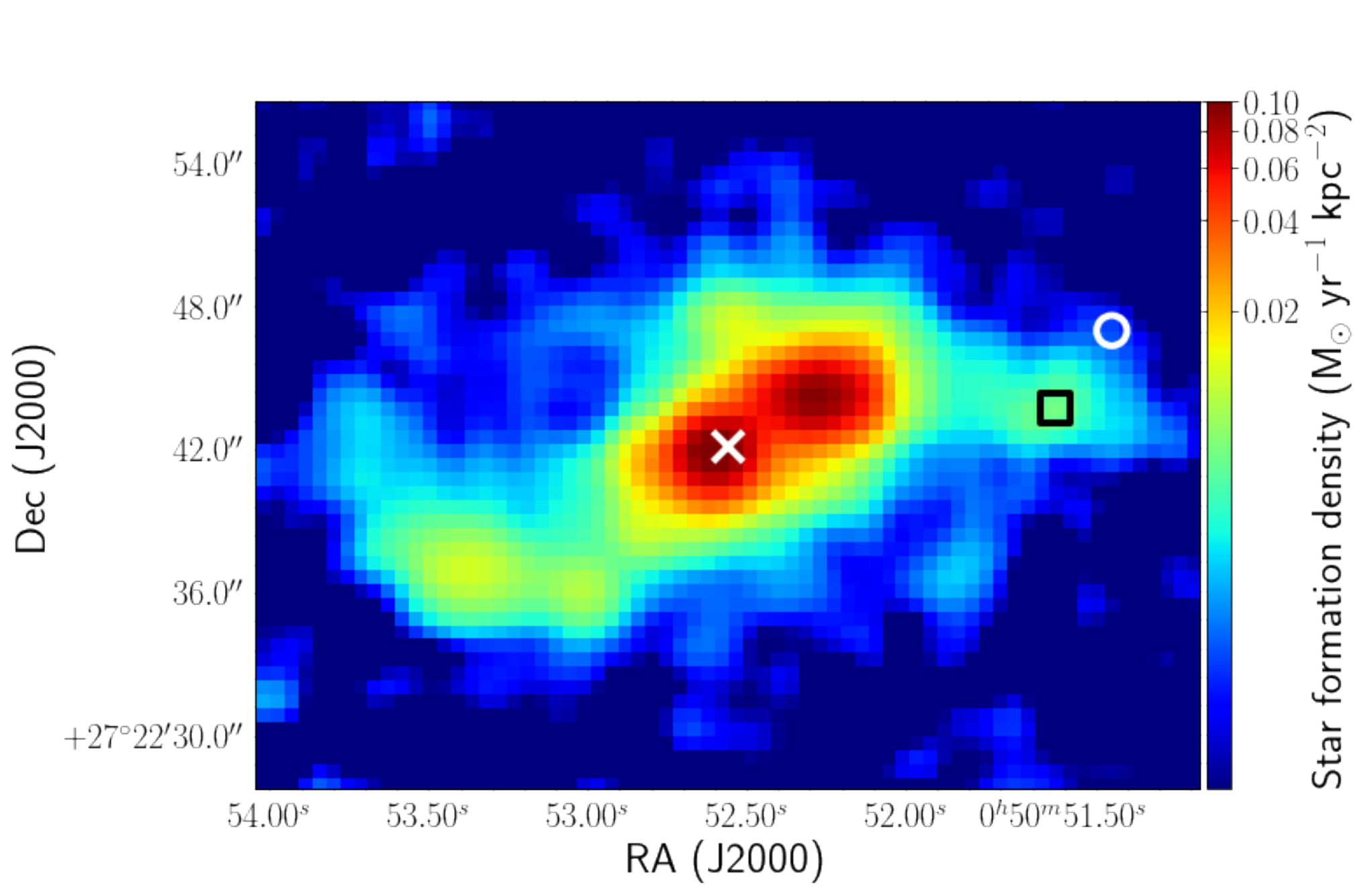}
\includegraphics[width=0.49\textwidth]{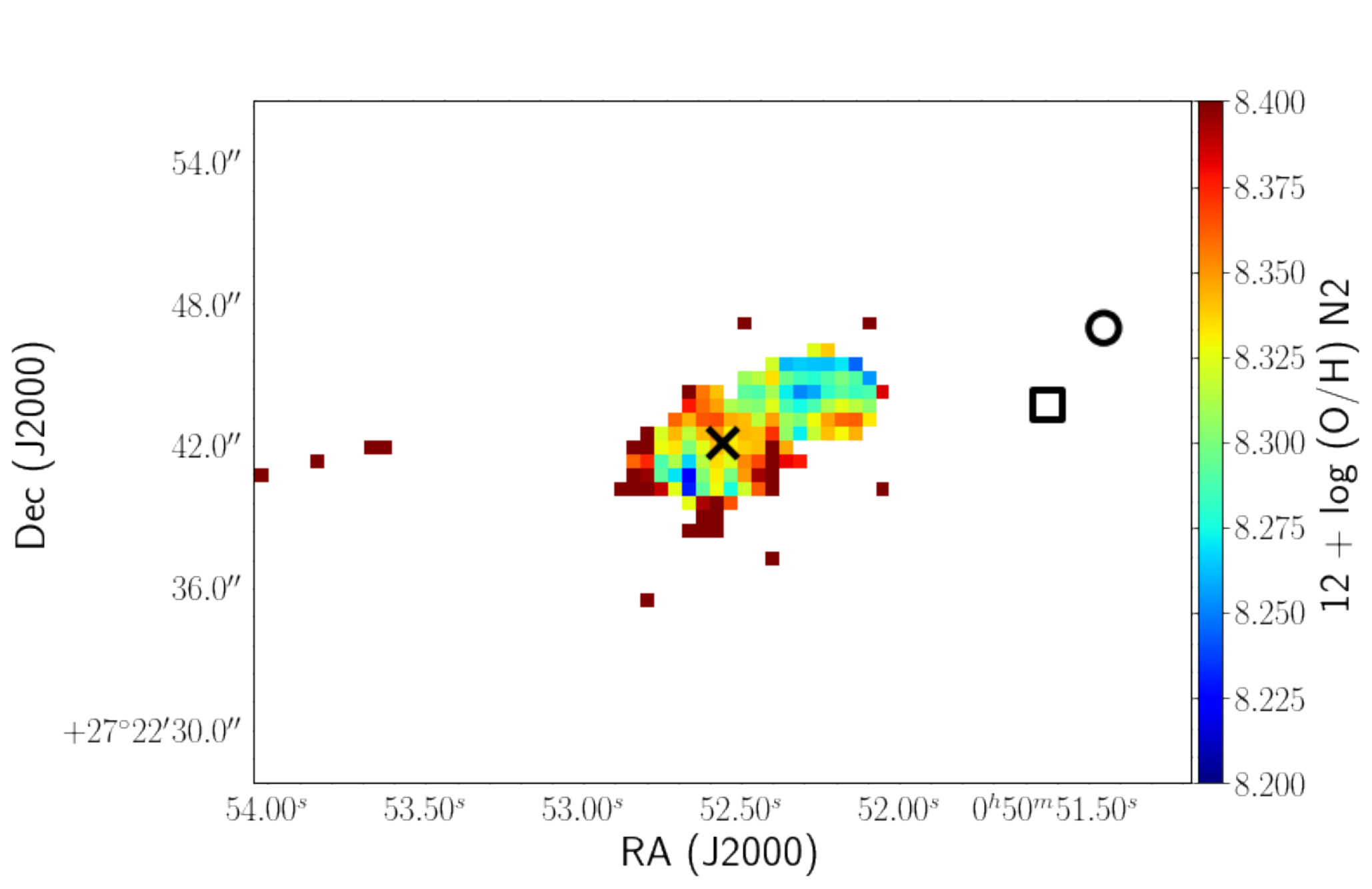}\\
\includegraphics[width=0.49\textwidth]{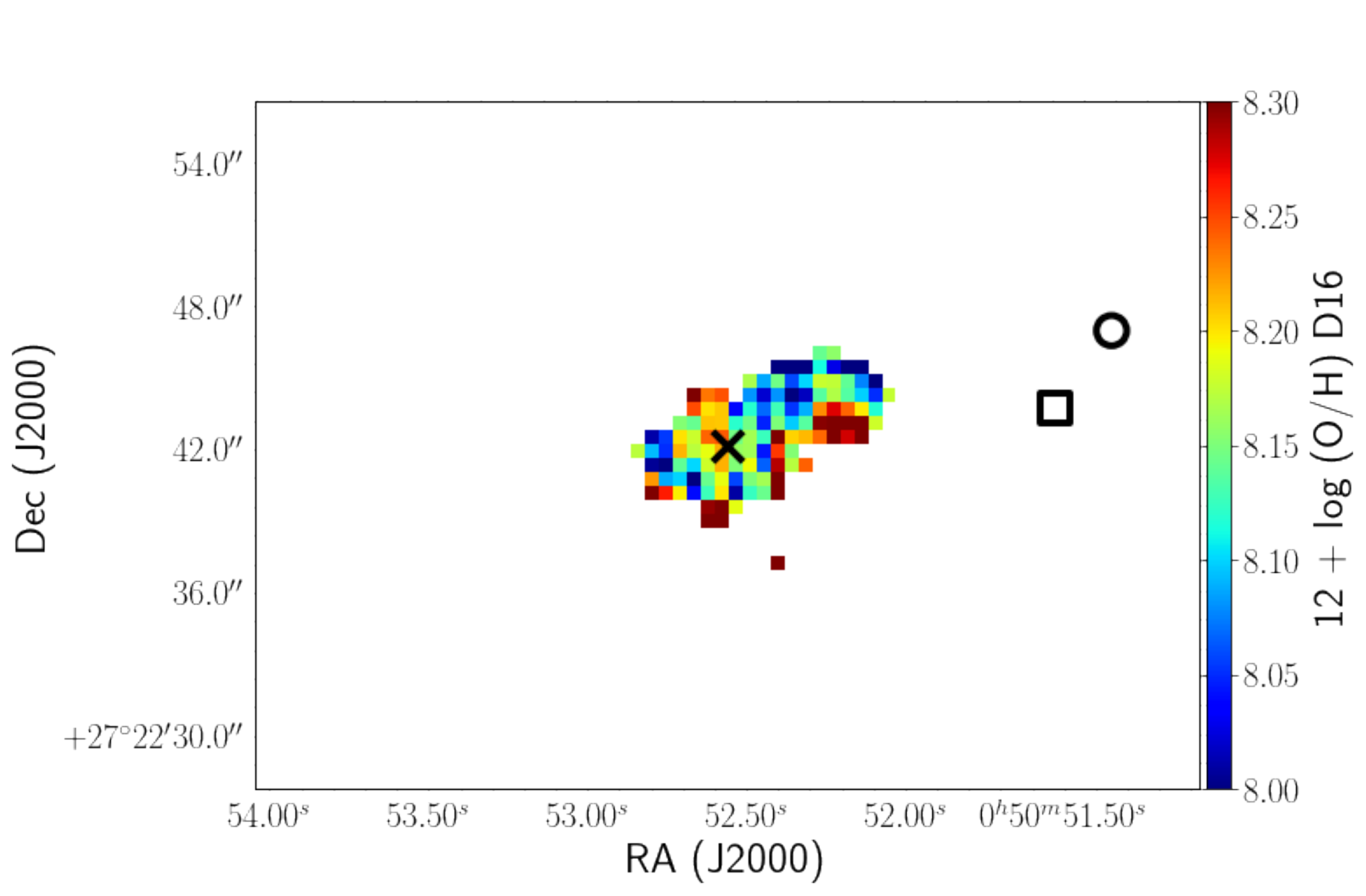}
\includegraphics[width=0.49\textwidth]{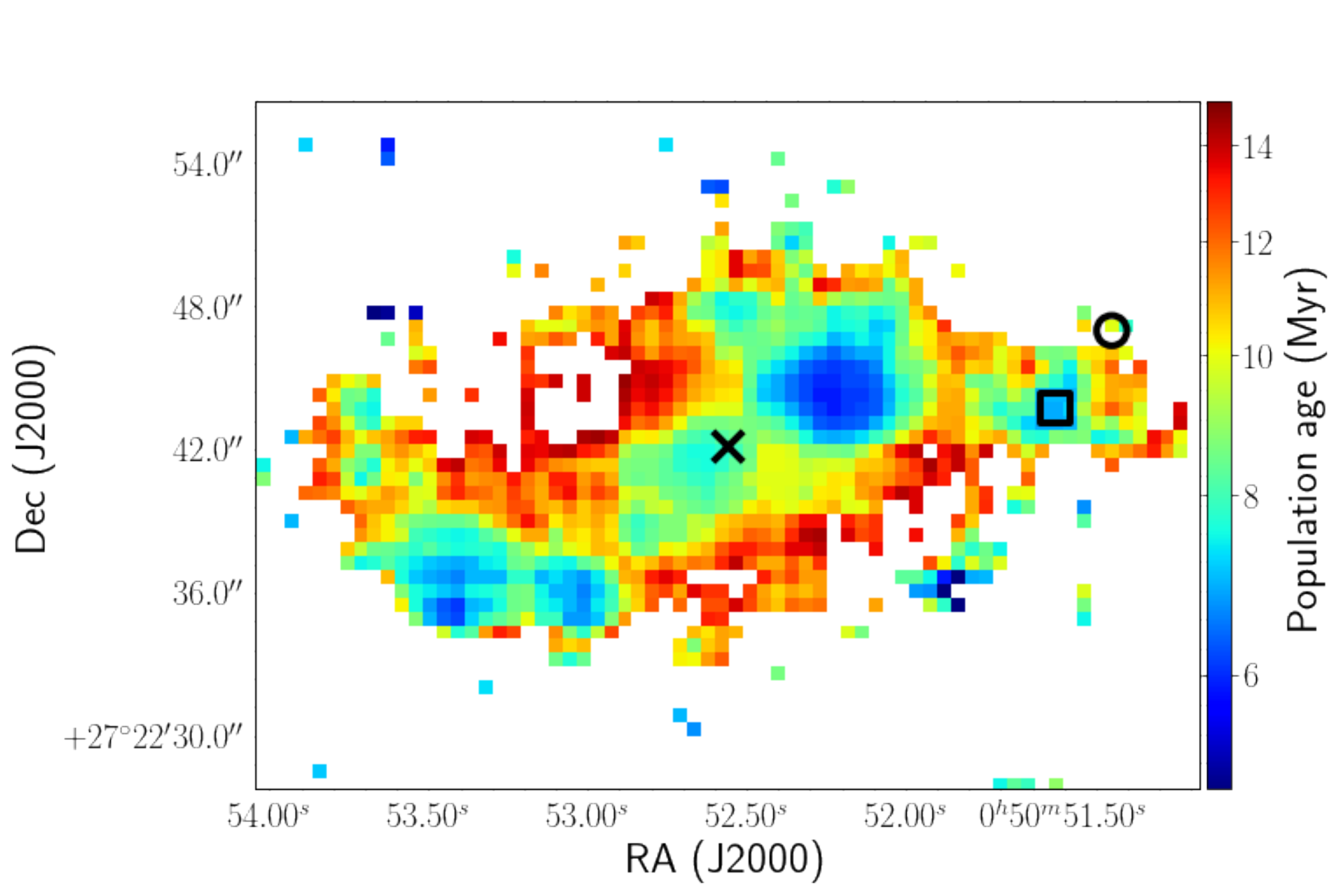}\\
\caption{Spatially resolved physical properties of the host, as derived from the IFU observations. (Top left) Star formation density derived from the H$\alpha$ flux map using the relations in \citealt{Kennicutt1998}. (Top right) Metallicity map of the host galaxy using the N2 index. (Bottom left) Metallicity map of the host galaxy using the D16 index (see text). (Bottom right) Population age inferred from the equivalent width of the H$\alpha$ emission line. In each of the panels, the cross denotes the cataloged position of the center of the host galaxy, the circle denotes the location of the transient and the square denotes the location of the nearest H-II region to the location of the transient.}
\label{fig:16hgs_IFUPhysMaps}
\end{figure*}

The metallicity maps of the host galaxy are shown in Figure \ref{fig:16hgs_IFUPhysMaps} for both the N2 index and the D16 scale. Owing to the relatively weaker strengths of the metal emission lines, the metallicity could be reliably measured only near the nucleus of the galaxy. In particular, it is easy to see that both the metallicity calibrations suggest that the host galaxy of iPTF\,16hgs is a low metallicity galaxy, consistent with our estimates from the spectrum of the host nucleus. There is an apparent offset between the D16 scale and the N2 index of $\approx 0.15$ dex, similar to the offset noted by \citealt{Kruhler2017} between the D16 scale and electron temperature based values. The gas phase metallicity exactly at the location of the transient could not be reliably measured in these observations due to the low flux of the metal emission lines. Nevertheless, given the observed metallicity gradients in galaxies \citep{Zaritsky1994}, the metallicity at the location of iPTF\,16hgs is likely to be lower than at the nucleus.\\

\subsubsection{Stellar population age}

Ages of the youngest stellar populations in a galaxy can be estimated using the equivalent widths of the Balmer emission lines. However, we note that these estimates can be heavily affected by various assumptions in the emission model (e.g. stellar multiplicity, metallicity). Regardless, such a comparison can provide estimates of the ages of the youngest stellar populations in the environment of the transient, which translates to an age estimate for the SN progenitor if the explosion was powered by the core-collapse of a massive star.\\

Following \citealt{Kuncarayakti2016,Kuncarayakti2017}, we use the simple stellar population (SSP) models from Starburst99 \citep{Leitherer1999} to translate our H$\alpha$ equivalent width measurements into an equivalent youngest stellar population age. Based on our metallicity measurements of the host galaxy (12 + $\log$(O/H) $\approx$ 8.2), we also fix the metallicity of the models to $Z = 0.008$ (the closest metallicity available in the model grid). The star formation is also assumed to occur in an instantaneous burst, with mass distributed according to a Salpeter IMF. The resulting maps of the stellar population age are shown in Figure \ref{fig:16hgs_IFUPhysMaps}.\\

Consistent with our expectations from the bright H$\alpha$ emission extending throughout the galaxy, the stellar population age maps consistently suggest that the host galaxy contains multiple clumps of young stellar populations coincident with the bright H II regions. In particular, the bright clump of H$\alpha$ emission located near the transient (offset by $\approx 1$ kpc and marked by a square) exhibits ages of $\lesssim 8$ Myr. The H$\alpha$ EW inferred age right at the location of the transient, however, suggests that iPTF\,16hgs exploded in an environment containing a stellar population with ages of $\gtrsim 11$ Myr. If iPTF\,16hgs originated in a core-collapse explosion of a massive star that was formed in its local explosion environment, the inferred population age puts a lower limit on the age of the progenitor of iPTF\,16hgs.\\

\subsection{Explosion site properties}
\label{sec:explSite}

In this section, we use both the IFU maps of the host galaxy as well as the deeper late-time LRIS spectrum at the location of the transient to measure the properties of the stellar population at the site of the explosion. First, the projected offset of the explosion site from the nucleus of the host galaxy (1.9 R$_{\textrm{eff}} \approx 6$ kpc) is typical of the host nucleus offsets of core-collapse SNe found by PTF (\citealt{Kasliwal2012a}; see also \citealt{Galbany2014}), while it is on the lower end of the distribution (cumulative fraction $\lesssim 50$\%) found in Type Ia SNe from PTF (\citealt{Lunnan2017}; L. Hangard et al. in prep.) . The host normalized offset is also  on the lower end (cumulative fraction $\lesssim 50$\%) of the distribution of both Type Ia SNe \citep{Lunnan2017} and short GRBs \citep{Fong2013}.  \\

Using the late-time LRIS spectrum which exhibits a number of galaxy emission lines, we measure a H$\alpha$ flux of $\approx 2.5 \times 10^{-17}$ ergs cm$^{-2}$ s$^{-1}$. This is consistent with the value estimated from the location of the transient in the IFU observations (Figure \ref{fig:16hgs_IFUmaps1}). At the redshift of the host galaxy, this corresponds to a H$\alpha$ luminosity of $\approx 1.6 \times 10^{37}$ ergs s$^{-1}$. Translated to an equivalent star formation rate using the relations in \citealt{Kennicutt1998}, we infer a star formation rate of $\approx 1.3 \times 10^{-4}$ M$_{\odot}$ yr$^{-1}$ within the 1\arcsec slit used for the observations. The measured H$\alpha$ luminosity is similar to the majority of H-II regions associated with Type Ib/c SNe \citep{Kuncarayakti2017,Crowther2013}. \\

As noted earlier, we find that the H$\alpha$ map of the host galaxy indicate the presence of a large H-II region with a young stellar population at a projected offset of $\approx 1$ kpc from the location of the transient (see square symbols in Figure \ref{fig:16hgs_IFUmaps1}). The integrated H$\alpha$ luminosity over this H-II region is $\approx 1.1 \times 10^{39}$ ergs s$^{-1}$, corresponding to a star formation rate of $\approx 0.01$ M$_{\odot}$ yr$^{-1}$. When compared to the typical H$\alpha$ luminosities of H-II regions hosting core-collapse SNe, this association lies on the brightest end of the observed distribution of all types of core-collapse SNe \citep{Kuncarayakti2017}. Hence, if iPTF\,16hgs was due to the core-collapse explosion of a massive star, it appears likely that the progenitor could have originated in this H-II region. Given the offset of $\approx 1$ kpc from the explosion site, the required systemic velocity of the progenitor would be $\sim 50$ km s$^{-1}$ assuming a progenitor lifetime of $\sim$ 20 Myr. Additionally, the observed offset would also be consistent (although at the higher end) with the distribution of offsets of stripped envelope SNe from their likely parent H-II regions \citep{Galbany2014}.\\

\section{Discussion}
\label{sec:discussion}

We have presented the discovery and follow-up observations of a double-peaked and fast evolving transient iPTF\,16hgs. Based on its photometric and spectroscopic properties, we have also shown that iPTF\,16hgs is unambiguously a member of the class of Ca-rich gap transients (as defined by \citealt{Kasliwal2012a}), as indicated by its rapid evolution, low peak luminosity, early nebular transition and [Ca II] emission dominated nebular phase. This makes the total number of such confirmed transients to nine. Nevertheless, there are some striking features of interest in this source that separate it from the other members of this class, in particular, its double-peaked light curve and its young star forming environment, and we discuss the implications of these below.\\

\subsection{The double-peaked light curve and implications on the nature of the explosion}

The overall light curve of iPTF\,16hgs (i.e. its main peak and subsequent decline) is well consistent with the other members of this class. In particular, our modeling suggests that the main peak can be well modeled by a $^{56}$Ni powered light curve with $\approx 0.4$ M$_{\odot}$ of ejecta and $\approx 8 \times 10^{-3}$ M$_{\odot}$ of radioactive $^{56}$Ni. This is similar to the explosion parameters estimated for other members of this class \citep{Kasliwal2012a,Valenti2014,Lunnan2017,Perets2010}. Despite its overall similarity to the class of Ca-rich gap transients, the double-peaked light curve of iPTF\,16hgs is unique amongst the members of this class. Hence, we now discuss the progenitor channels relevant for the various power sources that could power the first peak of the light curve.

\subsubsection{A thermonuclear detonation?}
We considered a radioactive powered scenario for the early peak in iPTF\,16hgs, and found that $\approx 0.01$ M$_{\odot}$ of $^{56}$Ni in the outer 0.05 M$_{\odot}$ of the ejecta can explain the early bump in the light curve. Interestingly, such configurations have been suggested for some physical scenarios relevant for the potential progenitors of Ca-rich gap transients. For example, double detonation models for Type Ia SNe invoke explosive ignition of a He layer on the surface of a carbon-oxygen WD that leads to the formation of iron group radioactive isotopes near its surface and subsequently, an explosive detonation of the entire star \citep{Kromer2010,Fink2010,Kromer2016}. The presence of such radioactive material close to the surface has been shown to lead bluer colors at early times \citep{Dessart2012, Piro2016, Shen2010}, consistent with the observed early blue colors of iPTF\,16hgs. \\

\begin{figure}
\includegraphics[width=\columnwidth]{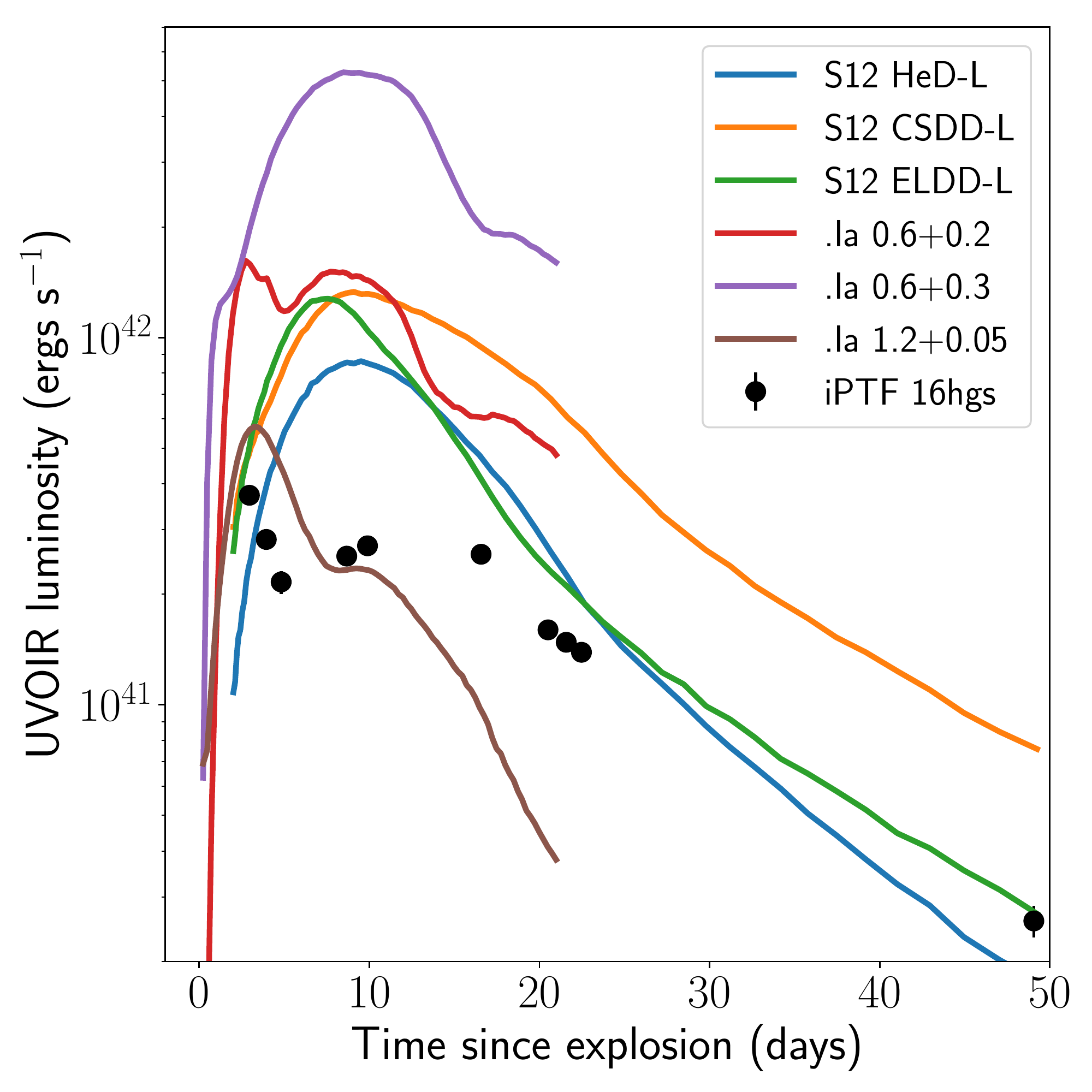}
\caption{Comparison of the pseudo-bolometric light curve of iPTF\,16hgs to the boloemtric light curve of models of He shell detonations (and double detonations) from \citealt{Sim2012} (denoted as S12 models), and with .Ia detonation models of \citealt{Shen2010} (denoted as .Ia models). For the \citealt{Sim2012} models, we show the low mass (L) models of only He shell detonations (HeD), converging shock (CS) double detonations and edge-lit (EL) double detonations. As shown, although the double-peaked behavior in iPTF\,16hgs is reminiscent of the .Ia detonation models, they are either too luminous or too rapidly evolving compared to the overall light curve of iPTF\,16hgs.}
\label{fig:16hgs_ddCompare}
\end{figure}

For the specific case of Ca-rich gap transients, a widely discussed progenitor channel involves the detonation of a He shell on the surface of a WD \citep{Perets2010,Waldman2011,Dessart2015a}. Such a configuration could arise from a close binary system with a CO WD that accretes He-rich matter from a He WD or a He-rich non-degenerate companion \citep{Bildsten2007,Shen2010,Waldman2011,Dessart2015a}. Numerical simulations for the expected optical signatures of these events were performed by \citealt{Shen2010} and \citealt{Sim2012}, and we show a comparison of these models to iPTF\,16hgs in Figure \ref{fig:16hgs_ddCompare}. Interestingly, as shown in Figure \ref{fig:16hgs_ddCompare}, \citealt{Shen2010} did find double peaked light curves in their simulations of He shell detonations for some combinations of core and shell masses. In particular,  their models suggest that the first peak arises out of radial stratification of short lived radioactive isotopes ($^{48}$Cr and $^{44}$Ti) in the outer ejecta, whereas the second peak is powered by $^{56}$Ni decay deeper in the ejecta.\\

Thus, qualitatively, the double peaked light curve of iPTF\,16hgs does appear to be consistent with some predictions of this model, noting that this would suggest that the first peak is likely powered by a different radioactive isotope than $^{56}$Ni. However, the .Ia detonation models presented in \citealt{Shen2010} involved detonations of typically low mass shells ($\lesssim 0.3$ M$_{\odot}$) so that overall, their light curves evolve faster (rising over $\approx 7 - 10$ days) and are brighter ($M_{peak} < -17$) than the light curve of iPTF\,16hgs. Even if the first peak is somewhat reproduced, the low shell masses lead to much faster evolving second peaks than iPTF\,16hgs. On the other hand, \citealt{Sim2012} did not find such double-peaked light curves in their 2D models. Although the timescales of the overall light curve of iPTF\,16hgs are similar to the \citealt{Sim2012} models, they are also brighter than that of iPTF\,16hgs. We thus find that while some features of the light curve of iPTF\,16hgs are reproduced in these models, they would require larger shell masses (than the \citealt{Shen2010} models) and smaller amounts of synthesized radioactive isotopes to explain the lower luminosity. \\ 

There are also a number of spectroscopic differences between the predictions of the He shell detonation models (as in \citealt{Shen2010} and \citealt{Sim2012}) and our observations. Spectroscopically, they find that the peak photospheric spectra are likely to be dominated by absorption lines of incomplete He burning products such as Ti II and Ca II (see also \citealt{Holcomb2013}), and most notably, lack lines of Si and Mg. This is different from our observations, where the spectra can be well modeled by prominent features of He I, Mg II, Si II and Ca II. While their spectra did not show He I lines, they did suggest that non-thermal excitation will likely lead to the production of He lines in these events (see also \citealt{Waldman2011,Dessart2015a}). For comparison, the .Ia detonation candidate OGLE-2013-SN-079 \citep{Inserra2015} did exhibit prominent Ti II and Ca II lines near peak light, unlike the Ca-rich transient SN 2005E, as noted in \citealt{Inserra2015}. While the highlighted differences may appear problematic to this interpretation, we caution that the .Ia detonation models shown for comparison involved simple simulations in 1-D spherical symmetry and the nucleosynthetic outcome may differ in more realistic 3-D simulations.\\

\subsubsection{A low luminosity core-collapse explosion?}

We showed that the early peak of iPTF\,16hgs can be well modeled by shock cooling of an extended progenitor star at the time of explosion. If the first peak was powered by shock cooling emission, the extended progenitor at the time of explosion would strongly argue for a core-collapse origin of the explosion. In fact, a core-collapse origin is plausible even if the early peak was radioactively powered, e.g., \citealt{Bersten2013} invoked a similar $^{56}$Ni clump near the surface to explain the double peaked light curve of SN 2008D, while \citealt{Drout2016} also suggested outward $^{56}$Ni mixing to explain the early blue bump in SN 2013ge.\\

In the case of a shock cooling first peak, the inferred parameters of the extended envelope ($M_e \approx 0.08$ M$_{\odot}$ and $R_e \approx 13$ R$_{\odot}$) provide important clues to the nature of the progenitor star. Since the \citealt{Piro2015} models used for this analysis is simplified and ignores the density structure of the envelope, these numbers are likely to be correct only to an order of magnitude \citep{Piro2017}. We note that such extended envelopes have indeed been previously inferred in several other stripped envelope SNe (e.g. \citealt{Arcavi2017,Taddia2016}), and are suggested to be associated with elevated mass loss prior to explosion, or formed due to binary interactions. In fact, studies of the pre-SN evolution of He stars suggest that they are capable of swelling significantly before core-collapse (up to radii $\sim 10 - 100$ R$_{\odot}$), consistent with such a scenario \citep{Woosley1995,Yoon2010}.\\

The main peak of the light curve suggests an ejecta mass of 0.4 M$_{\odot}$ and $^{56}$Ni mass of $\approx 8 \times 10^{-3}$ M$_{\odot}$, which is unusually low compared to the normal population of stripped envelope core-collapse SNe \citep{Drout2011,Taddia2017,Lyman2016a}. In particular, the low inferred ejecta mass would require significantly more stripping than observed in the typical population of stripped envelope SNe, either due to the presence of a compact companion or due to stripping by a companion in a very close orbit. Hence, we compare iPTF\,16hgs to models of ultra-stripped SNe arising from highly stripped massive star progenitors in close He star - neutron star (NS) binaries \citep{Tauris2013,Tauris2015}. \\

\citealt{Moriya2017} presented the expected light curves and spectra of ultra-stripped (Fe core-collapse) SNe in the context of systems that lead to double neutron star systems. However, their models did not explore SN explosions with ejecta masses as large as 0.4 M$_{\odot}$ (as in iPTF\,16hgs), although such ejecta masses are allowed by binary population synthesis models \citep{Tauris2015}. Thus, if iPTF\,16hgs originated in an ultra-stripped SN explosion from a He star - compact object binary, this would require either an initially more massive He star or a wider He star - NS binary separation than the systems simulated in \citealt{Moriya2017}, in order to explain the larger progenitor mass at the time of explosion. Nevertheless, we note that simulations of ultra-stripped explosions in \citealt{Suwa2015} did explore systems that produced $\approx 0.4$ M$_{\odot}$ of ejecta, and found synthesized $^{56}$Ni masses of $\approx 8 \times 10^{-3}$ M$_{\odot}$ in the explosion, very similar to our estimates for iPTF\,16hgs.\\

While the majority of ultra-stripped SNe are expected to be of Type Ic, more massive progenitors (as would be the case for iPTF\,16hgs) with larger He layers ($M_{He} \gtrsim 0.06$ M$_{\odot}$) may lead to He-rich Type Ib SNe \citep{Moriya2017,Hachinger2012}. Recently, \citealt{Yoshida2017} also showed that the nucleosynthesis in ultra-stripped explosions may produce ejecta that are particularly rich in isotopes of Ca, suggesting that the ultra-stripped interpretation may explain the Ca-rich nebular spectra as well. It is also important to note that the ejecta mass and He-rich spectra of iPTF\,16hgs would also be consistent with a core-collapse explosion in a close binary system of two non-degenerate massive stars \citep{Yoon2010}, where stripping by a close non-degenerate companion can lead to a similar highly stripped progenitor that retains a large amount of He in its outer layers. \\

Interestingly, the low peak luminosity of iPTF\,16hgs and the associated low inferred $^{56}$Ni mass, together with the peculiar signatures of nucleosynthesis (i.e. Ca-rich nebular spectra) is also reminiscent of models of electron capture SNe. Such SNe are initiated by the loss of pressure due to electron captures on to $^{24}$Mg and $^{20}$Ne in a degenerate O-Ne-Mg core of a massive star \citep{Nomoto1984}. While only single stars in the mass range of 8 - 12 M$_{\odot}$ are expected to undergo such an outcome, the mass range may be significantly extended when considering binary interactions \citep{Podsiadlowski2004}. Since stars in this mass range do not produce massive enough winds to remove their outer H layers, a stripped envelope SN such as iPTF\,16hgs would necessarily require a binary scenario to explain the observed SN. A similar scenario was also used to explain the Ca-rich SN 2005cz \citep{Kawabata2010}.\\

\citealt{Kitaura2006} presented simulations of such electron capture SNe and found explosion energies and $^{56}$Ni mass yields of $\sim 10^{50}$ ergs and $\sim 10^{-3}$ M$_{\odot}$. These are consistent with the properties of iPTF\,16hgs within a factor of a few. We also compare iPTF\,16hgs to the simulations of \citealt{Moriya2016} (hereafter M16), who investigated the expected signatures of stripped-envelope ECSNe with binary population synthesis models at solar and sub-solar metallicity. Specifically, they performed population synthesis simulations of binary massive stars that led to ECSN progenitors either via the merger of two initially less massive stars or due to close (Case B or Case C) stripping of an initially massive star by a non-degenerate companion. Note that \citealt{Tauris2015} also found ECSN progenitors in their simulations of He star - NS binaries, although the large stripping by the NS in a close orbit led to explosions that had significantly lower ejecta masses ($\lesssim 0.2$ M$_{\odot}$) than that inferred for iPTF\,16hgs and those presented in M16. \\ 

\begin{figure}
\includegraphics[width=\columnwidth]{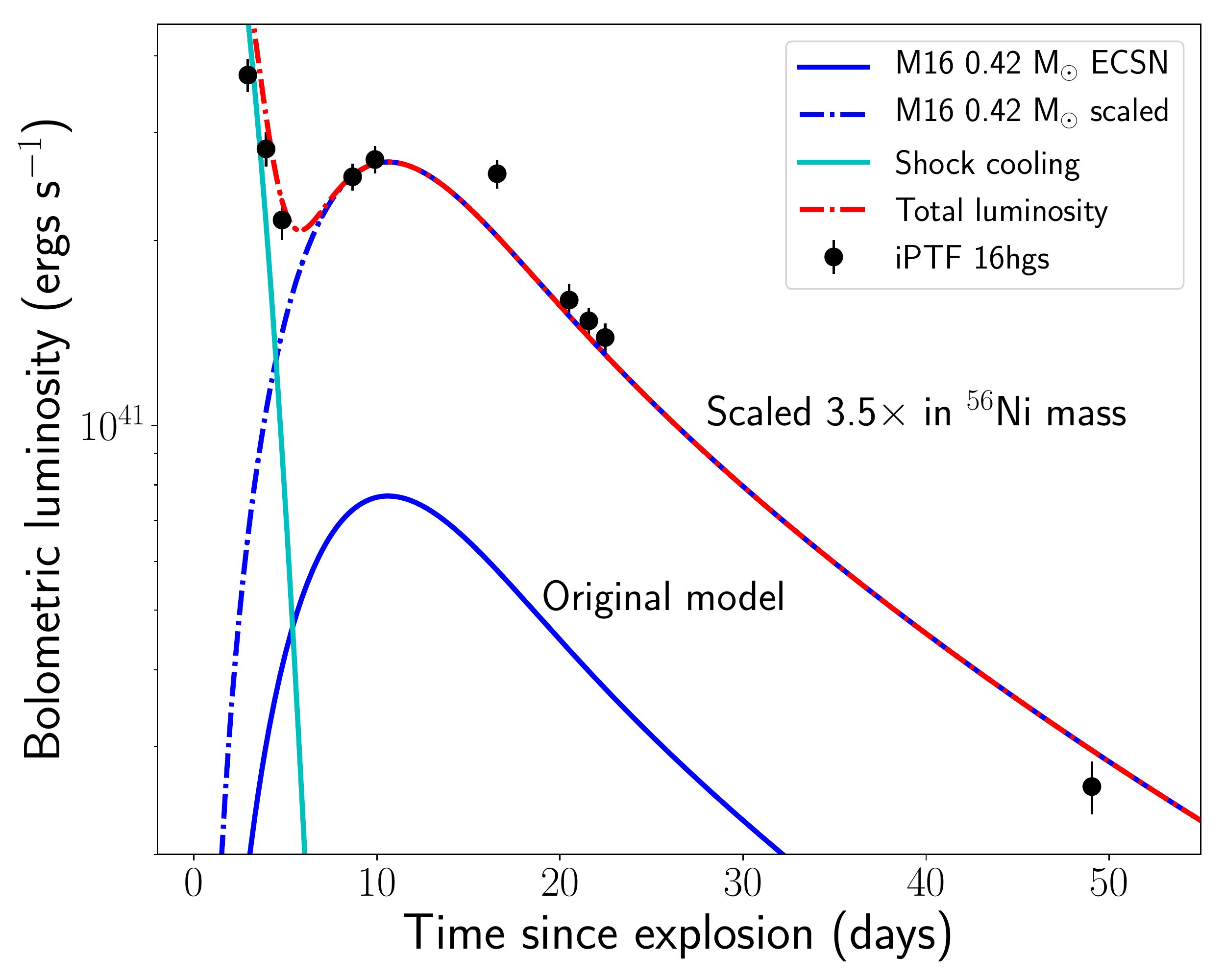}
\caption{Comparison of the bolometric light curve of iPTF\,16hgs to a model of a stripped envelope ECSNe at a sub-solar metallicity of $Z = 0.004$, as presented in \citealt{Moriya2016} (M16). The solid blue line shows the original model in M16 with an ejecta mass of 0.42 M$_{\odot}$ and $^{56}$Ni mass of $2.5 \times 10^{-3}$ M$_{\odot}$. The blue dashed line shows the same model scaled in luminosity by a factor of 3.5 to account for a higher $^{56}$Ni mass. The cyan line represents a shock cooling envelope model for an extended mass of $M_e = 0.05$ M$_{\odot}$ and radius of $R_e = 10$ R$_{\odot}$. The red line represents the total luminosity from the shock cooling and the scaled ECSN model.\\}
\label{fig:16hgs_ecsn}
\end{figure}

At sub-solar metallicity of $Z = 0.004$, the ejecta mass of iPTF\,16hgs is typical of the ejecta masses expected in these explosions, lying in the lower half of the distribution presented in M16. However, the peak bolometric luminosity ($\approx 3 \times 10^{41}$ ergs s$^{-1}$) is higher than that predicted by any of the models presented in M16, as the light curves presented in their work were much fainter (peak of $\approx 10^{41}$ ergs s$^{-1}$). We show a comparison of the bolometric light curves of stripped envelope ECSNe presented in M16 to iPTF\,16hgs in Figure \ref{fig:16hgs_ecsn}, for a model with 0.42 M$_{\odot}$ of ejecta (as derived from our Arnett fit) at a metallicity of $Z = 0.004$. As indicated earlier, the models presented in M16 included too little $^{56}$Ni ($2.5 \times 10^{-3}$ M$_{\odot}$ for the shown light curves) to account for the (main) peak luminosity of iPTF\,16hgs, so that the model prediction (solid blue curve) is much fainter than the data. We also show a comparison model to account for a higher $^{56}$Ni mass by scaling the luminosity of the original model by a factor of $\approx 3.5$. After adding a shock cooling component to the light curve (for $M_e = 0.05$ M$_{\odot}$ and $R_e = 10$ R$_{\odot}$) to account for the early declining emission, we find that the total predicted luminosity of the model (solid red curve) is consistent with the observations.\\

Taken at face value, we find that the explosion properties of iPTF\,16hgs are consistent (within a factor of a few) to the expected properties of highly stripped-envelope ECSNe in binary systems. Although the association may be reasonable given the uncertainties in the explosion properties of ECSNe \citep{Woosley2015}, we reiterate that current models of these explosions cannot explain the relatively high luminosity of iPTF\,16hgs. Nevertheless, \citealt{Kawabata2010} argued that such low luminosity core-collapse explosions (as SN 2005cz) from lower mass progenitors are likely associated with unique nucleosynthetic signatures, in particular, a higher abundance of Ca with respect to O, consistent with the Ca-rich nebular phase spectra of SN 2005cz.
Hence, the high [Ca II]/[O I] ratio in the nebular phase spectra of iPTF\,16hgs would be consistent with a low luminosity electron capture explosion of a stripped massive star \citep{Wanajo2013,Woosley2007,Nomoto2013,Sukhbold2016}.\\

\subsection{The local ISM constrained by radio observations}

We have used the non-detection of radio emission in iPTF\,16hgs to place stringent constraints on the environment of the progenitor using models of a spherical SN shock. Additionally, we have also used our radio limits to constrain the presence of a relativistic jet, as expected in some progenitor models for Ca-rich gap transients.\\

\subsubsection{A low density SN environment}

We discuss the CSM environment of the progenitor in the case where iPTF\,16hgs was purely powered by a spherical SN explosion. In this context, we note that core-collapse SNe exhibit a wide range of radio emission properties that reflect their diverse circumstellar mass-loss environments (e.g. \citealt{Soderberg2005,Weiler2011,vanDerHorst2011,Cao2013}). However, no Type Ia SNe, that likely share similar progenitor systems (i.e. WDs) as Ca-rich gap transients, have been detected in the radio band till date to very stringent limits. For instance, \citealt{Chomiuk2016} presented radio limits on a sample of Type Ia SNe and found that the radio non-detections constrained some of their environments to stringent limits of $\dot{M} \lesssim 10^{-9} \frac{v_{CSM}}{100 \textrm{km/s}}$ M$_{\odot}$ yr$^{-1}$. \citealt{Chomiuk2016} also presented radio limits on the Ca-rich gap transients SN 2005E \citep{Perets2010} and PTF 10iuv \citep{Kasliwal2012a} (in addition to some other `Ca-rich' transients), and constrained their mass loss environments to $\lesssim 4 \times 10^{-7}$ M$_{\odot}$ yr$^{-1}$ and $\lesssim 10^{-5}$ M$_{\odot}$ yr$^{-1}$ respectively. \\

\begin{figure}
\includegraphics[width=\columnwidth]{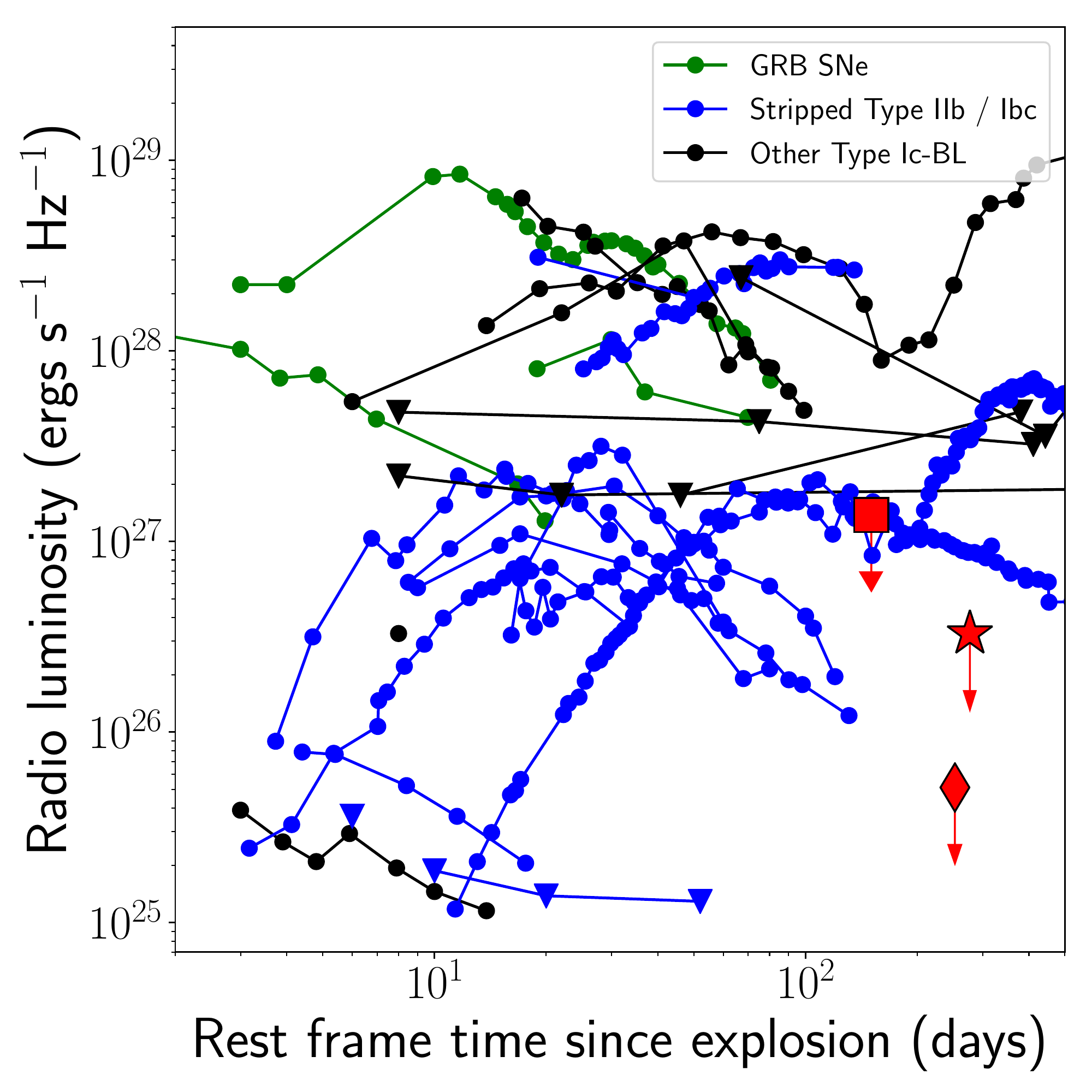}
\caption{Comparison of the radio luminosity limits on iPTF\,16hgs (symbols in red) to the radio light curves of core-collapse SNe. For the limits on iPTF\,16hgs, the red square shows the AMI limit, the red star shows the GMRT limit and the red diamond shows the VLA limit.\\}
\label{fig:16hgs_radioCompare}
\end{figure}

Taking standard values of the energy density fraction in the magnetic field of $\epsilon_B = 0.1$, we find that our radio limits constrain the mass loss environment of the progenitor to $\lesssim 2 \times 10^{-6} \frac{v_{CSM}}{100 \textrm{km/s}}$ M$_{\odot}$ yr$^{-1}$ for a wind-like CSM density profile, and to $n_e \lesssim 100$ cm$^{-3}$ for a constant density environment. While these limits are comparable to those obtained previously for Type Ia SNe and Ca-rich gap transients, they also imply a very low density environment in the context of a core-collapse explosion. We show a comparison of the radio limits on iPTF\,16hgs to the population of stripped and relativistic core-collapse SNe in Figure \ref{fig:16hgs_radioCompare}. As shown, these limits rule out radio emission similar to the majority of radio detected Type Ib/c SNe (see also Figure 16 in \citealt{Drout2016} and Figure 16 in \citealt{Milisavljevic2017}), that typically exhibit radio luminosities of $\gtrsim 10^{26}$ ergs s$^{-1}$ Hz$^{-1}$. However, these limits are consistent with the low density environments of stripped envelope SNe like the radio faint SN 2007gr \citep{Soderberg2010b} and SN 2002ap \citep{Berger2002}, along with the Type Ib/c SN 2013ge \citep{Drout2016} that remained undetected in the radio band. Additionally, these limits also rule out radio emission similar to that of late-time interacting events like SN 2007bg \citep{Salas2013} and SN 2014C \citep{Anderson2017}. \\

In the context of inferring the CSM environment of the progenitor, we also considered a scenario where the early peak in the optical light curve is powered by interaction of the ejecta with a dense CSM envelope around the progenitor. In order to power the early luminosity, we find that an ISM density $\rho \sim 3 \times 10^{-15}$ g cm$^{-3}$ ($n_e \sim 10^9$ cm$^{-3}$) is required for a constant CSM density environment, while a mass loss rate of $\dot{M} \sim 7 \times 10^{-5} \frac{v_{CSM}}{100 \textrm{km/s}}$ M$_{\odot}$ yr$^{-1}$ would be required for a constant mass loss environment. We note that this is at odds with CSM environment constraints derived from our late-time radio observations, which constrain the mass loss environment to $\dot{M} \lesssim 10^{-5}$ M$_{\odot}$ yr$^{-1}$ and $n_e \lesssim 10^3$ cm$^{-3}$ for a constant density environment, even for a $\epsilon_B = 0.01$. Indeed, this would be consistent with the elevated pre-explosion mass loss episodes inferred from very early observations of both H-rich and H-poor core-collapse SNe (\citealt{GalYam2014,Yaron2017}). \\

\subsubsection{Limits on a tidal disruption scenario}

Tidal detonations of low mass He WDs have been previously proposed as a potential progenitor channel for Ca-rich gap transients \citep{Sell2015}. In this context, \citealt{Foley2015} argued that the host offsets of Ca-rich gap transients appear to show a correlation with their radial velocity offsets (as inferred from the late-time nebular spectra) from their host galaxy, suggesting a nuclear origin for the progenitors of Ca-rich gap transients. This would be consistent, for instance, if a WD binary system (where the companion is a NS/BH) was hardened due to interaction with a central supermassive BH in the host galaxy, eventually leading to the tidal disruption of the WD \citep{Sell2015}. \\

However, \citealt{Milisavljevic2017} argue that the radial velocity argument (based on the blueshifts of the [Ca II] emission lines in \citealt{Foley2015}) may be flawed since the [O I] velocities do not show these systematic offsets even if the [Ca II] do exhibit them. In Figure \ref{fig:16hgs_caOProfiles}, we show the profiles of the [Ca II] $\lambda\lambda$ 7291, 7324 lines and the the [O I] $\lambda\lambda$ 6300, 6364 lines in the late-time spectra of iPTF\,16hgs, as a function of velocity offset from 7306 \AA\, and 6300 \AA\, respectively. It is interesting to note that the nebular emission features in iPTF\,16hgs show no evidence of a systematic offset in either the [Ca II] or [O I] lines.\\

We also compare the predictions of the models of \citealt{Macleod2016} to our data, who predicted the photometric and spectroscopic signatures of tidal disruptions of 0.6 M$_{\odot}$ WDs by an IMBH. In general, there are major differences between the predictions and our data, specifically that the time scales of the predicted light curves are longer while the predicted spectroscopic signatures are inconsistent. This is not surprising since the detonation of a 0.6 CO M$_{\odot}$ WD produces more massive ejecta (and hence a slower evolving light curve) and different nucleosynthetic signatures (dominated by Si) compared to the He-rich spectra of iPTF\,16hgs. Future modeling will be required to understand the properties of tidal disruptions of He-rich WDs, and if they can reproduce the observed signatures of iPTF\,16hgs.  Nevertheless, the first peak of the light curve remains to be explained in such a scenario, but could potentially be associated with an optical flare arising out of the initial rapid fall-back accretion on to the compact object.\\

As observed in some tidal disruption events of non-degenerate stars (e.g. \citealt{Gezari2012}), the production of a relativistic jet in the disruption was suggested to be a potential signature of such a progenitor channel \citep{Sell2015, Macleod2016}. Based on our multi-wavelength data set on iPTF\,16hgs we place stringent constrains on the presence of a relativistic jet in Section \ref{sec:radioJet}. In particular, we rule out the presence of a relativistic jet for near on-axis jet, unless the jet energy was very low ($\lesssim 10^{49}$ ergs, equivalent to $\lesssim 10^{-5}$ of the rest mass energy of a 0.4 M$_{\odot}$ WD). \\

Independently, we can try to place limits on the mass of the compact disrupting object using our X-ray limits on iPTF\,16hgs. The accretion flare coincident with the WD disruption is expected to be several orders of magnitude above the Eddington rate \citep{Rosswog2008, Macleod2014, Macleod2016}, and hence the X-ray emission along the axis of the jet is likely to be super-Eddington and tracing the $\propto t^{-5/3}$ accretion rate evolution of the BH \citep{Phinney1989}. Along all other directions, the X-ray luminosity is likely to be at least the Eddington luminosity $L_{Edd}$, where
\begin{equation}
L_{Edd} \approx 10^{41} \frac{M_{c}}{10^3 \textrm{M$_{\odot}$}} \textrm{ergs s$^{-1}$}
\end{equation}
and $M_c$ is the mass of the compact object. Hence, the non-detection of X-ray emission in our $Swift$ XRT observations constrains the mass of the compact object to be $M_c \lesssim 490$ M$_{\odot}$. However, such an interpretation may be complicated by possible effects of reprocessing layers in the accreting system, that can reprocess the high energy emission into optical / UV light (e.g. \citealt{Hung2017, Gezari2012, Miller2015}).  \\

\subsection{Ca-rich transients from multiple progenitor channels?}

Recent transient surveys have discovered a diverse population of events that can be classified as `Ca-rich' based on their nebular phase spectra. Apart from the members of the class of Ca-rich gap transients, notable examples include iPTF\,15eqv \citep{Milisavljevic2017} which exhibited a high [Ca II] / [O I] ratio in its nebular phase spectra and hence was suggested to be `Ca-rich', while its high luminosity and slow decline excludes it from the class of Ca-rich gap transients. \citealt{Fillipenko2003} also presented a set of events that appeared Ca-rich based on their nebular spectra, although no photometric information was available for comparison to Ca-rich gap transients.\\

\begin{figure}
\centering
\includegraphics[width=\columnwidth]{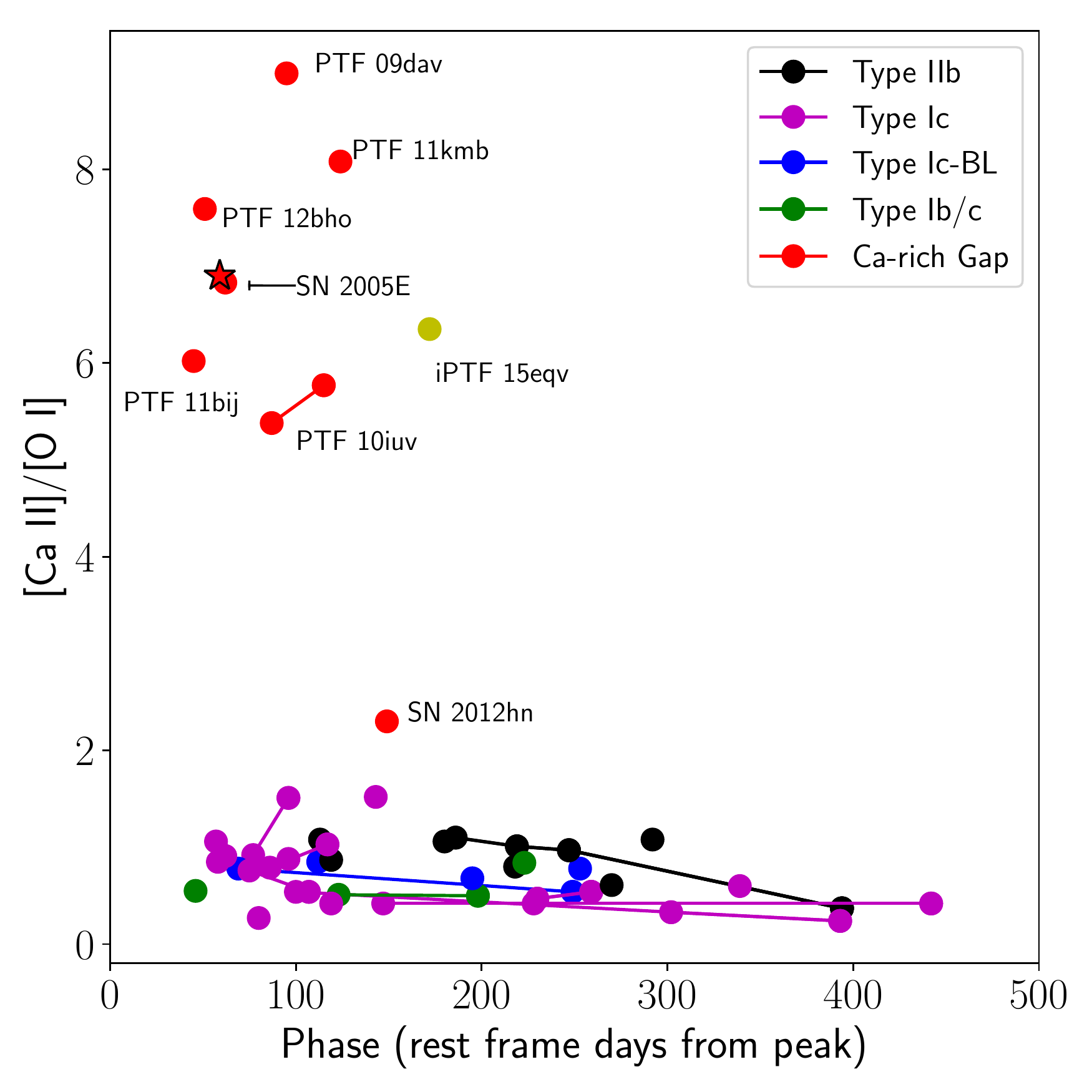}
\caption{Ca to O ratio as function of light curve phase for the PTF sample of stripped envelope SNe, together with previously known Ca-rich gap transients and iPTF\,16hgs (denoted by the star). The Ca-rich transients clearly occupy a unique phase space of high [Ca II]/[O I] in this plot.\\}
\label{fig:16hgs_CaO}
\end{figure}

Based on the general peculiarity of transients that exhibit strong [Ca II] emission in their nebular phase spectra, we searched the sample of spectroscopically classified stripped envelope SNe (Type IIb, Ib or Ic) discovered by PTF and iPTF (based on the compilation by Fremling et al. 2018) to look for objects that exhibit a high [Ca II] / [O I] ratio in their nebular phase spectra. We show our results in a plot of [Ca II]/[O I] ratio of the transient as a function of the phase from light curve peak in Figure \ref{fig:16hgs_CaO}, similar to Figure 11 in \citealt{Milisavljevic2017}. A major fraction of the events in this sample are Type Ic SNe, since these events were followed up to late times in the nebular phase. As shown, all the stripped envelope SNe in this sample exhibit low [Ca II]/[O I] ratio at all phases, consistent with the suggestion that the Ca-rich classification is not sensitively dependent on the phase of the transient \citep{Lunnan2017}.\\

The Ca-rich transients clearly stand out in this plot, occupying the phase space of high [Ca II]/[O I] ratio at early phases. Denoted as the star, iPTF\,16hgs is also consistent with the other members of this class. A notable exception is iPTF\,15eqv (denoted in yellow), which as mentioned earlier, exhibited a Ca-rich nebular phase spectrum but was not a Ca-rich gap transient photometrically. The lowest [Ca II]/[O I] ratio in the sample of Ca-rich gap transients is that of SN 2012hn \citep{Valenti2014}, which exhibited a [Ca II]/[O I] $\approx 2.3$ at a phase of $\approx 150$ days. In this context, \citealt{Milisavljevic2017} suggested that a [Ca II] / [O I] ratio of 2 separates the Ca-rich transients from the population of other stripped envelope SNe. \\

The class of Ca-rich gap transients, as defined by \citealt{Kasliwal2012a}, includes a unique subset of the broader class of Ca-rich transients that was empirically defined to include many faint and fast evolving SNe that exhibited conspicuous [Ca II] emission in their early nebular phase spectra. In particular, the definition did not constrain the properties of the photospheric phase spectra of these sources, which is likely to be an important indicator of the nature of the explosion. Hence, objects like PTF\,09dav \citep{Sullivan2011}
and PTF\,12bho \citep{Lunnan2017}
that exhibit significantly different photospheric phase spectra may arise from different progenitor channels than the majority of this class, which appear to exhibit He-rich Type Ib-like spectra near peak light, similar to the prototype event SN 2005E \citep{Perets2010}. \\

The discovery of iPTF\,16hgs adds further diversity to this class with its distinct double peaked light curve. If one were to invoke a similar progenitor channel for iPTF\,16hgs compared to the old inferred progenitor systems of other Ca-rich gap transients, we find that a thermonuclear detonation event from a WD binary system could still be a viable explanation that unifies this explosion to the rest of the class. In this case, the light curve of iPTF\,16hgs provides the first evidence of significant differences in the mixing of radioactive material in the detonation, providing strong constraints on the explosion mechanism. As the environment of the transient suggests its association with a young stellar population, this could perhaps also suggest that thermonuclear detonations occurring with shorter delay times lead to such highly mixed radioactive material in the explosion.\\

We also considered the scenario where the early emission in iPTF\,16hgs could be powered by the interaction shock between the ejecta and a non-degenerate companion. An attractive property of this interpretation is that it would naturally explain the uniqueness of iPTF\,16hgs in the class of Ca-rich gap transients, as this signature is expected to be prominent only if the observer is oriented along the direction of the companion. Based on this, \citealt{Kasen2010} estimates that only 10\% of Type Ia SNe should exhibit these early signatures. If Ca-rich gap transients arise from accreting WDs with non-degenerate companions, we would also expect only 1 in 10 of these transients to exhibit early excess emission, consistent with the discovery of iPTF\,16hgs. However, our analysis of the companion interaction model suggests that the color and luminosity evolution of iPTF\,16hgs are inconsistent with the models presented in \citealt{Kasen2010}. Future modeling will be required to understand if the assumptions of the companion interaction models for the massive ejecta in Type Ia SNe hold in low ejecta mass (and hence low optical depth) events with different ejecta compositions, as in iPTF\,16hgs.\\

Another widely discussed channel for the broader class of Ca-rich gap transients that may be relevant for iPTF\,16hgs is the disruption of a low mass WD by a NS or BH \citep{Macleod2016,Sell2015,Metzger2012,Margalit2016}. However, it has also been suggested that such a merger likely leads to an associated radio transient powered by the interaction of either a relativistic jet or a fast wind outflow with the surrounding CSM \citep{Macleod2016,Metzger2012,Margalit2016}. Based on our deep radio limits, we find that such a radio interaction signature can be hidden only if the explosion took place in a relatively clean environment ($n_e \lesssim 10^{-2}$ cm$^{-3}$) or if the jet energy was particularly low ($E \lesssim 10^{49}$ ergs), noting that the constraints are particularly stringent in the case of an on-axis event ($n_e \lesssim 10^{-6}$ cm$^{-3}$). If iPTF\,16hgs arises from a similar progenitor channel as those of other Ca-rich gap transients, our observations provide strong evidence against the formation of relativistic jet-like outflows in these explosions.\\

As noted in \citealt{Lunnan2017}, the host environments of Ca-rich gap transients are striking in light of their preference for hosts with old stellar populations in group and cluster environments. While iPTF\,16hgs was also found in a sparse galaxy group, its star forming dwarf host galaxy makes it stand out in this sample. Interestingly, the only other Ca-rich gap transient hosted in a star forming host galaxy was PTF\,09dav, which was also a spectroscopically peculiar event compared to the other members of this class. Unlike the majority of Ca-rich gap transients that do not show any evidence of underlying host systems down to stringent limits that rule out the presence of a dwarf galaxy or globular cluster \citep{Lyman2014,Lyman2016b,Lunnan2017}, iPTF\,16hgs clearly occurred inside its host galaxy, and is thus robustly associated with a host system.\\

Regardless, the low metallicity and young stellar population of the host galaxy still provides important clues to the nature of the progenitor star. In this context, we note that the remote locations of these transients in the outskirts of galaxies (where the metallicity is likely lower) have been used to argue for low metallicity progenitors of these explosions \citep{Yuan2013}. If the low metallicity is indeed a driving factor leading to these peculiar explosions, it is not surprising that the majority of Ca-rich gap transients are found in the metal-poor halos of old galaxies. Although iPTF\,16hgs was discovered at the smallest host offset of any known Ca-rich gap transient, it was found in a very low metallicity galaxy, and is thus consistent with the preference of these events for metal poor environments. In this context, it is also interesting to note that the host galaxy of SN 2005E \citep{Perets2010}, that was also discovered at a small projected offset, was a low metallicity galaxy (12 + $\log$(O/H) $\approx 8.4$; \citealt{Milisavljevic2017}), further strengthening the association of these transients to metal poor environments.\\

However, given that there are examples of core-collapse SNe like iPTF\,15eqv that exhibit high [Ca II]/[O I] ratio in their nebular phase spectra, and that iPTF\,16hgs occurred in a star forming environment, it is indeed plausible that iPTF\,16hgs is associated with a core-collapse explosion of a massive star. In this case, the low ejecta mass and low peak luminosity of the main peak makes it striking in the context of core-collapse explosions, suggesting its association with a highly stripped progenitor similar to those discussed in the context of stripped envelope electron capture explosions \citep{Yoon2010,Tauris2015,Moriya2017,Kitaura2006}. Although our radio limits suggest an unusually low density environment for the progenitor, the low final mass of the progenitor and the low metallicity of the progenitor environment may play a role in suppressing the formation of dense CSM environments by line-driven winds \citep{Crowther2013,Langer2012,Smartt2009}.  \\

\section{Summary \& Conclusions}
In this paper, we have presented the discovery and multi-wavelength follow-up observations of iPTF\,16hgs, a unique Ca-rich gap transient that exhibited a double-peaked light curve. The multi-wavelength properties of the transient can be summarized as follows.
\begin{itemize}
\item The light curve of the transient shows an initial blue decline, followed by rising to a second (main) peak on a timescale of $\approx 10$ days. The properties of the second peak are well consistent with other known Ca-rich gap transients.

\item The transient reached a peak $r$-band magnitude of $M_r \approx -15.65$, and to a bolometric luminosity of $\approx 3 \times 10^{41}$ ergs s$^{-1}$. On the other hand, the first peak was caught at discovery, declining from a bolometric luminosity of $\approx 4 \times 10^{41}$ ergs s$^{-1}$. 

\item Spectroscopically, the transient exhibited a He-rich peak photospheric spectrum with velocities of $\approx 8000 - 12000$ km s$^{-1}$ similar to a number of other Ca-rich gap transients, while it subsequently exhibited an early transition to a nebular phase dominated by strong [Ca II] emission.

\item No radio counterpart was detected in radio follow-up with the AMI, VLA and uGMRT down to 3$\sigma$ upper limits on the radio luminosity of $1.4 \times 10^{27}$ ergs s$^{-1}$ Hz$^{-1}$, $5.1 \times 10^{25}$ ergs s$^{-1}$ Hz$^{-1}$ and $3.3 \times 10^{26}$ ergs s$^{-1}$ Hz$^{-1}$ at 15 GHz, 10 GHz and 1.2 GHz respectively, at a phase of $\approx 250$ days after the explosion. These are the deepest radio limits published till date on radio emission from any Ca-rich gap transient.

\item X-ray follow-up with the \textit{Swift} XRT did not detect an X-ray counterpart to a 3$\sigma$ upper limit on the X-ray luminosity of $\approx 4.9 \times 10^{40}$ ergs s$^{-1}$.

\end{itemize}

Modeling the double-peaked light curve, we find that the main peak can be well described by a $^{56}$Ni powered light curve with $\approx 0.4$ M$_{\odot}$ of ejecta and $8 \times 10^{-3}$ M$_{\odot}$ of $^{56}$Ni. The first peak can be modeled either by $\approx 0.01$ M$_{\odot}$ of $^{56}$Ni mixed outwards into the surface of the progenitor star, or due to shock cooling of an extended envelope around the progenitor with a mass of $0.08$ M$_{\odot}$ and radius of $13$ R$_{\odot}$. We use the radio upper limits to constrain the environment of the progenitor, and find that the explosion occurred in a relatively `clean' environment ($\dot{M} \lesssim 2 \times 10^{-6}$ M$_{\odot}$ yr$^{-1}$ or $n_e \lesssim 150$ cm$^{-3}$) in the case of a standard spherical SN explosion. We also used these radio limits to place the first constraints on the presence of a relativistic jet as suggested in some models of Ca-rich gap transients, and find that the limits rule out a large parameter space of jet energies and ISM densities depending on the viewing angle of the observer.\\

We also presented the IFU observations of the host galaxy of iPTF\,16hgs, which are the first such observations of a Ca-rich gap transient. These observations suggest that iPTF\,16hgs occurred in the outskirts (at a projected offset of $\approx 6$ kpc $\approx 1.9$ R$_{\textrm{eff}}$) of a star forming spiral dwarf galaxy at a significantly sub-solar metallicity of $\approx 0.4$ Z$_{\odot}$. In particular, the young stellar population of the host galaxy and near the location of the transient is markedly different from the old environments of most Ca-rich gap transients. Hence, we find that both models of thermonuclear detonations on WDs (where the first peak is likely powered by outward mixed radioactive material) or the core-collapse of a highly stripped massive star (where the first peak is powered by cooling envelope emission or mixed radioactive material) are consistent with the observations and the host environment. \\

Taken together, we find that although the distinct properties of iPTF\,16hgs suggest that it may be an outlier in this observationally defined class, understanding the properties of this transient will be crucial to understand the physics of the broader class of Ca-rich explosions. Given the faint and fast evolving nature of these explosions, future and current wide-field transient surveys such as the Large Synoptic Survey Telescope and the Zwicky Transient Facility \citep{Bellm2017}
will be important to find more (and likely peculiar) examples of this intriguing class of explosions. This will shed light not only on their progenitors, but also help understand their distinct nucleosynthetic properties that likely play an important role in the chemical evolution of the universe \citep{Mulchaey2014,Frohmaier2018}.\\

\section*{Acknowledgements}
We thank C. Steidel, E. Kirby, K. Shen, T. Moriya, L. Bildsten, D. Kasen, A. Horesh and N. Stone for valuable discussions. We thank Ken Shen, Takashi Moriya and Stuart Sim for providing the comparison models presented in this paper. We thank Q. Ye, N. Blagorodnova, V. Ravi, S. Adams and R. Lau for assisting with the observations presented in the paper.\\

The Intermediate Palomar Transient Factory project is a scientific collaboration among the California Institute of Technology, Los Alamos National Laboratory, the University of Wisconsin, Milwaukee, the Oskar Klein Center, the Weizmann Institute of Science, the TANGO Program of the University System of Taiwan, and the Kavli Institute for the Physics and Mathematics of the Universe. This work was supported by the GROWTH (Global Relay of Observatories Watching Transients Happen) project funded by the National Science Foundation under PIRE Grant No 1545949. GROWTH is a collaborative project among California Institute of Technology (USA), University of Maryland College Park (USA), University of Wisconsin Milwaukee (USA), Texas Tech University (USA), San Diego State University (USA), Los Alamos National Laboratory (USA), Tokyo Institute of Technology (Japan), National Central University (Taiwan), Indian Institute of Astrophysics (India), Indian Institute of Technology Bombay (India), Weizmann Institute of Science (Israel), The Oskar Klein Centre at Stockholm University (Sweden), Humboldt University (Germany), Liverpool John Moores University (UK). \\

Some of the data presented herein were obtained at the W.M. Keck Observatory, which is operated as a scientific partnership among the California Institute of Technology, the University of California and the National Aeronautics and Space Administration. The Observatory was made possible by the generous financial support of the W.M. Keck Foundation. The authors wish to recognize and acknowledge the very significant cultural role and reverence that the summit of Mauna Kea has always had within the indigenous Hawaiian community. We are most fortunate to have the opportunity to conduct observations from this mountain. We thank the staff of the Mullard Radio Astronomy Observatory for their invaluable assistance in the commissioning and operation of AMI, which is supported by Cambridge University and the European Research Council under grant ERC-2012-StG-307215 LODESTONE. The National Radio Astronomy Observatory is a facility of the National Science Foundation operated under cooperative agreement by Associated Universities, Inc. We thank the staff of the GMRT that made these observations possible. The GMRT is run by the National Center for Radio Astrophysics of the Tata Institute of Fundamental Research. Part of this research was carried out at the Jet Propulsion Laboratory, California Institute of Technology, under a contract with the National Aeronautics and Space Administration. YCP is supported by a Trinity College JRF. \\

\facility{PO 1.2m, PO 1.5 m, DCT, Hale (DBSP, CWI), Keck-I (LRIS), Swift (XRT, UVOT), AMI, VLA, uGMRT}

\bibliographystyle{aasjournal}
\bibliography{iPTF16hgs}

\clearpage

\startlongtable
\begin{deluxetable*}{lcccccccc}
\tablecaption{Optical photometric follow-up of iPTF 16hgs (corrected for galactic extinction). Upper limits indicated are 5$\sigma$ upper limits in the respective bands.}
\tablehead{
\colhead{MJD} &
\colhead{Rest frame phase}  &
\colhead{Filter} &
\colhead{Magnitude} &
\colhead{Instrument}  &\\
\colhead{} &
\colhead{(days from explosion)} &
\colhead{} &
\colhead{} &
\colhead{} &
}
\startdata
57655.46 & $-35.50$ & $g$ & $>$ 20.42 & P48\\
57662.97 & $-28.11$ & $g$ & $>$ 21.30 & P48\\
57668.97 & $-22.21$ & $g$ & $>$ 20.84 & P48\\
57681.36 & $-10.03$ & $g$ & 18.57 $\pm$ 0.08 & P48\\
57682.37 & $-9.04$ & $g$ & 18.95 $\pm$ 0.09 & P48\\
57683.22 & $-8.20$ & $g$ & 19.27 $\pm$ 0.12 & P48\\
57684.36 & $-7.08$ & $g$ & 19.33 $\pm$ 0.17 & P48\\
57687.10 & $-4.39$ & $g$ & 19.42 $\pm$ 0.02 & P60\\
57688.40 & $-3.11$ & $g$ & 19.46 $\pm$ 0.04 & P60\\
57694.37 & 2.76 & $g$ & 19.74 $\pm$ 0.16 & P48\\
57695.15 & 3.53 & $g$ & 19.82 $\pm$ 0.05 & P60\\
57696.33 & 4.69 & $g$ & 19.98 $\pm$ 0.15 & P48\\
57697.36 & 5.70 & $g$ & 20.10 $\pm$ 0.16 & P48\\
57698.21 & 6.54 & $g$ & 20.43 $\pm$ 0.20 & P48\\
57699.08 & 7.39 & $g$ & $>$ 19.79 & P60\\
57699.31 & 7.62 & $g$ & 20.51 $\pm$ 0.16 & P48\\
57700.25 & 8.54 & $g$ & 20.70 $\pm$ 0.07 & P60\\
57700.35 & 8.64 & $g$ & 20.61 $\pm$ 0.28 & P48\\
57701.19 & 9.47 & $g$ & 20.86 $\pm$ 0.15 & P60\\
57701.19 & 9.47 & $g$ & $>$ 20.05 & P48\\
57702.18 & 10.44 & $g$ & $>$ 20.89 & P60\\
57704.22 & 12.45 & $g$ & $>$ 20.99 & P60\\
57706.10 & 14.30 & $g$ & $>$ 20.49 & P60\\
57709.30 & 17.44 & $g$ & $>$ 19.69 & P60\\
57710.09 & 18.22 & $g$ & $>$ 20.89 & P60\\
57717.16 & 25.17 & $g$ & $>$ 20.20 & P48\\
57722.16 & 30.09 & $g$ & $>$ 21.29 & P60\\
57723.16 & 31.07 & $g$ & $>$ 21.00 & P48\\
57728.23 & 36.06 & $g$ & $>$ 21.59 & P60\\
57734.17 & 41.90 & $g$ & $>$ 19.29 & P60\\
57655.50 & $-35.46$ & $r$ & $>$ 20.23 & P48\\
57663.01 & $-28.07$ & $r$ & $>$ 21.09 & P48\\
57669.01 & $-22.17$ & $r$ & $>$ 20.63 & P48\\
57681.32 & $-10.07$ & $r$ & 18.81 $\pm$ 0.10 & P48\\
57682.33 & $-9.08$ & $r$ & 18.94 $\pm$ 0.09 & P48\\
57683.26 & $-8.16$ & $r$ & 19.18 $\pm$ 0.12 & P48\\
57687.09 & $-4.40$ & $r$ & 18.99 $\pm$ 0.02 & P60\\
57687.32 & $-4.17$ & $r$ & 18.98 $\pm$ 0.11 & P48\\
57688.39 & $-3.12$ & $r$ & 18.85 $\pm$ 0.03 & P60\\
57694.33 & 2.72 & $r$ & 18.76 $\pm$ 0.06 & P48\\
57695.27 & 3.65 & $r$ & 18.80 $\pm$ 0.07 & P48\\
57696.30 & 4.66 & $r$ & 18.86 $\pm$ 0.08 & P48\\
57698.25 & 6.58 & $r$ & 19.11 $\pm$ 0.07 & P48\\
57699.08 & 7.39 & $r$ & 19.26 $\pm$ 0.11 & P60\\
57699.28 & 7.59 & $r$ & 19.24 $\pm$ 0.09 & P48\\
57700.25 & 8.54 & $r$ & 19.40 $\pm$ 0.03 & P60\\
57700.26 & 8.55 & $r$ & 19.38 $\pm$ 0.03 & P60\\
57700.30 & 8.59 & $r$ & 19.44 $\pm$ 0.12 & P48\\
57701.18 & 9.46 & $r$ & 19.42 $\pm$ 0.04 & P60\\
57701.23 & 9.51 & $r$ & 19.46 $\pm$ 0.13 & P48\\
57704.21 & 12.44 & $r$ & 20.05 $\pm$ 0.10 & P60\\
57706.10 & 14.30 & $r$ & 20.17 $\pm$ 0.19 & P60\\
57709.29 & 17.43 & $r$ & $>$ 19.96 & P60\\
57710.09 & 18.22 & $r$ & 20.46 $\pm$ 0.22 & P60\\
57711.13 & 19.24 & $r$ & 20.61 $\pm$ 0.08 & P60\\
57717.20 & 25.21 & $r$ & $>$ 20.90 & P48\\
57722.16 & 30.09 & $r$ & $>$ 21.06 & P60\\
57723.20 & 31.11 & $r$ & $>$ 21.03 & P48\\
57726.17 & 34.03 & $r$ & 20.84 $\pm$ 0.31 & P48\\
57728.23 & 36.06 & $r$ & 21.57 $\pm$ 0.27 & P60\\
57734.17 & 41.90 & $r$ & $>$ 18.96 & P60\\
57687.10 & $-4.39$ & $i$ & 18.79 $\pm$ 0.03 & P60\\
57688.40 & $-3.11$ & $i$ & 18.63 $\pm$ 0.03 & P60\\
57695.15 & 3.53 & $i$ & 18.48 $\pm$ 0.03 & P60\\
57699.08 & 7.39 & $i$ & 18.94 $\pm$ 0.11 & P60\\
57700.25 & 8.54 & $i$ & 18.90 $\pm$ 0.03 & P60\\
57700.26 & 8.55 & $i$ & 18.92 $\pm$ 0.03 & P60\\
57701.18 & 9.46 & $i$ & 18.95 $\pm$ 0.04 & P60\\
57702.17 & 10.43 & $i$ & 19.08 $\pm$ 0.05 & P60\\
57704.21 & 12.44 & $i$ & 19.61 $\pm$ 0.08 & P60\\
57706.10 & 14.30 & $i$ & 19.64 $\pm$ 0.10 & P60\\
57709.30 & 17.44 & $i$ & $>$ 19.69 & P60\\
57710.09 & 18.22 & $i$ & 20.11 $\pm$ 0.15 & P60\\
57711.13 & 19.24 & $i$ & 20.22 $\pm$ 0.07 & P60\\
57722.16 & 30.09 & $i$ & 20.76 $\pm$ 0.20 & P60\\
57728.23 & 36.06 & $i$ & 20.87 $\pm$ 0.15 & P60\\
57734.17 & 41.90 & $i$ & $>$ 18.39 & P60
\enddata
\label{tab:lc}
\end{deluxetable*}

\begin{table*}
\centering
\begin{tabular}{lccccc}
\hline
Observation Date & MJD & Rest frame phase & Telescope + Instrument & Range\\
& & (days from $r$ peak) & & (Observed {\AA})\\
\hline
2016 Oct 22.37 & 57683.36 & $-8.06$ & DCT + DeVeny & 3600 -- 7900\\
2016 Oct 26.21 & 57687.21 & $-4.28$ & P200 + DBSP & 3500 -- 10000\\
2016 Oct 31.29 & 57692.29 & 0.72 & Keck I + LRIS & 3500 -- 10200\\
2016 Nov 28.32 & 57720.32 & +28.28 & Keck I + LRIS & 3500 -- 10000\\
2016 Dec 29.71* & 57751.70 & +59.13 & Keck I + LRIS & 3500 -- 9900\\
\hline
\end{tabular}
\caption{Summary of spectroscopic observations of iPTF 16hgs. The observation marked by * was in the form of a slit mask to measure the redshifts of potential host galaxies near the transient.}
\label{tab:spectra}
\end{table*}

\begin{table*}
\centering
\begin{tabular}{ccc}\hline
Emission line & Rest wavelength (\AA) & Flux (10$^{-15}$ ergs cm$^{-2}$ s$^{-1}$) \\
\hline
H$\alpha$ & 6563 & 9.50 $\pm$ 0.37 \\
H$\beta$ & 4865 & 1.58 $\pm$ 0.27 \\
$[$S II$]$ & 6716 & 1.40 $\pm$ 0.08\\
$[$S II$]$ & 6731 & 1.08 $\pm$ 0.12\\
$[$N II$]$ & 6584 & 1.07 $\pm$ 0.14\\
$[$O II$]$ & 3727 & 8.27 $\pm$ 0.40\\
$[$O III$]$ & 5007 & 6.25 $\pm$ 0.41\\
$[$O III$]$ & 4959 & 1.86 $\pm$ 0.20\\
\hline
\end{tabular}
\caption{Extinction corrected emission lines fluxes in the spectrum of the nucleus of the host galaxy of iPTF\,16hgs. The fluxes were calculated by fitting a Gaussian profile to the emission line profiles, measuring the integrated flux under the profile.}
\label{tab:16hgs_hostFlux}
\end{table*}

\begin{table*}
\centering
\begin{tabular}{lcccc}
\hline
Object name & $\alpha$ (J2000) & $\delta$ (J2000) & Redshift & Offset (")\\
\hline
Obj1 & \ra{00}{50}{51.16} & \dec{27}{22}{45.19} & 0.362 & 4.12 \\
Obj2 & \ra{00}{50}{53.29} & \dec{27}{22}{36.53} & 0.017 & 27.70\\
Obj3 & \ra{00}{50}{53.26} & \dec{27}{23}{1.66} & 0.195 & 28.36\\
Obj4 & \ra{00}{50}{45.90} & \dec{27}{22}{49.64} & 0.160 & 73.17\\
Obj5 & \ra{00}{50}{52.56} & \dec{27}{24}{3.70} & 0.290 & 77.30\\
Obj6 & \ra{00}{50}{55.20} & \dec{27}{23}{55.47} & 0.210 & 84.43\\
Obj7 & \ra{00}{50}{55.52} & \dec{27}{24}{6.34} & 0.107 & 95.70\\
Obj8 & \ra{00}{50}{52.00} & \dec{27}{24}{26.22} & 0.107 & 98.57\\
Obj9 & \ra{00}{50}{45.33} & \dec{27}{21}{45.31} & 0.175 & 102.26\\ 
Obj10 & \ra{00}{50}{34.34} & \dec{27}{21}{45.60} & 0.170 & 235.51\\
Obj11 & \ra{00}{50}{33.34} & \dec{27}{21}{31.34} & 0.171 & 252.40\\
Obj12 & \ra{00}{50}{29.08} & \dec{27}{21}{33.40} & 0.170 & 306.36\\

\hline
\end{tabular}
\caption{Redshifts of galaxies near iPTF 16hgs as measured from our spectroscopic mask observation.}
\label{tab:16hgs_hostSpec}.
\end{table*}

\end{document}